\documentclass[12pt,a4paper]{article}
\usepackage{amsfonts,amsthm,amsmath,amssymb,upgreek,bm}
\usepackage{hyperref}
\usepackage[paper=letterpaper,margin=.85in]{geometry}
\usepackage{graphicx}
\usepackage{units}
\usepackage{float}
\usepackage[export]{adjustbox}
\usepackage{xcolor}
\usepackage[shortlabels]{enumitem}
\usepackage[mathscr]{euscript}
\usepackage[normalem]{ulem}
\usepackage{bm}
\usepackage{pgfplots}
\usepackage{slashed}
\usepackage{mathtools}
\usepackage{comment}
\usepackage{cite}

\newcommand{\myquad}[1][1]{\hspace*{#1em}\ignorespaces}
\renewcommand{\i}{\mathrm{i}}

\newcommand{\ra}{\rangle}
\newcommand{\lab}{\left|}
\newcommand{\rab}{\right|}

\numberwithin{equation}{section}

\def\bR {\mathbb{R}}
\def\bZ {\mathbb{Z}}

\def\ASTM{$\textrm{AdS}_3$$\times$$\textrm{S}^3$$\times$$\mathcal{M}_4$}

\newcommand{\pa}{\partial}

\def\be{\begin{equation}}
\def\ee{\end{equation}}
\def\ie{\begin{equation}\begin{aligned}}
\def\fe{\end{aligned}\end{equation}}
\def\ieg{\begin{equation}\begin{gathered}}
\def\feg{\end{gathered}\end{equation}}

\newcommand{\cH}{{\mathcal H}}

\newcommand{\cJ}{{\mathcal J}}
\newcommand{\cK}{{\mathcal K}}

\newcommand{\cN}{{\mathcal N}}

\newcommand{\cS}{{\mathcal S}}

\newcommand{\cZ}{{\mathcal Z}}

\newcommand{\AdS}{\mathrm{AdS}}
\newcommand{\Tr}{{\rm Tr\,}}

\newcommand{\adsS}{\mathrm{AdS}_3 \times \mathrm{S}^3}
\newcommand{\adsM}{\mathrm{AdS}_3 \times \mathrm{S}^3 \times \mathcal{M}_4}
\newcommand{\adsT}{\mathrm{AdS}_3 \times \mathrm{S}^3 \times \mathbb{T}^4}
\newcommand{\adsK}{\mathrm{AdS}_3 \times \mathrm{S}^3 \times \mathrm{K3}}

\newcommand{\symT}{\textrm{Sym}^N(\mathbb{T}^4)}
\newcommand{\symK}{\textrm{Sym}^N(\mathrm{K3})}
\newcommand{\symM}{\textrm{Sym}^N(\mathcal{M}_4)}

\newcommand{\orbifold}{(\mathrm{AdS}_3 \times \mathrm{S}^3)/\mathbb{Z}_k}
\newcommand{\orbifoldM}{(\mathrm{AdS}_3 \times \mathrm{S}^3)/\mathbb{Z}_k \times \mathcal{M}_4}
\newcommand{\orbifoldT}{(\mathrm{AdS}_3 \times \mathrm{S}^3)/\mathbb{Z}_k \times \mathbb{T}^4}
\newcommand{\orbifoldK}{(\mathrm{AdS}_3 \times \mathrm{S}^3)/\mathbb{Z}_k \times \mathrm{K3}}

\newcommand{\Kthree}{\mathrm{K3}}
\newcommand{\Tfour}{\mathbb{T}^4}
\newcommand{\Mfour}{\mathcal{M}_4}

\newcommand{\Res}{\mathop{\mathrm{Res}}}
\newcommand{\qb}{\bar{q}}
\newcommand{\yb}{\bar{y}}
\newcommand{\Lb}{\bar{L}}
\newcommand{\Jb}{\bar{J}}
\newcommand{\lb}{\bar{l}}
\newcommand{\hb}{\bar{h}}
\newcommand{\jb}{\bar{j}}
\newcommand{\PE}{\mathrm{PE}}

\newcommand{\Zhat}{\widehat{Z}}

\DeclareRobustCommand \ccdots{\mathinner{\cdotp\mkern-3mu\cdotp\mkern-3mu\cdotp}}

\begin{document}
\thispagestyle{empty}

\vspace*{2cm}
\begin{center}

{\bf {\LARGE AdS$_3$ Quantum Gravity and \\ Finite $N$ Chiral Primaries}}

\begin{center}

\vspace{1cm}

{\bf Ji Hoon Lee$^{a}$ and Wei Li$^{b,c}$}\\
 \bigskip \rm

\bigskip 

${}^a$Institut f\"{u}r Theoretische Physik, ETH Zurich,\\ CH-8093 Z\"urich, Switzerland

\bigskip

${}^b$Institute of Theoretical Physics, Chinese Academy of Sciences,\\ 100190 Beijing, China

\bigskip

${}^c$Peng Huanwu Center for Fundamental Theory, \\ 230026 Hefei, Anhui, China

\rm
  \end{center}

\vspace{1.5cm}
{\bf Abstract}
\end{center}
\begin{quotation}
\noindent

String theory on AdS$_3$ $\times$ S$^3$ $\times$ $\mathcal{M}_4$ provides a well-studied realization of AdS$_3$/CFT$_2$ holography, but its non-perturbative structure at finite $N \sim 1/G_N^{(3)}$ is largely unknown. A long-standing puzzle concerns the stringy exclusion principle: what bulk mechanism can reproduce the boundary expectation that the chiral primary Hilbert space of the symmetric orbifold contains only a finite number of states at finite $N$? In this work, we present a bulk prescription for computing the finite $N$ spectrum of chiral primary states in symmetric orbifolds of $\mathbb{T}^4$ or K3. We show that the integer spectrum at any $N$ is reproduced exactly by summing over one-loop supersymmetric partition functions of the IIB theory on (AdS$_3$ $\times$ S$^3$)/$\mathbb{Z}_k$ $\times$ $\mathcal{M}_4$ orbifolds and their spectral flows. Using the worldsheet in the tensionless limit, we verify that the terms appearing in our proposal coincide with the partition functions of these orbifold geometries and their asymmetric generalizations. These partition functions contribute with alternating signs due to BPS modes with negative conformal dimensions and charges in twisted sectors. The resulting alternating sum collapses via large cancellations to the finite $N$ polynomials observed in symmetric orbifold CFTs, providing a bulk explanation of the stringy exclusion principle. We identify different Stokes sectors where different infinite subsets of these geometries contribute to the path integral, and propose a classification as functions of the chemical potentials.

\end{quotation}

\setcounter{page}{0}
\setcounter{tocdepth}{2}
\setcounter{footnote}{0}

\newpage

\parskip 0.1in
 
\setcounter{page}{2}

\tableofcontents

\section{Introduction and summary}

The D1-D5 system, in the large $N$ limit, provides a well-studied realization of AdS$_3$/CFT$_2$ holography \cite{Maldacena:1997re}.

Specifically, the AdS$_3$/CFT$_2$ duality says that type IIB string theory on $\adsM$ with $Q_1, Q_5$ units of Ramond-Ramond flux is dual to a CFT whose moduli space contains the symmetric orbifold $\symM = \Mfour^{ N}/S_N$ with $N = Q_1 Q_5$ for $\Mfour = \Tfour$ and $N = Q_1 Q_5 + 1$ for $\Mfour = \Kthree$. This system can be rotated via the U-duality group $SO(5,5,\mathbb{Z})$ to a frame with $Q_1$ fundamental strings and $Q_5$ NS5-branes and, in particular, to a frame with $N = Q_1 Q_5$ fundamental strings and one NS5-brane.

Note that due to various subtleties concerning the global structure of the moduli space, it is not clear whether for a given $N$ the theory at the different points 
above are completely equivalent.
For example, the theory on a pure RR background might not be smoothly connected to the theory on a pure NS-NS background in the moduli space; and the theory with only one unit of NS flux has some structural differences from the one with higher number of NS fluxes.

The BPS spectrum is nevertheless expected to be protected and hence invariant across the moduli space. 
This is certainly known to be true at large $N$ \cite{Maldacena:1998bw,deBoer:1998kjm,Larsen:1998xm,Deger:1998nm} and has served as one of the first non-trivial checks of AdS$_3$/CFT$_2$.\footnote{See e.g.\ \cite{Gaberdiel:2025eaf} for a recent discussion of subtleties in relating the algebraic structure of the BPS spectrum on pure RR and NS-NS backgrounds at finite $N$.} For instance, chiral primary states in $\symK$ and Kaluza-Klein states in $\adsK$ share the large $N$ spectrum
\be \label{eq: KK intro}
Z_{\infty} = \prod_{n=1}^\infty \frac{1}{\left(1 - y^{n+1} \bar{y}^{n-1}\right) \left(1 - y^{n-1} \bar{y}^{n+1}\right) \left(1 - y^{n+1} \bar{y}^{n+1}\right) \left(1 - y^{n} \bar{y}^{n}\right)^{21}}
\ee
where the states are weighted by their R-charges $j,\jb$. In $\adsK$, these states are harmonic forms of the $\Kthree$ Laplacian $\Delta_{\Kthree}$. In the symmetric orbifold in the NS sector, the left-right chiral primary states are states whose dimension and R-charge satisfy $h = j$ and $\hb = \jb$.

However, non-perturbative aspects of the duality at finite values of $N \sim 1/G_N^{(3)}$ remain largely unknown. Let us highlight a well-appreciated, but yet unresolved, puzzle in AdS$_3$/CFT$_2$ at finite $N$ that can be stated at the level of the chiral primary spectrum \cite{Maldacena:1998bw,deBoer:1998us}:

In any 2D $\cN = (2,2)$ superconformal field theory, the unitarity bounds imply that there is an absolute cutoff
\be \label{eq: unitarity bound}
h, \, \hb, \, j, \, \jb \ \leq \ \frac{c}{6}
\ee
on the dimensions/charges of its left-right chiral primary states \cite{Boucher:1986bh,Lerche:1989uy}. Since our $\cN=(4,4)$ symmetric orbifold has $c=6N$, the quantum numbers of these states are bounded by $\leq N$ and, in particular, the Hilbert space $\cH_{cc}$ of chiral primary states in $\symM$ is finite-dimensional at finite $N$. The holographic expectation, that \textit{some} non-perturbative effect in string theory must place an upper bound on the AdS$_3$ string spectrum, is known as the stringy exclusion principle. Then the question for AdS$_3$/CFT$_2$ holography is the following: What is the \textit{bulk} explanation for the presence of only a finite total number of states in the chiral primary Hilbert space, in accordance with CFT expectations at finite $N$?

We can sharpen the puzzle by examining the structure of chiral primary spectra with respect to $N$, following the work of de Boer \cite{deBoer:1998us}. Let us define the partition function over the chiral primary Hilbert space $\cH_{cc}$ of $\symM$ in the NS sector as
\be \label{eq: chiral primary part fn intro}
Z_N = \Tr_{\cH_{cc}} \left( y^{2 J_0} \yb^{2 \Jb_0} \right)  = \lim_{q,\qb \to 0} \Tr_{\symM} \left( q^{L_0 - J_0} \bar{q}^{\bar{L}_0 - \bar{J}_0} y^{2 J_0} \yb^{2 \bar{J}_0} \right).
\ee
The large $N$ spectrum $Z_{\infty}$ for $\Mfour = \Kthree$ in \eqref{eq: KK intro} is
\be
Z_{\infty} = 1+23 y^2+300 y^4+2876 y^6+22450 y^8+150606 y^{10}+897464 y^{12}+4856776 y^{14} + \cdots 
\ee
where we set $y = \yb$ for convenience. By contrast, there are only a finite number of states in the BPS partition function $Z_N$ of $\symK$ at finite $N$:
\ie \label{eq: finite N spectra intro}
Z_1 &= 1 +  22 y^2+y^4 \\
Z_2 &= 1+23 y^2 +  276 y^4+23 y^6+y^8 \\
Z_3 &= 1+23 y^2+299 y^4 + 2554 y^6+299 y^8+23 y^{10}+y^{12} \\
Z_4 &= 1+23 y^2+300 y^4 +2852 y^6 + 19298 y^8+2852 y^{10}+300 y^{12}+23 y^{14}+y^{16} \\
Z_5 &= 1+23 y^2+300 y^4 +2875 y^6+22127 y^8 + 125604 y^{10}+22127 y^{12} \\
&\myquad[17]+2875 y^{14}+300 y^{16}+23 y^{18}+y^{20}
\fe
and so on. The BPS partition functions are palindromic polynomials because the chiral primary states of $\symK$ consist of fermion bilinears.

Two features are worth noting: (1) Observe that $Z_N$ agrees with the Kaluza-Klein spectrum $Z_\infty$ up to energies
\be
E = h+\bar{h} = j+\bar{j} \leq \frac{N}{2},
\ee
which is of order the central charge $c = 6 N \sim 1/G_N^{(3)}$. From the bulk perspective, the fact that the discrepancy occurs at order $c = 6N$ suggests that the bulk contributions that account for the difference between finite $N$ and large $N$ partition functions must be heavy enough to backreact on AdS$_3$. (2) A subset of the states at the center of the palindrome are of distinguished status. The cosmic censorship bound in AdS$_3$ to have charged black holes without naked singularity is \cite{Cvetic:1998xh}
\ie \label{eq: black hole threshold}
4 N \left( L_0 - J_0 \right) - (2 J_0 - N)^2 &\geq 0 \\
4 N \left( \Lb_0 - \Jb_0 \right) - (2 \Jb_0 - N)^2 &\geq 0
\fe
in terms of NS sector charges. For left-right chiral primary states, this bound is satisfied only if a state has $h = \hb = j = \jb = \frac{N}{2}$, in which case
\be
E = h+\bar{h} = j+\bar{j} = N.
\ee
These are the quantum numbers of states at the center of the palindrome, so (a subset of) these states should correspond to ``small'' chiral primary black holes. Their large $N$ degeneracy is given by $\exp (2 \sqrt{2} \pi \sqrt{N})$ for $\Tfour$ and by $\exp (4 \pi \sqrt{N})$ for $\Kthree$ \cite{Maldacena:1998bw}.

Combining the two, we find that the finite $N$ spectrum $Z_N$ deviates from $Z_{\infty}$ at energy $E = N/2$ of order the central charge $c$ but that it does so well before we reach the energy $E = N$ to excite small black holes. There is thus a growing wedge $N/2 < E < N$ of chiral primary states which are explained neither in terms of Kaluza-Klein states nor in terms of black holes. These are present in addition to the states in the even more mysterious wedge $N < E \leq 2 N$ whose upper bound is set by unitarity \eqref{eq: unitarity bound}.

What bulk contributions do we need to consider in addition to the Kaluza-Klein states, if not black holes? And what bulk mechanism can reproduce the boundary expectation that the chiral primary Hilbert space $\cH_{cc}$ of the symmetric orbifold contains only a finite number of states at finite $N$?

\subsection{Summary of results}

In this work, we present a bulk prescription for computing the finite $N$ spectrum $Z_N$ of chiral primary states in the symmetric orbifold.

We propose that the integer spectrum of chiral primary states of $\symM$, at any integer $N$, is reproduced exactly in terms of an infinite sum over the one-loop supersymmetric partition functions $\Zhat_{k}^{\mu}$ of a set of asymptotically-AdS$_3$ geometries:
\be \label{eq: main proposal intro}
Z_N(y,\yb) = \sum_{\mu \in \cS_{y,\yb}} \sum_{k=1}^\infty  \Zhat_{k}^{\mu}(y,\bar{y}).
\ee
We identify these geometries as $\orbifoldM$ and their asymmetric generalizations \cite{Martinec:2001cf,Gaberdiel:2023dxt}, which admit an interpretation in terms of spectral flows of $\orbifoldM$ that preserve NS-NS fermion boundary conditions.

We find that the finite $N$ partition function $Z_N$ receives contributions from different infinite sets $\cS_{y,\yb}$ of $\orbifoldM$ geometries, where the set depends discontinuously on the region in the $y,\yb$ fugacity space one works in. The bulk interpretation is that the supersymmetric partition function $Z_N$ of the IIB theory on asymptotically $\adsM$ backgrounds contains different Stokes sectors $\cS_{y,\yb}$ in which different infinite subsets of the spectrally-flowed $\orbifoldM$ saddles contribute to the path integral. We propose a classification of the Stokes sectors $\cS_{y,\yb}$ of the finite $N$ supersymmetric partition function $Z_N(y,\yb)$ of the IIB theory on $\adsM$ backgrounds in Section \ref{sec: finite N saddles}.

The sum \eqref{eq: main proposal intro} over these geometries results in a finite total number of states for the following reason: The BPS partition functions $\Zhat_{k}^{\mu}$ contain overall signs $(-1)^{k-1}$ that alternate with the orbifold number $k$. This happens because a finite number of BPS modes in the $\bZ_k$-twisted sector of the $\orbifoldM$ orbifold can be found to have, somewhat surprisingly, negative conformal dimensions and charges. While this feature is studied using worldsheet methods in our work, we expect the signs $(-1)^{k-1}$ to be produced, in a one-loop gravitational path integral computation, via the need to Wick rotate the integration contours for the negative modes.\footnote{As explained below, this phenomenon has a direct analogy in the context of giant graviton branes in higher-dimensional AdS/CFT \cite{Lee:2024hef}.} The alternating infinite sum over these saddles collapses, as a consequence of large cancellations, to the polynomials $Z_N$ at finite $N$ seen in \eqref{eq: finite N spectra intro}.

The result is that, non-perturbatively, one finds vastly fewer states than what one would expect from a semiclassical analysis around any given background. This provides a bulk explanation of the stringy exclusion principle for chiral primary states in AdS$_3$/CFT$_2$.

A central observation that enabled us to arrive at \eqref{eq: main proposal intro} was that the one-loop partition functions $\Zhat_{k}^{\mu}$ of geometries that sum up to the finite $N$ answer $Z_N$ are contained in the analytic data of the grand-canonical partition function
\be
\cZ(p,y,\yb) = \sum_{N=0}^\infty p^N Z_N(y,\yb)
\ee
of symmetric orbifolds. We find that properties of the sum over geometries, such as Stokes phenomena therein, can be stated in terms of the analytic properties of $\cZ$ on the grand-canonical $p$-plane as a function of the fugacities $y,\yb$.

Connections between the residues of grand-canonical partition functions and gravitational saddles were observed previously in \cite{DeLange:2018wbz,Eberhardt:2021jvj} in the AdS$_3$ context. Our approach was motivated by analogous observations in the context of giant graviton expansions in higher-dimensional AdS/CFT \cite{Gaiotto:2021xce,Lee:2022vig}.

\subsection{Comparison to higher-dimensional AdS/CFT}

It is useful to compare the situation here with that in AdS$_5$/CFT$_4$.

In $U(N)$ $\cN = 4$ super Yang-Mills, the finite $N$ spectrum starts deviating from that at large $N$ at energy $E$ of order $N \sim \sqrt{c}$. The discrepancy occurs due to the presence of trace relations between $U(N)$-invariants formed from the $N \times N$ matrix-valued fields. From the bulk perspective, the scaling $\sqrt{c}$ suggests that the bulk object responsible for the discrepancy is not heavy enough to backreact on AdS$_5$, at least until $E \sim c \sim N^2$ in the spectrum when their number would reach order $N$.

The bulk contributions that capture the effect of trace relations between BPS states at finite $N$ are, in AdS$_5$ $\times$ S$^5$, supersymmetric one-loop partition functions of D3 giant graviton branes in AdS$_5$ $\times$ S$^5$ \cite{Lee:2023iil}. The DBI+CS action of this brane has an even number of negative modes whose path integral contours need to be Wick-rotated. This causes the brane saddles to contribute to the full BPS partition function with signs $(-1)^k$ that alternate with the number $k$ of coincident giants \cite{Lee:2024hef} (see also \cite{Lee:2022vig,Eleftheriou:2023jxr,Beccaria:2024vfx,Gautason:2024nru,Deddo:2025lfm}). The supersymmetric one-loop partition functions of these branes can be matched with the spectrum of finite $N$ trace relations in the BPS sector and, in this sense, giant graviton branes can be viewed as the bulk manifestation of finite $N$ null states in $\cN = 4$ SYM. The resulting formula, known as the giant graviton expansion, expresses the finite $N$ BPS index of $\cN = 4$ SYM in terms of a sum over contributions from giant graviton branes as well as the Kaluza-Klein spectrum of AdS$_5$ $\times$ S$^5$ \cite{Imamura:2021ytr,Gaiotto:2021xce}.

Our formula \eqref{eq: main proposal intro} is the AdS$_3$/CFT$_2$ analog of the giant graviton expansion in higher-dimensional AdS/CFT. A key difference in AdS$_3$/CFT$_2$ is that, because the energy at which the finite $N$ relations of the symmetric orbifold starts to play a role is
\be
E \sim N \sim c \sim 1/G_N^{(3)},
\ee
the relevant contributions are full-fledged geometries $\orbifold$ each with its own Kaluza-Klein spectrum. We will argue in Section \ref{sec: states dual to orbifolds} that the one-loop supersymmetric partition functions of the IIB theory on $\orbifoldM$ backgrounds encode the spectrum of finite $N$ null states in the BPS sector of $\symM$.

\subsection{Relation to work in the literature}

\noindent \textit{Fuzzballs and geometric quantization}

Left-right chiral primary states in AdS$_3$/CFT$_2$ have been studied previously from the bulk perspective in terms of a set of so-called Lunin-Mathur geometries \cite{Lunin:2001jy,Lunin:2002bj,Kanitscheider:2007wq} and their geometric quantization \cite{Rychkov:2005ji,Krishnan:2015vha}. While the count of these classical BPS solutions has been shown to reproduce, e.g., the leading entropy $4 \pi \sqrt{N}$ for $\Mfour = \Kthree$, it is expected that an explicit consideration of quantum effects is required to apprehend subleading, fine-grained information associated to the chiral primary black hole. The problem, however, is that it is a priori unclear what bulk analysis one should even perform in order to find a microscopic result valid at finite $N$.

Our work provides an answer to this problem in the chiral primary sector of AdS$_3$/CFT$_2$. The quantum effects at one-loop and via the sum over geometries \eqref{eq: main proposal intro} qualitatively change the result. For example, it was known for some time that the $\orbifold$ solutions have, classically, $h = \hb = j = \jb = \frac{N}{2}(1 - \frac{1}{k})$ approaching the quantum numbers of the chiral primary black hole in the $k \to \infty$ limit \cite{Maldacena:2000dr,Balasubramanian:2000rt,Martinec:2001cf}. It will turn out that the aforementioned negative modes in the $\bZ_k$-twisted sector of $\orbifoldK$ orbifolds introduce one-loop quantum corrections to the quantum numbers:
\be
h = \hb = j = \jb = \frac{N}{2} \left( 1- \frac{1}{k} \right) + \frac{1}{4}(k-1)
\ee
such that $\orbifoldK$ geometries with $k \gtrsim \sqrt{2 N}$ have energies/charges that exceed those of the chiral primary black hole (see Section \ref{subsec: higher k}). The ``staggering'' of charges is important not only for recovering the exact degeneracies of chiral primary black holes in terms of $\orbifold$ but also for the effective truncation of the BPS spectrum beyond those charges.

\noindent \textit{Farey-tail expansion} 

It is known that the elliptic genus of $\symK$ admits a so-called Farey-tail expansion, which provides a bulk interpretation for the ``non-polar'' degeneracies on or above the black hole threshold \eqref{eq: black hole threshold} as an infinite, convergent sum over an $SL(2, \bZ)$ family of BTZ black hole geometries \cite{Dijkgraaf:2000fq,Manschot:2007ha,Kraus:2006nb}. Note, however, that the Farey-tail expansion does not provide a bulk interpretation for the ``polar'' states below the bound \eqref{eq: black hole threshold} describing many states under our consideration. The polar degeneracies are an input for constructing the Farey-tail, and it had been an open question to address whether polar states can be found purely via bulk methods \cite{Dijkgraaf:2000fq}.

The interesting case is, again, the chiral primary black hole. Its count seems to be retrieved both (1) in terms of the Farey-tail, i.e. a Rademacher-type expansion computing the degeneracies as an infinite convergent sum of non-integer terms, and (2) in terms of our formula \eqref{eq: main proposal intro}, which computes the degeneracies at each charge as a finite sum of integer terms. (1) has the property of retaining manifest modular covariance but the state-counting interpretation is absent due to the fact that truncating the sum gives a decimal approximation. (2) retains the state-counting interpretation at every step, where each $\orbifold$ in the sum contributes an integer number of states. Truncating the sum gives a partial integer count rather than a decimal approximation. Given that the proposed formula \eqref{eq: main proposal intro} reproduces the non-polar degeneracy of the chiral primary black hole, it is natural to ask whether our methods can be extended to the states above the black hole threshold. This point clearly deserves further investigation \cite{IP}.

\noindent \textit{Strings in the grand-canonical ensemble} 

In recent years, it has been proposed that string theory on $\adsT$ with one unit of NS5-brane flux is dual to a grand-canonical ensemble of symmetric orbifold CFTs, rather than being dual to a specific CFT of fixed central charge \cite{Eberhardt:2020bgq,Eberhardt:2021jvj,Aharony:2024fid}. 
For example, the exponentiated worldsheet partition function computes the grand-canonical partition function $\cZ = \sum_{N=0}^\infty p^N Z_N$ of $\symT$, and correlators in the symmetric orbifold receives contributions from worldsheets of different genera at a given order in the $1/N$-expansion. On the other hand, our formula \eqref{eq: main proposal intro} for the finite $N$ BPS spectrum of $\symM$ appears as that which would result from a computation involving the gravitational path integral with $N \sim 1/G_N^{(3)}$. While we do not provide a resolution, it is conceivable that the ensemble that is described by string theory on $\adsM$ changes abruptly as a function of the moduli.

\subsection{Outline of the paper}

Section \ref{sec: review} reviews relevant notions in the symmetric orbifold and its worldsheet dual.

In Section \ref{sec: grand canonical}, we describe the main observation relating the analytic structure of the grand-canonical partition function $\cZ$ to the sum over geometries for $Z_N$.

In Section \ref{sec: large N and spectral flows}, we show that the lowest grand-canonical residues correspond to the partition functions of spectrally-flowed $\adsM$.

In Section \ref{sec: orbifolds}, we show that the grand-canonical residues at higher $k$ coincide with the supersymmetric partition functions of strings on $\orbifoldM$ backgrounds and their asymmetric generalizations, under the assumption of a certain Gauss constraint. The asymmetric orbifolds admit an interpretation in terms of the spectral flows of $\orbifoldM$.

In Section \ref{sec: states dual to orbifolds}, we argue that a path integral quantization of BPS fluctuations of $\orbifold$ geometries, defined with rotated contours for the negative modes, produces bulk states that are holographically dual to the chiral primary states of $\symM$ that become null at a finite value of $N$.

In Section \ref{sec: finite N saddles}, we state our main proposal and present a classification for the Stokes sectors $\cS_{y,\yb}$ of the finite $N$ supersymmetric partition function $Z_N$ of the IIB theory on $\adsM$ backgrounds. We provide a derivation of the proposed formula and demonstrate a few explicit checks. We conclude with a list of open questions in the Discussion.

In Appendix \ref{appssec:N4SCA}, we review the $\cN = (4,4)$ superconformal algebra and its realization in terms of the $\mathbb{T}^4$ fields. 
In Appendix \ref{app: relevant formulas}, we write explicitly the grand-canonical residues identified with supersymmetric one-loop partition functions of the IIB theory on $\orbifoldM$ backgrounds in the NS sector. 
In Appendix \ref{app: further checks}, we provide further checks in different Stokes sectors $\cS_{y,\yb}$.

\section{Review: Symmetric orbifolds of $\mathcal{M}_4$} \label{sec: review}

In this section we review basic aspects of the symmetric orbifold of $\mathcal{M}_4$ that will be necessary for later.
For more complete reviews see e.g.\ \cite{Pakman:2009zz,Lunin:2000yv,Lunin:2001pw,David:2002wn,Dei:2019iym}.
\subsection{$\mathcal{M}_4$ and its symmetric orbifold}

Before orbifolding, the seed theory is a 2D sigma model with small $\mathcal{N}=(4,4)$ superconformal symmetry and target space $\mathcal{M}_4$.\footnote{In this paper, we focus on the small $\mathcal{N}=4$ theories, which correspond to $\mathcal{M}_4$ being $\mathbb{T}^4$ or K3. 
There is also a similar story with \textit{large} $\mathcal{N}=4$ superconformal symmetry, corresponding to $\mathcal{M}_4$ being $S^3\times S^1$.} 
For a review on the $\mathcal{N}=(4,4)$ superconformal algebra see App.~\ref{appssec:N4SCA}.
The small $\mathcal{N}=(4,4)$ superconformal symmetry requires that the target space $\mathcal{M}_4$ to be hyper-K\"ahler, namely
\begin{equation}
\mathcal{M}_4= \mathbb{T}^4 
\qquad \textrm{or} \qquad
\textrm{K3}\,,
\end{equation}
with Hodge diamond
\begin{equation}
\begin{array}{cccccc}
&& h^{0,0} && \\
& h^{1,0} && h^{0,1} & \\
h^{2,0} && h^{1,1} && h^{0,2} \\
& h^{2,1} && h^{1,2} & \\
&& h^{2,2} &&
\end{array}
\ = \ \begin{array}{cccccc}
&& 1 && \\
& 2 && 2 & \\
1 && 4 && 1 \\
& 2 && 2 & \\
&& 1 &&
\end{array} 
\quad (\mathbb{T}^4) \,, 
\quad 
\begin{array}{cccccc}
&& 1 && \\
& 0 && 0 & \\
1 && 20 && 1 \\
& 0 && 0 & \\
&& 1 &&
\end{array} 
\quad (K3) \,.
\end{equation}

For $\mathbb{T}^4$, the seed CFT consists of four bosons and four fermions, which generate the $\mathcal{N}=(4,4)$ superconformal symmetry with central charge $c_\textrm{L}=c_{\textrm{R}}=6$, see the review in App.~\ref{appssec:N4SCA}. 
The K3 can be viewed as the orbifold $\mathbb{T}^4/\mathbb{Z}_2$.

The symmetric orbifold of $\mathcal{M}_4$ is a 2D CFT obtained by taking the tensor product of $N$ copies of the seed theory and taking the quotient by the symmetric group $S_N$.
Equivalently, it can be viewed as a sigma model with target space
\begin{equation}
\textrm{Sym}^{N}(\mathcal{M}_4) =(\mathcal{M}_4)^{\otimes N}/S_N  \,.
\end{equation}

The field content of the symmetric orbifold of $\mathbb{T}^4$ is
\begin{equation}
\begin{aligned}
\textrm{Bosons:}\qquad &\vec{X}^{\beta A}=(X^{(1)\beta A},X^{(2)\beta A},\dots, X^{(N)\beta A})   \\
\textrm{Fermions:}\qquad &\vec{\psi}^{\alpha A}=(\psi^{(1)\alpha A},\psi^{(2)\alpha A}_2,\dots, \psi^{(N)\alpha A})   \\ &\vec{\bar{\psi}}^{\bar{\alpha}\bar{A}}=(\bar{\psi}^{(1)\bar{\alpha}\bar{A}},\bar{\psi}^{(2)\bar{\alpha}\bar{A}}_2,\dots, \bar{\psi}^{(N)\bar{\alpha}\bar{A}})  \,, \\   
\end{aligned}
\end{equation}
where $\alpha$ (resp.\ $\bar{\alpha}$) $=\pm$ is the spinor index of the $\mathfrak{su}(2)_\textrm{R}$ R-symmetries of the left-(resp.\ right-) moving  $\mathcal{N}=4$ SCA, $\beta$ (resp.\ $\bar{\beta}$) $=\pm$ is the spinor index of the $\mathfrak{su}(2)_{\textrm{o}}$ outer-automorphism of the left- (resp.\ right-) moving $\mathcal{N}=4$ SCA, and finally $A,\bar{A}=\pm$ are the flavor indices; for a more detailed explanation of the spinor index structure see App.~\ref{appssec:N4SCA}. 

The $\mathcal{N}=(4,4)$ superconformal symmetry is generated by currents built out of these free fields. 
For the left-movers, 
\begin{equation}
\begin{aligned}
T&=\tfrac{1}{2}\epsilon_{\beta_1\beta_2}\epsilon_{ A_1 A_2}:\partial \vec{X}^{\beta_1 A_1}\cdot \partial \vec{X}^{\beta_2 A_2}: 
+\tfrac{1}{2}\epsilon_{\alpha_1\alpha_2}\epsilon_{A_1 A_2}:(\partial\vec{\psi}^{\alpha_1 A_1}) \cdot \vec{\psi}^{\alpha_2 A_2}:\\
G^{\alpha\beta}&= \epsilon_{A_1 A_2}:\vec{\psi}^{\alpha A_1}\cdot \partial \vec{X}^{\beta A_2}:\\
J^{a}&=\tfrac{1}{2}\epsilon_{\alpha_3\alpha_1}\epsilon_{A_1 A_2}D^{(1/2)}(t^a)^{\alpha_3}{}_{\alpha_2}:\vec{\psi}^{\alpha_1 A_1}\cdot \vec{\psi}^{\alpha_2 A_2}: \,,
\end{aligned}
\end{equation}
where $\epsilon_{+-} = -\epsilon_{-+} =+1$.
The expressions for the right-movers are similar.

The bosons have the mode expansion
\begin{equation}
\partial X^{(i)\beta A}(z)=\sum_{n\in\mathbb{Z}}\frac{\mathtt{a}^{(i)\beta A}_n}{z^{n+1}} 
\,, \qquad 
\bar{\partial} X^{(i)\bar{\beta}\bar{A}}(\bar{z})=\sum_{n\in\mathbb{Z}}\frac{\bar{\mathtt{a}}^{(i)\bar{\beta}\bar{A}}_n}{\bar{z}^{n+1}} \,.
\end{equation}
The fermions can have two possible periodicities along the $S^1$ of the worldsheet cylinder: 
\begin{equation}
\psi^{(i)\alpha A}(e^{2\pi i}z)=e^{2\pi i \nu} \psi^{(i)\alpha A}(z)
\end{equation} 
with $\nu=0$ (R) or $\nu=\tfrac{1}{2}$ (NS), and similarly for $\bar\psi^{(i)\alpha A}$ with periodicity labeled by $\bar{\nu}$. 
The mode expansion for the fermions depends on whether they are in the NS or R sector:
\begin{equation}
\psi^{(i)\alpha A}(z)=\sum_{r\in\mathbb{Z}+\nu}\frac{\psi^{(i)\alpha A}_r}{z^{r+1/2}}
\,, \qquad
\bar{\psi}^{(i)\bar{\alpha}\bar{A}}(\bar{z})=\sum_{r\in\mathbb{Z}+\bar{\nu}}\frac{\bar{\psi}^{(i)\bar{\alpha}\bar{A}}_r}{\bar{z}^{r+1/2}}\,.
\end{equation}

The bosons and fermions  transform under the $\mathcal{N}=(4,4)$ SCA as given in \eqref{eq:N4a} and \eqref{eq:N4psi1}.
In particular, for later purposes we will need \begin{equation}\label{eq:N4psi}
[L_0\,,\, \psi^{(i)\alpha A}_{-r}]=r\, \psi^{(i)\alpha A}_{-r} 
\,, \qquad  
[J^{3}_0\,,\, \psi^{(i)\alpha A}_{-r}]=\tfrac{1}{2} \alpha \psi^{\alpha A}_{-r} \,,
\end{equation}
and similarly for the right-moving sector.
From now on, we will denote
\begin{equation}
J_0:=J^{3}_0
\,, \qquad
\bar{J}_0:=\bar{J}^{3}_0\,.
\end{equation}
We will also use $h$ and $j$ to denote the eigenvalues of $L_0$ and $J_0$:
\begin{equation}
L_0|\phi\rangle =h|\phi\rangle
\,, \quad
J_0|\phi\rangle =j|\phi\rangle\,,
\end{equation} 
and similarly for the right-moving sector.

The $\mathcal{N}=4$ superconformal algebra has an important automorphism 
called spectral flow. 
We will denote the spectral flow action on the $\mathcal{N}=4$ SCA in the left-moving sector of the symmetric orbifold CFT by $\sigma^{\eta}$ labelled by $\eta\in \mathbb{Z}$, and its action on the left-moving $\mathcal{N}=4$ SCA generators is given by 
\begin{equation}\label{eq:SpectralFlow}
\begin{aligned}
\sigma^{\eta}(L_n) &= L_n + \eta J^3_n + \tfrac{c}{24} \, \eta^2 \delta_{n,0} \,,\\ 
\sigma^{\eta}(J^3_n) & = J^3_n + \tfrac{c}{12} \eta \, \delta_{n,0} 
\,, \qquad 
\sigma^{\eta}(J^\pm_n)  = J^\pm_{n\pm \eta}  \,, \\
\sigma^\eta(G^{\alpha\beta}_r)  &=G^{\alpha\beta}_{r+\frac{\alpha}{2}\eta} \,.
\end{aligned}
\end{equation}
Similarly for the right-moving sector, with spectral flow labeled by $\bar{\eta}$. 
In particular, the spectral flow action on the left-moving zero modes is
\begin{equation} \label{eq: spectral flow zeromode}
\sigma^{\eta}(L_0) =  L_0 + \eta J_0 + \tfrac{c}{24} \eta^2 
\,,
\qquad 
\sigma^{\eta}(J_0) = J_0 + \tfrac{c}{12} \eta
\end{equation}
and similarly for the right-moving sector, with the spectral labeled by $\bar{\eta}$. 
The spectral flow $(\eta,\bar{\eta})$ acts on the modes of the fermions by
\begin{equation}
\sigma^\eta(\psi^{(i)\alpha A}_r)  =\psi^{(i)\alpha A}_{r+\frac{\alpha}{2}\eta}\,,
\,,\qquad
\sigma^{\bar{\eta}}(\bar{\psi}^{(i)\bar{\alpha}\bar{A}}_r)  =\bar{\psi}^{(i)\bar{\alpha}\bar{A}}_{r+\frac{\bar{\alpha}}{2}\eta}\,,
\end{equation}
and it doesn't affect the boson modes.
Therefore, spectral flow by 
even integers $\eta,\bar{\eta}$ preserves the periodicity of the fermions along $S^1$ of the worldsheet cylinder, whereas the flow by odd integers $\eta,\bar{\eta}$ modifies the periodicity around the spatial circle (${\rm R} \leftrightarrow {\rm NS}$).

\subsection{Untwisted and twisted sectors
}

We will consider orbifolding by the symmetric group $S_N$.
Let us first review basic facts about $S_N$.

The symmetric group $S_N$ is defined as the group of all permutations of $N$ distinct elements. 
The order of the group is $|S_N|=N!$. 
A $w$-cycle $\sigma_w \in S_N$ (or, a cycle of length $w$) is a special element of $S_N$ that cyclically permutes $w\leq N$ elements while leaving the remaining $N-w$ elements fixed. 
Any permutation $\rho \in S_N$ can be decomposed as a product of disjoint cycles, which do not share any elements:
\begin{equation}
\rho=(\sigma_{w_1})(\sigma_{w_2})\dots (\sigma_{w_I})
\,, \qquad
\sigma_{w_i}\in \bZ_{w_i}
\,, \quad 
\sum^{I}_{i=1} w_i=N \,.
\end{equation} 
This decomposition of $\rho$ is unique up to the ordering of the cycles and the cyclic rotation of the entries within each cycle. 
Each conjugacy class $[\rho]$ is uniquely characterized by its \emph{cycle shape}, namely the lengths and multiplicities of cycles in $\rho$: 
\begin{equation}\label{eq:rhoCC}
[\rho]=(1)^{N_1}(2)^{N_2}\dots (M)^{N_M}= \prod^{M}_{n=1}(n)^{N_n}
\qquad \textrm{with}\quad
\sum^{M}_{n=1}\, n N_n=N   \,, 
\end{equation}
where the notation $(w)$ refers to a $w$-cycle $(\sigma_w)$ for which we have ``forgotten'' the labels specifying which $w$ elements out of $N$ are being permuted.

The Hilbert space of the orbifolded theory contains both untwisted and twisted sectors, and all of them are invariant under the $S_N$ action. 
Different sectors are characterized by different conjugacy classes $[\rho]$ of $S_N$.
Recall that in gauge theory, the single-particle (resp.\ multi-particle) spectrum consists of single-trace (resp.\ multi-trace) operators.
In the symmetric orbifold, single-particle (resp.\ multi-particle) states correspond to $[\rho]$ consisting of only one (resp.\ more than one) non-trivial cycle(s).

Let us first focus on the single-particle spectrum. 
(The multi-particle spectrum can then be obtained from the single-particle one in a straightfoward manner.) 
The $w$-twisted sector corresponds to the conjugacy class 
\begin{equation}
[\rho]=(w)=(12\dots w)\,.
\end{equation}
In the $w$-twisted sector, for each field $\phi$ in the seed theory, there are now $w$ fields:
\begin{equation}
\phi^{(1)}(z)\,, \phi^{(2)}(z)
\,, \dots \,,
\phi^{(w)}(z)\,,
\end{equation}
one from each copy, with their mode expansions inherited from the one in the seed theory:
\begin{equation}
\phi^{(i)}(z)=\sum_{n \in \mathbb{Z}+\nu} \frac{\phi^{(i)}_{n}}{z^{n+h}}	\,,
\end{equation}
where $h$ is the conformal dimension of $\phi^{(i)}(z)$ and
\begin{equation}
\begin{aligned}
\nu=0: \qquad& \textrm{bosons, fermions in R sector}\\
\nu=\tfrac{1}{2}: \qquad& \textrm{fermions in NS sector} \,.
\end{aligned}
\end{equation}
In the untwisted sector, i.e. $(w) = 1$, the periodicity for the fields are 
\begin{equation}
\phi(e^{2 \pi \i} z)=(e^{2 \pi \i})^{-h-\nu}\phi(z)\,.
\end{equation}
However, in the $w$-twisted sector, the periodicity is modified to
\begin{equation}
\phi^{(i)}(e^{2\pi \i}z)=(e^{2 \pi \i})^{-h-\nu} \phi^{(\rho(i))}(z) \,,
\end{equation}
where $\rho=(w)=(12\dots w)$.

Out of these $w$ fields, define $w$ linear combinations:
\begin{equation}
\Phi^{j}(z):=\sum^{w}_{i=1} (\mathtt{q}_w)^{j (i-1)} \phi^{(i)}(z)
\,, \qquad
j=0,1,\dots, w-1 \,,
\end{equation}
where 
\begin{equation}
\mathtt{q}_w:=e^{2\pi \mathtt{i}/w}
\end{equation}
is the primitive $w$-th root of unity. 
The fields in the new basis satisfy the boundary condition 
\begin{equation}
\Phi^{j}(e^{2\pi \mathtt{i} }z)=(\mathtt{q}_w)^{-j-w(h+\nu)}
\Phi^{j}(z) \,.
\end{equation}
As a result, they have the mode expansion
\begin{equation}
\Phi^j(z)=\sum_{r\in \mathbb{Z}+ \frac{j}{w}+\nu} 	\frac{\Phi^{j}_{r}}{z^{r+h}} \,.
\end{equation}

Let us consider the $\mathbb{T}^4$ fermions in the $w$-twisted sector.
The fermions in the new basis are defined as
\begin{equation}
\Psi^{j\alpha A}(z):=\sum^{w}_{i=1} (\mathtt{q}_w)^{j (i-1)} \psi^{(i)\alpha A}(z) \,, \qquad
j=0,1,\dots, w-1\,.
\end{equation}
In the NS sector, the fermions in the new basis have periodicity
\begin{equation}
\Psi^{j\alpha A}(e^{2\pi i }z)=(\mathtt{q}_w)^{-j}
\Psi^{j\alpha A}(z) \,.
\end{equation}
Hence they have the mode expansion
\begin{equation}
\Psi^{j\alpha A}(z)=\sum_{r\in \mathbb{Z}+ \frac{j}{w}+\frac{1}{2}} 	\frac{\Psi^{j\alpha A}_{r}}{z^{r+\frac{1}{2}}} \,.
\end{equation}
In the end, the fermion modes are
\begin{equation}
\chi^{\alpha A}_{m-\frac{1}{2}+\frac{j}{w}}:=\Psi^{j\alpha A}_{m-\frac{1}{2}+\frac{j}{w}}
\,, \qquad
j=0,1,\dots, w-1
\,, \quad m\in \mathbb{Z} \,.
\end{equation}
In particular, the modes
\begin{equation}\label{eq:chimodes}
\chi^{\alpha A}_{-\frac{1}{2}+\frac{n}{w}} \qquad
\begin{cases}
\begin{aligned}
&n=0,1,2, \dots, \tfrac{w-1}{2} \quad  &w\textrm{ odd}\\
&n=0,1,2, \dots, \tfrac{w}{2}-1 \quad  &w\textrm{ even}\\
\end{aligned}
\end{cases}
\end{equation} 
change the eigenvalue of $L_0$ by
\ie
h \ &= \ \tfrac{1}{2}, \, \tfrac{1}{2}-\tfrac{1}{w}, \, \tfrac{1}{2}-\tfrac{2}{w}, \, \cdots, \, \tfrac{1}{2w} \qquad w\textrm{ odd}\\
h \ &= \ \tfrac{1}{2}, \, \tfrac{1}{2}-\tfrac{1}{w}, \, \tfrac{1}{2}-\tfrac{2}{w}, \, \cdots, \, \tfrac{1}{w} \qquad w\textrm{ even} \,.
\fe
We will use these modes to build chiral primary states in the NS sector.

\subsection{Chiral primary spectrum
} \label{subsec: chiral spectrum}

Consider the vacuum in the Ramond sector, with 
\begin{equation}
h - \tfrac{c}{24} =\bar{h} - \tfrac{c}{24}= 0
\end{equation} 
and arbitrary $j, \bar{j}$. 
The spectral flows with $\eta,\bar{\eta}=\pm 1$ bring the RR-vacuum to the following four types of $\tfrac{1}{2}$-BPS states in the NS-NS sector:
\begin{equation}
\begin{aligned}
(c,c): \qquad & h-j =\bar{h}-\bar{j}=0 \\
(c,a):\qquad & h-j =\bar{h}+\bar{j}=0 \\
(a,c):\qquad & h+j =\bar{h}-\bar{j}=0 \\
(a,a):\qquad & h+j =\bar{h}+\bar{j}=0
\end{aligned}
\end{equation}
where $c$ and $a$ stand for chiral and anti-chiral, respectively. 

Let us explain how to construct chiral primary states in Sym$^{N}(\mathbb{T}^4)$. 
Start from the NS ground state in the $w$-twisted sector:
\begin{equation}
|w\ra_{\textrm{NS}} := |w\rangle_{\textrm{L,NS}}\otimes     
|w\rangle_{\textrm{R,NS}}\,.
\end{equation}
We first explain the left-moving sector and the right-moving counterpart will work similarly. 
Let us first consider the left-moving part $|w\ra_{\textrm{L}} := |w\rangle_{\textrm{L,NS}}$, with\footnote{The conformal dimension in \eqref{eq:hforwL} can be computed in various ways, e.g.\ by applying an S-transformation of the twining character with an insertion of the $w$-cycle $\sigma_w$, Tr$[\sigma_w q^{L_0-\frac{c}{24}}]$, see e.g.\ \cite[Appendix A]{Gaberdiel:2018rqv}.}
\begin{equation}\label{eq:hforwL}
|w\rangle_{\textrm{L}}: \qquad 
\begin{cases}
\begin{aligned}
& h=\tfrac{w}{4}-\tfrac{1}{4w} \,, 
\quad \,\, 
j=0\,,  \qquad w \textrm{ odd} \qquad \\
& h=\tfrac{w}{4} \,, 
\quad\quad  \qquad 
j=\tfrac{1}{2}\,, \qquad w \textrm{ even}\,.
\end{aligned}    
\end{cases}
\end{equation}
Note that here the state $|w\rangle$ is the highest weight state of the spin-$j$ representation of the su$(2)_{\textrm{R}}$, e.g.\ for $w$ even, it is the state with magnetic quantum number $\tfrac{1}{2}$ in the su$(2)_{\textrm{R}}$ doublet.
Then we apply the left-moving fermions modes from $\mathbb{T}^4$ with the mode number satisfying
\begin{equation}
\begin{aligned}
\chi^{+\pm}_{-\frac{1}{2}+\frac{n}{w}} 
\quad 
&\quad \textrm{with} \quad
n=1,2,\dots,\tfrac{w-1}{2}\,, \quad w \textrm{ odd}
\\  \chi^{+\pm}_{-\frac{1}{2}+\frac{n}{w}} 
\quad  
&\quad \textrm{with} \quad
n=1,2,\dots,\tfrac{w}{2}-1 \,,\quad w \textrm{ even}\,.
\end{aligned}
\end{equation}
(Note that this collection of the mode numbers almost matches  the list \eqref{eq:chimodes}, with only $\chi^{+\pm}_{-\frac{1}{2}}$ modes missing.)  
Each such mode has conformal dimension and su$(2)_{\textrm{R}}$ charge
\begin{equation}
h=\frac{1}{2}-\frac{n}{w}\,, \qquad j=\frac{1}{2} \,.   
\end{equation}
The resulting state is
\begin{equation}
|w\rangle_{\rm L}^{\textrm{BPS}}=\begin{cases}
\begin{aligned}
(\chi^{++}_{-\frac{1}{2}+\frac{1}{w}} \chi^{+-}_{-\frac{1}{2}+\frac{1}{w}})  
\dots
(\chi^{++}_{-\frac{1}{2w}} \chi^{+-}_{-\frac{1}{2w}}) |w\rangle_{\textrm{L}} \qquad w \textrm{ odd}\,,\\
(\chi^{++}_{-\frac{1}{2}+\frac{1}{w}} \chi^{+-}_{-\frac{1}{2}+\frac{1}{w}})   
\dots
(\chi^{++}_{-\frac{1}{w}} \chi^{+-}_{-\frac{1}{w}}) |w\rangle_{\textrm{L}} \qquad w \textrm{ even}\,.\\
\end{aligned}    
\end{cases}
\end{equation}
Collecting all the conformal dimensions and the charges, we can check that for both $w$ even and odd, we have
\begin{equation}
h=j=\frac{w-1}{2}   \,. 
\end{equation}
Namely, this is a left chiral primary, and it is a highest weight state of the spin-$\tfrac{w-1}{2}$ representation of $\mathfrak{su}(2)_{\mathrm{R}}$.

Recall that from the list \eqref{eq:chimodes}, we still have 
\begin{equation}
\chi^{++}_{-\frac{1}{2}} 
\quad\textrm{and}\quad
\chi^{+-}_{-\frac{1}{2}} \,,
\end{equation}
which are chiral primary modes that we can apply on $|w\rangle_{\rm L}^{\textrm{BPS}}$. 
In total, we have four chiral primaries in the left-moving sector:
\begin{equation}\label{eq:BPSquartetSymN}
\begin{array}{ccc}
&\chi^{++}_{-\frac{1}{2}} 
\chi^{+-}_{-\frac{1}{2}}|w\rangle_{\textrm{L}}^{\textrm{BPS}}: \quad  h=j=\frac{w+1}{2} & \\
\chi^{++}_{-\frac{1}{2}} |w\rangle_{\textrm{L}}^{\textrm{BPS}}:\quad
h=j=\frac{w}{2}&  
& \chi^{+-}_{-\frac{1}{2}} |w\rangle_{\textrm{L}}^{\textrm{BPS}}:\quad h=j=\frac{w}{2}  \\
&|w\rangle_{\textrm{L}}^{\textrm{BPS}}:\quad h=j=\frac{w-1}{2}&  \,.
\end{array}     
\end{equation}

The right-moving sector proceeds in complete parallel, with the bottom state given by 
\begin{equation}
|w\rangle_{\rm R}^{\textrm{BPS}}=\begin{cases}
\begin{aligned}
(\bar\chi^{++}_{-\frac{1}{2}+\frac{1}{w}} \bar\chi^{+-}_{-\frac{1}{2}+\frac{1}{w}})  
\dots
(\bar\chi^{++}_{-\frac{1}{2w}} \bar\chi^{+-}_{-\frac{1}{2w}}) |w\rangle_{\textrm{R}} \qquad w \textrm{ odd}\,,\\
(\bar\chi^{++}_{-\frac{1}{2}+\frac{1}{w}} \bar\chi^{+-}_{-\frac{1}{2}+\frac{1}{w}})   
\dots
(\bar\chi^{++}_{-\frac{1}{w}} \bar\chi^{+-}_{-\frac{1}{w}}) |w\rangle_{\textrm{R}} \qquad w \textrm{ even}\,.\\
\end{aligned}    
\end{cases}
\end{equation}
After applying either
\begin{equation}
\bar{\chi}^{++}_{-\frac{1}{2}} 
\quad\textrm{or}\quad
\bar{\chi}^{+-}_{-\frac{1}{2}} \,,
\end{equation}
or both, we obtain the right-moving analogue of \eqref{eq:BPSquartetSymN}.
For later purpose, it will be useful to define the bottom component 
\be
|w\ra^{\rm BPS} = |w\ra_{\rm L}^{\rm BPS} \otimes |w\ra_{\rm R}^{\rm BPS} 
\,.
\ee

We will be interested in the state that is both left and right chiral primary. 
For $\mathbb{T}^4$, there are $4\times 4=16$ such BPS states for each $w$. 
For K3=$\mathbb{T}^4/\mathbb{Z}_2$, the $\mathbb{Z}_2$ projects out the odd-even and even-odd states,\footnote{The odd states are those in the middle line in \eqref{eq:BPSquartetSymN} and similarly for the right-movers.} therefore there are only $8$ such BPS states for each $w$, plus $16$ that come from the ($\mathbb{Z}_2$) twisted sector. 
For both $\mathbb{T}^4$ and K3, the top component in the left or right sectors of $\symT$ can be viewed as the result of applying the  R-current modes $J_{-1}^+$ or $\bar{J}_{-1}^+$  on $|w\ra^{\rm BPS} $:
\be
J_{-1}^+ |w\ra^{\rm BPS} = \chi_{-\frac{1}{2}}^{++} \chi_{-\frac{1}{2}}^{+-} |w\ra^{\rm BPS} , \quad \Jb_{-1}^+ |w\ra^{\rm BPS} = \bar\chi_{-\frac{1}{2}}^{++} \bar\chi_{-\frac{1}{2}}^{+-}|w\ra^{\rm BPS} \,,
\ee
and the top-top component
\begin{equation}
J_{-1}^+ \Jb_{-1}^+|w\ra^{\rm BPS} = (\chi_{-\frac{1}{2}}^{++} \chi_{-\frac{1}{2}}^{+-})( \bar\chi_{-\frac{1}{2}}^{++} \bar\chi_{-\frac{1}{2}}^{+-})|w\ra^{\rm BPS}\,.  
\end{equation}
These states are shared by $\mathbb{T}^4$ and $\Kthree$.

\subsection{Dijkgraaf-Moore-Verline-Verlinde formula
}

Let
\ie\label{eq:ZRseedDef}
\mathtt{z}_{\mathcal{M}_4}^{\rm R}(q,\bar{q},y,\bar{y}) &:= \Tr_{\mathcal{M}_4}^{\textrm{R}} \left[ (-1)^{F}q^{L_0 - \frac{c}{24}} \bar{q}^{\bar{L}_0 - \frac{c}{24}} y^{2 J_0} \bar{y}^{2\bar{J}_0} \right] 
\fe
be the partition function of a seed theory on $\mathcal{M}_4 = \mathbb{T}^4$ or K3 in the Ramond sector, where $F$ is the total fermion number.\footnote{Bose-Fermi cancellations do not occur in \eqref{eq:ZRseed} because $(-1)^F = e^{2 \pi i (J_0 + \Jb_0)}$ and states with different R-charges are refined by $y^{2 J_0} \bar{y}^{2\bar{J}_0}$.}
Suppose we can evaluate the seed theory partition function and obtain 
\begin{equation}\label{eq:ZRseed}
\mathtt{z}_{\mathcal{M}_4}^{\rm R}(q,\bar{q},y,\bar{y}) 
=\sum_{h,\bar{h},j,\bar{j}} c(h,\bar{h},j,\bar{j}) q^h \bar{q}^{\bar{h}} y^{2 j} \bar{y}^{2 \bar{j}} \,,
\end{equation}
where the left- and right-moving conformal dimensions $h, \bar{h}$ were taken to be the eigenvalues of $L_0 - \tfrac{c}{24}$ and $\bar{L}_0 - \tfrac{c}{24}$. 
We can then apply the formula of Dijkgraaf-Moore-Verlinde-Verlinde (DMVV) \cite{Dijkgraaf:1996xw} to obtain the partition function of $\symM$ 
\be\label{eq:ZRsymDef}
\begin{aligned}
Z_N^{\rm R}(q,\bar{q},y,\bar{y}) &:= \Tr_{\symM}^{\textrm{R}} \left[\varepsilon^F q^{L_0 - \frac{c}{24}} \bar{q}^{\bar{L}_0 - \frac{c}{24}} y^{2J_0} \bar{y}^{2\bar{J}_0} \right] \,,
\end{aligned}
\ee
where $\varepsilon$ is introduced for later convenience. Namely, the grand canonical partition function, defined as
\ie\label{eq:ZRGCEDef}
\cZ_{\textrm{R}}(p;q,\bar{q}, y,\bar{y}) &:= \sum_{N=0}^\infty p^N Z_N^{\rm R}(q,\bar{q},y,\bar{y}) \,,
\fe
can be expressed in terms of  the degeneracies $c(h,\bar{h},j,\bar{j})$ of the seed theory (see \eqref{eq:ZRseed}) as
\ie\label{eq:ZRGCE}
\cZ_{\textrm{R}}(p;q,\bar{q}, y,\bar{y}) &= \prod_{n=1}^\infty \prod_{\substack{h,\bar{h},j,\bar{j} \\ h - \bar{h} \in n \bZ}} \frac{1}{(1 -(-\varepsilon)^{\frac{1}{2}(1-\textrm{Sgn}[ c(h,\bar{h},j,\bar{j})])} p^n q^{\frac{h}{n}} \bar{q}^{\frac{\bar{h}}{n}} y^{2j} \bar{y}^{2\bar{j}})^{c(h,\bar{h},j,\bar{j})}} \,.
\fe
One can then obtain the finite-$N$ canonical partition function $Z_N^{\rm R}(q,\bar{q},y,\bar{y})$ as the coefficient of $p^N$ of the  grand-canonical partition function $\cZ_{\textrm{R}}(p;q,\bar{q}, y,\bar{y})$.

The DMVV formula \eqref{eq:ZRGCE} is of central importance to the symmetric orbifold theory and it has a nice interpretation in terms of second-quantized string theory. 
Therefore let us briefly review its derivation and in particular show how the structure of the symmetric orbifold gives rise to this simple-looking formula.

Denote the Hilbert space of the seed theory by $\mathcal{H}_{\mathcal{M}_4}$, then the Hilbert space of the symmetric orbifold theory is
\begin{equation}
\mathcal{H}_{\textrm{Sym}^N(\mathcal{M}_4)}    =\bigoplus_{[\rho]} \mathcal{H}^{C_{[\rho]}}_{[\rho]}\,,
\end{equation}
where $[\rho]$ is a conjugacy class of $S_N$ (see \eqref{eq:rhoCC}) and $\mathcal{H}^{C_{[\rho]}}_{[\rho]}$ is the subsector of the twisted sector Hilbert space $\mathcal{H}_{[\rho]}$ that is invariant under the centralizer $C_{[\rho]}$ of $[\rho]$.
For a general $[\rho]$ given in \eqref{eq:rhoCC}, its centralizer $C_{\rho}$ is
\begin{equation}\label{eq:Centralizer}
C_{[\rho]}=\bigotimes^M_{n=1} \left(S_{N_n} \rtimes (\mathbb{Z}_{n})^{N_n}\right)\,.
\end{equation}
Therefore the Hilbert space $\mathcal{H}^{C_{[\rho]}}_{[\rho]}$ can be decomposed into
\begin{equation}\label{eq:HSdecompose}
\mathcal{H}^{C_{[\rho]}}_{[\rho]}=\bigotimes^M_{n=1}S^{N_n}(\mathcal{H}^{\mathbb{Z}_n}_{(n)} ) \,, 
\end{equation}
where $\mathcal{H}_{(n)}$ is the Hilbert space of a single string on $\mathcal{M}_4\times S^1$ with winding number $n$ along $S^1$,  $\mathcal{H}^{\mathbb{Z}_n}_{(n)}$ is the $\mathbb{Z}_n$ invariant subsector of $\mathcal{H}_{(n)}$, and $S^N(\mathcal{H})$ is the symmetric tensor product of $\mathcal{H}$.
Namely, for each conjugacy class $[\rho]$ given in \eqref{eq:rhoCC}, the corresponding twisted sector  Hilbert space $\mathcal{H}^{C_{[\rho]}}_{[\rho]}$ is decomposed according to the decomposition of $[\rho]$ into disjoint cycles $(n)$. 
This corresponds to the breaking of a long string of length $N$ into shorter ones, with each factor $(n)$ corresponding to a shorter string winding around the $S^1$ $n \leq N$ times.

The Hilbert space $\mathcal{H}^{\mathbb{Z}_n}_{(n)}$ is the building block of the symmetric orbifold Hilbert space, therefore we will first compute its partition function. 
Given the partition function of the seed theory in \eqref{eq:ZRseed}, namely on the Hilbert space $\mathcal{H}$, the partition function on the Hilbert space  $\mathcal{H}_{(n)}$ is
\begin{equation}\label{eq:ZRHn}
\mathtt{z}_{\mathcal{H}_{(n)}}^{\rm R}(q,\bar{q},y,\bar{y}) 
=\sum_{h,\bar{h},j,\bar{j}} c(h,\bar{h},j,\bar{j}) q^{\frac{h}{n}} \bar{q}^{\frac{\bar{h}}{n}} y^{2 j} \bar{y}^{2 \bar{j}}\,,
\end{equation}
where the fractional modes $\tfrac{h}{n}$ appear since $\tau\rightarrow\tau/n$ due to string winding $n$ times along $S^1$.
Next, the $\mathbb{Z}_n$ projection is implemented by the insertion of the projection operator $\tfrac{1}{n}\sum^{n-1}_{k=0} \hat{\rho}^k$ in the trace, where $\hat{\rho}$ acts by $\phi_i\rightarrow \phi_{i+1}$ for $i=1,2,\dots,n$, where $\phi_i$ are the fields in the $i^{\textrm{th}}$ sector in the Hilbert space $\mathcal{H}_{(n)}$ --- in terms of $(q,\bar{q})$, $\hat{\rho}$ acts by $(q,\bar{q})\rightarrow  (e^{2\pi i }q ,  e^{-2\pi i }\bar{q})$.
Then using the fact that 
\begin{equation}
\frac{1}{n}\sum^{n-1}_{k=0} e^{2 \pi i (h-\bar{h})k/n} =\delta_{h-\bar{h}, n m}\qquad m\in \mathbb{Z}\,, 
\end{equation}
we see that the $\mathbb{Z}_n$ invariant subsector of $\mathcal{H}_{(n)}$ is given by those states for which the difference between the sum of left-moving fractional mode numbers and the sum of right-moving ones is an integer, and hence
\begin{equation}\label{eq:ZRHnZn}
\mathtt{z}_{\mathcal{H}^{\mathbb{Z}_n}_{(n)}}^{\rm R}(q,\bar{q},y,\bar{y}) 
=\sum_{\substack{h,\bar{h},j,\bar{j} \\ h - \bar{h} \in n \bZ}}
c(h,\bar{h},j,\bar{j}) q^{\frac{h}{n}} \bar{q}^{\frac{\bar{h}}{n}} y^{2 j} \bar{y}^{2 \bar{j}}\,.
\end{equation}

Finally, we need to compute the partition function on the symmetric tensor product Hilbert space $S^{N_n}(\mathcal{H}^{\mathbb{Z}_n}_{(n)} )$ in \eqref{eq:HSdecompose}.
If the canonical ensemble partition function of a theory with Hilbert space $\mathcal{H}$ is 
$\mathtt{z}_{\mathcal{H}}(\mathtt{q})=\sum_\mathtt{m} c(\mathtt{m}) \mathtt{q}^\mathtt{m}$,
where $\mathtt{q}$ and  $\mathtt{m}$ denote the fugacities and modes collectively, then the grand canonical ensemble partition function of the symmetric tensor product theory is
\begin{equation}\label{eq:onecycle}
\sum^{\infty}_{N=0} \zeta^N \mathtt{Z}_{S^N(\mathcal{H})}(\mathtt{q})
=\prod_{\mathtt{m}}\frac{1}{(1-(-\varepsilon)^{\frac{1}{2}(1-\textrm{Sgn}[ c(\mathtt{m})])}\zeta \mathtt{q}^{\mathtt{m}})^{c(\mathtt{m})}}   \,.
\end{equation}
Now, apply \eqref{eq:onecycle} to the space $S^{N_n}(\mathcal{H}^{\mathbb{Z}_n}_{(n)} )$: namely in \eqref{eq:onecycle}, substitute $N$ by $N_{n}$, $\mathcal{H}$ by $\mathcal{H}^{\mathbb{Z}_n}_{(n)}$, $\zeta$ by $p^{n}$, and $c(\mathtt{m})$ by $c(h,\bar{h},j,\bar{j})$ of \eqref{eq:ZRHnZn}. Finally, tensoring all sectors labeled by $n$ together, we obtain the grand canonical ensemble partition function for the symmetric orbifold theory \eqref{eq:ZRGCE}.\footnote{We neglect sectors of the torus CFT with non-trival momentum and winding.} 

The full partition function $\cZ^{\textrm{R}}(p;q,\bar{q}, y,\bar{y})$ in the grand-canonical ensemble can be projected onto that $\cZ^{\textrm{R}}(p;y,\bar{y})$ computed only over the Ramond \textit{ground states} by taking the limit 
\be
q, \bar{q} \to 0 \,.
\ee
Since the eigenstates of the Laplacian on $\mathcal{M}_4 = \mathbb{T}^4$ or K3 with zero eigenvalue have multiplicities given by the Hodge numbers $h^{r,s}$, we can write the grand-canonical partition function/index over the ground states explicitly:
\be
\cZ^{\textrm{R}}(p;y,\bar{y}) = \prod_{n=1}^\infty \prod_{r,s=0}^2 \frac{1}{\left(1 - (-\varepsilon)^{r+s} p^{n} y^{r-1} \bar{y}^{s-1}\right)^{(-1)^{r+s} h^{r s}}}\,,
\ee
where $\varepsilon=\pm 1$ for the partition function and index, respectively, and we have used the fact that $\textrm{Sgn}[c]=(-1)^{r+s}$. In the large $N$ limit, the ground states of the symmetric orbifold in the Ramond sector become infinitely degenerate.

We are interested in the NS sector of the symmetric orbifold, whose ground state is unique. 
The NS ground state is holographically dual to the $\adsM$ background \cite{Maldacena:1998bw}. Under spectral flow by $\eta = \bar\eta = -1$, the R ground states map to the left-right chiral primary states in the NS sector.
This spectral flow amounts to shifting $p \to p y \bar{y}$ in the partition function. The grand-canonical partition function/index of left-right chiral primaries in the NS sector is then
\begin{equation} \label{eq: NS grand canonical}
\cZ^{\rm NS}(p;y,\bar{y}) = \sum_{N=0}^\infty p^N Z_N^{\rm NS}(y,\bar{y}) = \prod_{n=1}^\infty \prod_{r,s=0}^2 \frac{1}{\left(1 - (-\varepsilon)^{r+s} p^{n} y^{n+r-1} \bar{y}^{n+s-1}\right)^{(-1)^{r+s} h^{r, s}}} \,.
\end{equation}
The remainder of this work will concern the finite $N$ partition function $Z_N(y,\bar{y}) := Z_N^{\rm NS}(y,\bar{y})$ of chiral primary states in the NS sector.

\subsection{Worldsheet dual of symmetric orbifold}
\label{ssec:WorldSheet}

The BPS spectrum of string theory in \ASTM\ can be analysed using either the symmetric orbifold CFT or the worldsheet CFT for the \ASTM\ background with one unit of NS flux \cite{Eberhardt:2018ouy}.
Given that these two CFT's are rather different, the spectra from the two sides match in a rather non-trivial manner.

In this subsection, we briefly review this worldsheet theory, in order to demonstrate this match. The key point is that the twist $w$ in the symmetric orbifold CFT corresponds to the worldsheet spectral flow by $w$ units. 
This will be important for computing the worldsheet spectrum in the $\orbifoldM$ background in Section \ref{subsec: strings on orbifolds}, where $w$ takes fractional values $w\in\frac{1}{k}\mathbb{N}_{\geq 1}$.

We will not use the worldsheet CFT in the next few sections. The reader may skip this subsection for now and refer to it later when reaching Sections \ref{subsec: strings on orbifolds} and \ref{subsec: orbifold spectral flows}.

\subsubsection*{Tensionless limit}

The AdS length is related to the NS5 charge by $\ell^2_{\textrm{AdS}_3}/\alpha'=Q_5$. 
Therefore, when there is only one unit of the NS flux, the AdS length takes the minimal value in string units.  This is the so-called tensionless limit of string theory, which corresponds to the high curvature regime in the bulk gravity.

In this limit, the RNS formalism breaks down, and one needs to use the hybrid formalism of Berkovits-Vafa-Witten \cite{Berkovits:1999im}.
A description of this worldsheet CFT was proposed in \cite{Eberhardt:2018ouy} in terms of a free field realization of the current algebra $\mathfrak{psu}(1,1|2)_1$,\footnote{
Note that here the level of the current algebra takes the minimal value, namely one, since in the tensionlesss limit, the AdS$_3$ radius takes the minimal value in string units.
} and shown to capture the \textit{entire}\footnote{In particular, it doesn't suffer from the problem of ``gaps'' \cite{Seiberg:1999xz, Argurio:2000tb,Raju:2007uj}  in matching the spectrum of the worldsheet theory in the RNS formalism to the symmetric orbifold spectrum (for a more recent review of this problem see 
\cite{Eberhardt:2017pty}). 
}
single-particle spectrum, including the non-BPS part, of the  free symmetric orbifold of $\mathbb{T}^4$ at $N\rightarrow \infty$.\footnote{Further, it was suggested in \cite{Eberhardt:2020bgq} that the worldsheet CFT should actually correspond to the grand canonical ensemble of Sym$^{N}(\mathcal{M}_4)$ instead of the $N\rightarrow \infty$ limit.} The set of free fields consists of $4$ symplectic bosons and $2$ complex fermions, all with conformal weight $h=\tfrac{1}{2}$. 
The neutral bilinears of them generate the current algebra $\mathfrak{u}(1,1|2)_1$, which after some further projection gives the current algebra $\mathfrak{psu}(1,1|2)_1$, for more details see \cite[App.\ C]{Eberhardt:2018ouy} and \cite{Dei:2020zui}.\footnote{It is remarkable that a free-field realization of the $\mathfrak{psu}(1,1|2)_1$ WZW model exists, given that it correspond to the high curvature regime. 
It also makes the detailed check of holography feasible.
However, since we will focus on the BPS spectrum in this paper, we will not need the explicit free field realization.}

\subsubsection*{$\mathfrak{psu}(1,1|2)_1$ and its representations}

The superalgebra $\mathfrak{psu}(1,1|2)$ is the global subalgebra of the small $\mathcal{N}=4$ superconformal algebra, which we review in Appendix \ref{appssec:N4SCA}.
We will now consider its affine version at the smallest level $\mathsf{k}=1$.
The super affine algebra $\mathfrak{psu}(1,1|2)_1$ 
has eight supercharges and its bosonic subalgebra is
\begin{equation}
\mathfrak{sl}(2,\mathbb{R})_{1}\oplus \mathfrak{su}(2)_{1}\,.
\end{equation}
We will denote the currents of  $\mathfrak{sl}(2,\mathbb{R})_{1}$ by $\mathcal{J}^{3,\pm}_m$, and those of $\mathfrak{su}(2)_{1}$ by $\mathcal{K}^{3, \pm}_m$;\footnote{
Recall that $\mathfrak{sl}(2,\mathbb{R})_{\mathsf{k}}=\mathfrak{su}(1,1)_{\mathsf{k}}=\mathfrak{su}(2)_{-\mathsf{k}}$, and $\mathfrak{su}(2)_{\mathsf{k}}$ is defined as \begin{equation}
{}[J^3_m,J^3_n]  = \frac{\mathsf{k}}{2} \, m 
,\delta_{m+n,0} 
\,, \quad 
{}[J^3_m,J^\pm_n]  =  \pm J^\pm_{m+n} 
\,, \quad 
{}[J^+_m,J^-_n]  = 2  J^3_{m+n} + \mathsf{k} \, m \, \delta_{m+n,0}\,.\nonumber
\end{equation}
}
in adddtion, we denote the contribution to the stress-energy tensor from these two factors as $\mathcal{L}^{\mathfrak{sl}(2,\mathbb{R})}_m$ and $\mathcal{L}^{\mathfrak{su}(2)}_m$, respectively.
For $\mathsf{k}=1$, instead of the Sugawara expression, the total  stress-energy tensor can be expressed purely in terms of the bosonic currents of $\mathfrak{psu}(1,1|2)_1$:
\begin{equation}
\mathcal{L}=\mathcal{L}^{\mathfrak{sl}(2,\mathbb{R})}_{}
+\mathcal{L}^{\mathfrak{su}(2)}_{}  \,,
\end{equation}
and in particular, the zero modes $\mathcal{L}^{\mathfrak{sl}(2,\mathbb{R})}_0$ and $\mathcal{L}^{\mathfrak{su}(2)}_0$ can be computed in terms of the quadratic Casimir: 
\ie
\mathcal{L}^{\mathfrak{sl}(2,\mathbb{R})}_0 &=\frac{1}{\mathsf{k}-2}\mathcal{C}^{\mathfrak{sl}(2,\mathbb{R})}
=-j^{\mathfrak{sl}(2,\mathbb{R})}(1-j^{\mathfrak{sl}(2,\mathbb{R})}) \\
\mathcal{L}^{\mathfrak{su}(2)}_0 &=\frac{1}{\mathsf{k}+2}\mathcal{C}^{\mathfrak{su}(2)}=\frac{1}{3}\ell^{\mathfrak{su}(2)}(1+\ell^{\mathfrak{su}(2)})
\fe
where $j^{\mathfrak{sl}(2,\mathbb{R})}$ and $\ell^{\mathfrak{su}(2)}$ are the $\mathfrak{sl}(2,\mathbb{R})$ and $\mathfrak{su}(2)$ spins, respectively.

The physical states satisfy the on-shell condition $\mathcal{L}_0 = 0$ and their $\mathcal{J}^3_0$ and $\mathcal{K}^3_0$ charges are related to the spacetime charges $L_0, J_0$ as
\begin{equation}\label{eq:chargeSTvsWS}
\mathcal{J}^3_0 =L_0, \qquad \mathcal{K}^3_0=J_0,
\end{equation}
and similarly in the right-moving sector.

One important reason that the worldsheet CFT with $Q_5=1$ is much simpler than those with higher $Q_5$ is that it 
corresonds to the minimal level $\mathsf{k}=1$ in the current algebra  $\mathfrak{psu}(1,1|2)_{\mathsf{k}}$ of the WZW model, and the representation theory of $\mathfrak{psu}(1,1|2)_{\mathsf{k}}$ drastically simplifies at $\mathsf{k}=1$.
In terms of its  bosonic subalgebra, the  representation of $\mathfrak{psu}(1,1|2)_{\mathsf{k}}$ can be labeled as $(J^{\mathfrak{sl}(2,\mathbb{R})},\lambda|J^{\mathfrak{su}(2)})$, with
\ie
J^{\mathfrak{sl}(2,\mathbb{R})} &: \ \mathfrak{sl}(2,\mathbb{R}) \ \textrm{spin} \\
\lambda &: \
\textrm{fractional part of }\mathcal{J}^{3}_0 \textrm{ eigenvalue} \\
J^{\mathfrak{su}(2)} &: \ \mathfrak{su}(2) \ \textrm{spin} 
\fe
of the top component of the
$\mathfrak{psu}(1,1|2)_{\mathsf{k}}$ multiplet, which in general is the  16-dimensional long multiplet generated by $4$ fermionic raising operators (out of the 8 supercharges).
For $\mathsf{k}\geq 2$, $\mathfrak{psu}(1,1|2)_{\mathsf{k}}$ has both long and short representations, and each can be either discrete (with $J^{\mathfrak{sl}(2,\mathbb{R})}\in \mathbb{R}$ and $\tfrac{1}{2}\leq  J^{\mathfrak{sl}(2,\mathbb{R})} \leq \tfrac{\mathsf{k}+1}{2}$) or continuous (with $J^{\mathfrak{sl}(2,\mathbb{R})}=\tfrac{1}{2}+i p$) representations. 
But at the minimal level $\mathsf{k}=1$, since the only unitary representations of $\mathfrak{su}(2)_{\mathsf{k}=1}$ are those with $J^{\mathfrak{su}(2)}=0,\tfrac{1}{2}$, this forces $J^{\mathfrak{sl}(2,\mathbb{R})}$ to be $\tfrac{1}{2}$, and the resulting representations are ultra-short representations that has $3$ components instead of $16$:
\begin{equation}\label{eq:unltrashort}
(\tfrac{1}{2},\lambda|\tfrac{1}{2})
\,,\qquad
(1,\lambda+\tfrac{1}{2}|0)
\,,\qquad
(0,\lambda+\tfrac{1}{2}|0)\,.
\end{equation}
Later, we will use this ultra-short represenation at $\lambda=\tfrac{1}{2}$ to reproduce the chiral primary spectrum of the spacetime symmetric orbifold CFT.

\subsubsection*{Worldsheet spectral flow}
Let us now desribe the spectral flows of the worldsheet CFT.
At the algebraic level, the $w$-spectral flow $\Sigma^{w}$, for $w\in \mathbb{R}$, is an outer-automorphism of the $\mathfrak{psu}(1,1|2)_1$ algebra, and it acts on the bosonic part of the $\mathfrak{psu}(1,1|2)_1$ generators via
\ie \label{eq:SFonworldsheet}
&\Sigma^{w}(\mathcal{L}^{\mathfrak{sl}(2,\mathbb{R})}_n) = \mathcal{L}^{\mathfrak{sl}(2,\mathbb{R})}_n - w \mathcal{J}^3_n - \tfrac{1}{4} \, w^2 \delta_{n,0} \,, \\ 
&\Sigma^{w}(\mathcal{J}^3_n) = \mathcal{J}^3_n + \tfrac{1}{2} w \, \delta_{n,0} \,, 
 \qquad 
\Sigma^{w}(\mathcal{J}^\pm_n)  = \mathcal{J}^\pm_{n\mp w}  \,, \\
&\Sigma^{w}(\mathcal{L}^{\mathfrak{su}(2)}_n)= \mathcal{L}^{\mathfrak{su}(2)}_n + w \mathcal{K}^3_n + \tfrac{1}{4} \, w^2 \delta_{n,0} \,,\\ 
&\Sigma^{w}(\mathcal{K}^3_n) = \mathcal{K}^3_n + \tfrac{1}{2} w \, \delta_{n,0} \,, \qquad 
\Sigma^{w}(\mathcal{K}^\pm_n)  = \mathcal{K}^\pm_{n\pm w} \,.
\fe
We omitted the spectral flow on the fermionic fields. 
At the algebraic level, $w$ can take any real values. 
But to describe strings in $\adsM$, we have $w=1,2,3,\cdots$; and we will see later in Section \ref{subsec: orbifold spectral flows} that to describe string in $\orbifoldM$, we have $w=\frac{n}{k}$ with $n=1,2,3,\cdots$.

Physically, the spectral flow operation causes the strings to wind $w$ times around the spatial circle of AdS$_3$ \cite{Maldacena:2000hw}. Consider a string that winds once around the spatial $S^1$. This means that the fields that correspond to the AdS$_3$ coordinates in the boundary directions, namely $t,\phi$, has boundary condition: 
\ie
t(\tau, \sigma+2\pi) &= t(\tau,\sigma) \\
\phi(\tau, \sigma+2\pi) &=\phi(\tau, \sigma)+2\pi.
\fe
After the spectral flow $\Sigma^{w}$, we obtain  a string that winds $w$ times around the $S^1$ cycle that is the spatial boundary of AdS$_3$,  
which means that $t,\phi$ now has boundary condition: 
\ie\label{eq:BCoriginal}
t(\tau, \sigma+2\pi) &= t(\tau,\sigma) \\
\phi(\tau, \sigma+2\pi) &=\phi(\tau, \sigma)+2\pi w .
\fe

The effect of the spectral flow on the total worldsheet stress-energy tensor is 
\begin{equation}\label{eq:SFonworldsheetL}
\Sigma^{(w)}(\mathcal{L}_n)=\mathcal{L}_n+w(\mathcal{J}^3_n-\mathcal{K}^3_n) \,.
\end{equation}
In particular, if one starts with a physical state that is BPS, namely, satisfying $\mathcal{L}_0=0$ and $\mathcal{J}^{3}_0=\mathcal{K}^3_0=h=j$, then a worldsheet spectral flow will produce another BPS physical state since
\ie
&\Sigma^{(w)}(\mathcal{L}_0)=0 \,,\\
&\Sigma^{(w)}(\mathcal{J}^{3}_0)=\Sigma^{(w)}(\mathcal{K}^3_0)= h+\frac{w}{2} =j+\frac{w}{2} \,.
\fe

\subsubsection*{Chiral primary spectrum} 

Let us now see how to reproduce the spectrum of left-right chiral primary states from the worldsheet CFT.

Let us first look at the unflowed sector, with $w=0$.
In the continous ultrashort representation given in \eqref{eq:unltrashort},  take its top component, and consider $\lambda=\frac{1}{2}$.
This state has $\mathfrak{psu}(1,1|2)$ charge $(\frac{1}{2},\frac{1}{2}|\frac{1}{2})$ and thus has the spacetime charges $\mathcal{L}_0=0$ and
\be\label{eq:BPSchargew1}
\mathcal{J}^3_0=\mathcal{K}^3_0=\frac{1}{2} \qquad (w=0) \,.
\ee
Therefore it corresponds to a BPS state with
$h=j=\frac{1}{2}$. However, this state is not part of the spectrum, see \cite[Section 5.4]{Eberhardt:2018ouy}.

Let us now apply the spectral flow $\Sigma^{(w)}$ on these BPS states. Using \eqref{eq:SFonworldsheet} and \eqref{eq:SFonworldsheetL}, the charges of the spectrally flowed states are $\mathcal{L}_0=0 $ and
\be \label{eq:BPSchargew}
\mathcal{J}^3_0=\mathcal{K}^3_0=\frac{w+1}{2}
\ee
correspond to a BPS physical state with $
h=j=\frac{w+1}{2}$. Therefore, applying the $w$-spectrally flow on the BPS states in the unflowed sector $w=0$, with charges \eqref{eq:BPSchargew1}, we generate new BPS states with charges \eqref{eq:BPSchargew}. So we have obtained all the top components of the BPS spectrum from the symmetric orbifold, summarized in \eqref{eq:BPSquartetSymN}. 

To obtain the other three lower states in \eqref{eq:BPSquartetSymN}, we have two fermion zeromodes $\mathcal{S}^{++\pm }_0$ to apply,  we can apply either of them, or both of them, and obtain the other three states. 
In total, the chiral primary states of the left-moving $w$-spectrally flowed sector are \cite{Eberhardt:2018ouy}
\begin{equation}\label{eq:BPSquartetWS}
\begin{array}{ccc}
&
h=j=\frac{w+1}{2} & \\
h=j=\frac{w}{2} &  & h=j=\frac{w}{2}  \\
& h=j=\frac{w-1}{2}&
\end{array}     \,.
\end{equation}
And we have another copy for the right-moving sector. Comparing \eqref{eq:BPSquartetSymN} and \eqref{eq:BPSquartetWS}, we see that the chiral primary spectrum of the $w$-spectrally flowed sector of the worldsheet CFT matches precisely with the (single-particle part of the) chiral primary spectrum of the $w$-twisted sector of the symmetric orbifold of $\mathbb{T}^4$.

The full worldsheet partition function is then found by summing over all the spectrally-flowed sectors:
\begin{equation}
\mathsf{z}^{\textrm{WS}}_{\textrm{\ASTM}}(q)=\sum^{\infty}_{w=1} \mathsf{z}^{(w)}_{\textrm{\ASTM}}(q) \,.
\end{equation}
Note that unlike in the RNS formalism, there is no need to sum over spin-structures in the hybrid formalism. The worldsheet partition function and index in the chiral primary sector of $\adsT$ is
\begin{equation}
\mathsf{z}^{\rm WS}(y,\yb) = \sum_{w=1}^\infty \left| y^{w-1}(1 + 2\varepsilon  y + y^2) \right|^2 -1.
\end{equation}
The worldsheet partition function in the tensionless limit on $\adsK$ can also be found by working at the orbifold point $\Kthree = \Tfour/\bZ_2$ in the moduli space of $\Kthree$ and including contributions from the 16 blowup modes \cite{schmidthesis}. The worldsheet partition function in the chiral primary sector of $\adsK$ is
\be \label{eq: K3 worldsheet spectrum}
\mathsf{z}^{\rm WS}(y,\yb) = \sum_{w=1}^\infty \bigg[ \frac{1}{2} \left| y^{w-1}(1 + 2 y + y^2) \right|^2 + \frac{1}{2} \left| y^{w-1}(1 - 2 y + y^2) \right|^2 + 16 \left| y^{w-1} y \right|^2 \bigg] - 1.
\ee

\section{Bulk saddles as grand-canonical residues}
\label{sec: grand canonical}

With the DMVV formula at hand, we are ready to state the central observation: The residues at poles in the $p$-plane, of the grand-canonical partition function of symmetric orbifolds
\be
\cZ(p,y,\yb) = \sum_{N=0}^\infty p^N Z_N(y,\yb),
\ee
can be identified with the one-loop partition functions of asymptotically $\adsM$ geometries that sum to the finite $N$ answer $Z_N(y,\yb)$.

This observation will allow us to determine the set of bulk saddles that would contribute to the IIB supergravity path integral for asymptotically $\adsM$ spacetimes in the chiral primary sector. While this observation may hold in a more general context, such as for the elliptic genus of the $\Kthree$ symmetric orbifold or for the partition function at the symmetric orbifold point, the scope of the current work will be limited to the chiral primary sector of $\symM$. 

Let us suppose, for the moment, that the above assertion is true. Then one may wish to express the partition function of chiral primaries at finite $N$ as a sum over residues in the complex $p$-plane
\ie \label{eq: sumoverresidues}
Z_N(y,\bar{y}) &= \oint_{p=0} \frac{dp}{2 \pi i} p^{-N-1} \cZ(p;y,\bar{y}) \\
&\stackrel{?}{=} - \sum_{i: \, p_i \neq 0} \, 
\Res_{p = p_i} \, p^{-N-1} \cZ(p;y,\bar{y}).
\fe
The grand-canonical partition function/index of $\symM$ in the chiral primary sector
\be
\cZ(p;y,\bar{y}) = \prod_{n=1}^\infty \prod_{r,s=0}^2 \frac{1}{\left(1 - (-\varepsilon)^{r+s} p^{n} y^{n+r-1} \bar{y}^{n+s-1}\right)^{(-1)^{r+s} h^{r, s}}}
\ee
has four infinite sequences of simple poles
\ie \label{eq: four towers}
(0,0)_k^m:& \quad p \, = \, e^{2 \pi i \frac{m}{k}} y^{-\left(1 - \frac{1}{k}\right)} \yb^{-\left(1 - \frac{1}{k}\right)} \\
(2,0)_k^m:& \quad p \, = \, e^{2 \pi i \frac{m}{k}} y^{-\left(1 + \frac{1}{k}\right)} \yb^{-\left(1 - \frac{1}{k}\right)} \\
(0,2)_k^m:& \quad p \, = \, e^{2 \pi i \frac{m}{k}} y^{-\left(1 - \frac{1}{k}\right)} \yb^{-\left(1 + \frac{1}{k}\right)} \\
(2,2)_k^m:& \quad p \, = \, e^{2 \pi i \frac{m}{k}} y^{-\left(1 + \frac{1}{k}\right)} \yb^{-\left(1 + \frac{1}{k}\right)}.
\fe
These poles are located at $k$-th roots of unity on the complex $p$-plane, where
\ie
k &= 1,2,3 \cdots \\
m &=0,1,2 \cdots, k-1.
\fe
These sequences originate from terms in $\bZ$ corresponding to the four corners of the Hodge diamond $h^{0,0} = h^{2,0} = h^{0,2} = h^{2,2} = 1$. The labels ``$0$'' and ``$2$'' will acquire an interpretation in terms of $\cN = 4$ spectral flows in the next section. Importantly, $\cZ$ also has a wall of essential singularities on the circle of radius
\be
|p|=|y^{-1} \yb^{-1}|
\ee
on the $p$-plane.

It is clear from the structure of its singularities that we cannot apply the residue theorem \eqref{eq: sumoverresidues} to $\cZ$. The poles of $\cZ$ accumulate on the circle of radius $|p|=|y^{-2} \yb^{-2}|$ at large $k$, and there is an impenetrable wall of essential singularities at the same radius. In the Ramond sector, this wall becomes that of the function
\be
\frac{1}{\eta(p)^{h^{1,1}}} \sim \frac{1}{\prod_{n=1}^\infty (1 - p^n)^{h^{1,1}}}
\ee
on the unit circle $|p|=1$. The accumulating poles and the wall of essential singularities are generic features shared by grand-canonical partition functions of symmetric orbifolds.

We nevertheless find that there is a version of the residue formula that applies to the grand-canonical partition function $\cZ$ in the chiral primary sector. We provide a derivation of the residue formula for $\cZ$ in Section \ref{sec: finite N saddles}. Interestingly, the set of poles that contribute to the finite $N$ answer turn out to depend, in a discontinuous manner, on the domain of fugacities under consideration. Moreover, while the essential singularities affect the expression for the residue at any simple pole, they do not contribute explicitly to the sum \eqref{eq: sumoverresidues}.

In Sections \ref{sec: large N and spectral flows} and \ref{sec: orbifolds}, we focus on the individual residues
\be \label{eq: res definition}
\Zhat_k^{(0/2,0/2)} (y,\yb) = - \sum_{m=0}^{k-1} \ \Res_{p = (0/2,0/2)_k^m} \, p^{-N-1} \cZ(p;y,\yb)
\ee
at the poles \eqref{eq: four towers} labelled by $k$ and show that they coincide with the one-loop partition functions of
\be
\orbifoldM
\ee
backgrounds and their asymmetric generalizations, which can be understood as spectral flows thereof. We defer the question of which subset of these residues contributes to the finite $N$ answer $Z_N$ to Section \ref{sec: finite N saddles}.

\section{The large $N$ limit and spectral flows of $\adsS$} \label{sec: large N and spectral flows}

\subsection{Large $N$ limit} \label{subsec: large N}

In this section, we will be interested in the physical interpretation associated with the residues \eqref{eq: res definition}, labelled by $k=1$, of the grand-canonical partition function of chiral primary states in symmetric orbifolds. Explicitly, the grand-canonical partition functions/indices of chiral primary states of $\symT$ and of $\symK$ are
\ie
\cZ_{\Tfour} &= \prod_{n=1}^\infty \frac{\left(1 + \varepsilon p^{n} y^{n} \bar{y}^{n-1}\right)^2 \left(1 + \varepsilon p^{n} y^{n-1} \bar{y}^{n}\right)^2  \left(1 + \varepsilon p^{n} y^{n} \bar{y}^{n+1}\right)^2 \left(1 + \varepsilon p^{n} y^{n+1} \bar{y}^{n}\right)^2}{\left(1 - p^{n} y^{n-1} \bar{y}^{n-1}\right) \left(1 - p^{n} y^{n+1} \bar{y}^{n-1}\right) \left(1 - p^{n} y^{n-1} \bar{y}^{n+1}\right) \left(1 - p^{n} y^{n+1} \bar{y}^{n+1}\right) \left(1 - p^{n} y^{n} \bar{y}^{n}\right)^4} \\
\cZ_{\Kthree} &= \prod_{n=1}^\infty \frac{1}{\left(1 - p^{n} y^{n-1} \bar{y}^{n-1}\right) \left(1 - p^{n} y^{n+1} \bar{y}^{n-1}\right) \left(1 - p^{n} y^{n-1} \bar{y}^{n+1}\right) \left(1 - p^{n} y^{n+1} \bar{y}^{n+1}\right) \left(1 - p^{n} y^{n} \bar{y}^{n}\right)^{20}}
\fe
in the NS sector, where $\varepsilon = \pm1$ respectively for the partition function and index. We are interested in the residues at the $k=1$ poles
\ie
(0,0)_{1}^0:& \quad p \, = \, 1 \\
(2,0)_{1}^0:& \quad p \, = \, y^{-2} \\
(0,2)_{1}^0:& \quad p \, = \, \bar{y}^{-2} \\
(2,2)_{1}^0:& \quad p \, = \, y^{-2} \bar{y}^{-2}.
\fe
that sit at the bottom of the four towers \eqref{eq: four towers}.

It is a well appreciated fact that the behavior of the partition function $Z_N$ in the strict large $N$ limit is governed by the pole of the grand-canonical partition function $\cZ$ that is closest to the origin on the $p$-plane \cite{deBoer:1998us}. Assuming $|y|,|\yb| < 1$, this pole is located at $p=1$ and has the label $(0,0)_{1}^0$. Consequently, the large $N$ limit of $Z_N$ is given by the residue
\be
\Zhat_1^{(0,0)} = -\Res_{p=1} \, p^{-N-1} \cZ(p;y,\yb).
\ee
Explicitly, the large $N$ limit of the partition function $Z_N$ for $\symT$ is
\be \label{eq: large N T4}
\Zhat_1^{(0,0)} = \prod_{n=1}^\infty \frac{\left(1 + \varepsilon y^{n} \bar{y}^{n-1}\right)^2 \left(1 + \varepsilon y^{n-1} \bar{y}^{n}\right)^2  \left(1 + \varepsilon y^{n} \bar{y}^{n+1}\right)^2 \left(1 + \varepsilon y^{n+1} \bar{y}^{n}\right)^2}{\left(1 - y^{n+1} \bar{y}^{n-1}\right) \left(1 - y^{n-1} \bar{y}^{n+1}\right) \left(1 - y^{n+1} \bar{y}^{n+1}\right) \left(1 - y^{n} \bar{y}^{n}\right)^5}
\ee
and for $\symK$ is
\be \label{eq: large N K3}
\Zhat_1^{(0,0)} = \prod_{n=1}^\infty \frac{1}{\left(1 - y^{n+1} \bar{y}^{n-1}\right) \left(1 - y^{n-1} \bar{y}^{n+1}\right) \left(1 - y^{n+1} \bar{y}^{n+1}\right) \left(1 - y^{n} \bar{y}^{n}\right)^{21}}.
\ee
The agreement between the large $N$ spectrum \eqref{eq: large N T4}, \eqref{eq: large N K3} of chiral primary states in symmetric orbifolds and the spectrum of bulk Kaluza-Klein states in $\adsM$ was shown in early works on AdS$_3$/CFT$_2$ \cite{Maldacena:1998bw,Deger:1998nm,Larsen:1998xm,deBoer:1998kjm}.

The truncation of ten-dimensional IIB supergravity on $\adsM$ to the massless modes on $\Tfour$ (or $\Kthree$) produces a six-dimensional supergravity on $\adsS$ with 5 (or 21) tensor multiplets. The degeneracies of chiral primary contributions from the tensor and graviton multiplets are reflected in the partition functions \eqref{eq: large N T4} and \eqref{eq: large N K3} upon a further decomposition on $S^3$.

We recognize the pole $(0,0)_{1}^{0}$ as that which has the residue $\Zhat_{1}^{(0,0)}$ corresponding to the Kaluza-Klein spectrum of $\adsM$.

\subsection{Bulk spectral flows} \label{subsec: bulk spectral flows}

What of the other three poles $(2,0)$, $(0,2)$, and $(2,2)$ at $k=1$? The residues at the three other poles are related to $\Zhat_{1}^{(0,0)}$ as
\ie\label{eq: three other}
\Zhat_{1}^{(2,0)}(y,\bar{y}) \, &= \, y^{2N} \Zhat_{1}^{(0,0)}(y^{-1},\bar{y}) \\
\Zhat_{1}^{(0,2)}(y,\bar{y}) \, &= \, \bar{y}^{2N} \Zhat_{1}^{(0,0)}(y,\bar{y}^{-1}) \\
\Zhat_{1}^{(2,2)}(y,\bar{y}) \, &= \, y^{2N} \bar{y}^{2N} \Zhat_{1}^{(0,0)}(y^{-1},\bar{y}^{-1}).
\fe
It is clear that the bulk interpretations of $\Zhat_{1}^{(2,0)}$, $\Zhat_{1}^{(0,2)}$, and $\Zhat_{1}^{(2,2)}$, if they exist, must be closely related to the ordinary Kaluza-Klein spectrum of $\adsM$. We will find that the residues \eqref{eq: three other} are partition functions of chiral primary Kaluza-Klein states on spectral flows of $\adsM$ backgrounds.

Six-dimensional supergravity on $\adsS$ may be truncated to Chern-Simons supergravity on $\AdS_3$ based on the gauge group $SU(1,1|2) \times SU(1,1|2)$ \cite{Knizhnik:1986wc,Bershadsky:1986ms,Sevrin:1993vv,Achucarro:1986uwr,Witten:1988hc,Howe:1995zm,deBoer:1998kjm,Henneaux:1999ib}. The global $\AdS_3$ solution of this theory is given by the metric and the $SU(2) \times SU(2)$ gauge fields $A,\bar{A}$
\ieg \label{eq: global before flow}
ds^2 = -(1 + r^2) dt^2 + \frac{dr^2}{1 + r^2} + r^2 d\phi^2 \\
A = 0, \quad \bar{A} = 0,
\feg
where $\phi \sim \phi + 2 \pi$. The spinors are anti-periodic around the spatial $\phi$-circle because this circle is contractible in global $\AdS_3$. The asymptotic symmetry generators of the $SU(1,1|2) \times SU(1,1|2)$ Chern-Simons supergravity form a $\cN = (4,4)$ superconformal algebra. As explained in Section \ref{sec: review}, the $\cN = (4,4)$ algebra admits a family of spectral flow automorphisms labelled by the parameters $(\eta, \bar\eta)$. The spectral flow by $\eta$ acts on the generators in the left-moving sector as
\ie \label{eq: spectral flow recall}
L_0 \ &\to \ L_0 + \eta J_0 + \tfrac{c}{24} \eta^2 \\
J_0 \ &\to \ J_0 + \tfrac{c}{12} \eta
\fe
and similarly in the right-moving sector. Spectral flows of global $\AdS_3$ that preserve the NS boundary condition of spinors are those given by even units of $\eta, \bar\eta$.

The spectral flow operation is realized in $\AdS_3$ supergravity in terms of large gauge transformations of $A, \bar{A}$ that shift their values at the asymptotic boundary by the constants $\eta, \bar\eta$. For instance, the global $\AdS_3$ solution after spectral flow is
\ieg \label{eq: global after flow}
ds^2 = -(1 + r^2) dt^2 + \frac{dr^2}{1 + r^2} + r^2 d\phi^2 \\
A = \eta \sigma^3 d\phi, \quad \bar{A} = \bar\eta \sigma^3 d\phi.
\feg
These gauge fields are related to the R-charges of the dual CFT as
\be
J_0 = \frac{c}{12} A_\phi^3, \qquad \Jb_0 = \frac{c}{12} \bar{A}_\phi^3.
\ee
So spectral flows by $(\eta,\bar\eta)=(0,0),(2,0),(0,2),(2,2)$, which preserve the NS boundary conditions of spinors in global $\AdS_3$, would induce
\ie \label{eq: flow ads charges}
(0,0):& \quad j = 0, \quad \ \jb = 0 \\
(2,0):& \quad j = N, \quad \jb = 0 \\
(0,2):& \quad j = 0, \quad \ \jb = N \\
(2,2):& \quad j = N, \quad \jb = N.
\fe
These geometries would then contribute to the chiral primary partition function with the weights $1$, $y^{2N}$, $\yb^{2N}$, and $y^{2N} \yb^{2N}$ seen in $\Zhat_{1}^{(0,0)}$, $\Zhat_{1}^{(2,0)}$, $\Zhat_{1}^{(0,2)}$, and $\Zhat_{1}^{(2,2)}$. The analysis of Section \ref{subsec: orbifold spectral flows} will show that these solutions indeed satisfy the conditions for a chiral primary.

Let us address why only the solutions \eqref{eq: global after flow} that are related to vacuum $\AdS_3$ by the spectral flows $(\eta,\bar\eta)=(0,0),(2,0),(0,2),(2,2)$ are allowed to contribute as saddles in a bulk computation of the chiral primary partition function $Z_N$. In other words, we need to explain why other solutions that are related to vacuum $\AdS_3$ by arbitrary even units of the spectral flow parameters do not appear as contributions to $Z_N$. Recall our definition
\be \label{eq: chiral primary part fn recall}
Z_N = \Tr_{\cH_{cc}} \left( y^{2 J_0} \yb^{2 \Jb_0} \right)  = \lim_{q,\qb \to 0} \Tr_{\symM} \left( q^{L_0 - J_0} \bar{q}^{\bar{L}_0 - \bar{J}_0} y^{2 J_0} \yb^{2 \bar{J}_0} \right)
\ee
of the chiral primary partition function $Z_N$ of the symmetric orbifold at finite $N$. Our goal is to understand how to compute this quantity from a bulk perspective.

In a gravitational path integral computation, one is instructed to sum over the contributions from all backgrounds consistent with the asymptotic boundary conditions. The saddles contributing to $Z_N$ would need to, at the least, (1) have NS-NS boundary conditions for spinors around the asymptotic spatial circle and (2) survive the projection $q,\qb \to 0$ on to the chiral primary Hilbert space.\footnote{The Euclidean continuation of the solution \eqref{eq: global after flow} is real and the thermal circle is not contractible in the bulk. Therefore, there is no subtlety in going between the real-time canonical and thermal path integral formalisms as long as we are considering states resulting from the quantization of perturbative fluctuations around a fixed saddle. We use this fact in the following discussion.} We would like to determine which of the geometries \eqref{eq: global after flow} labelled by $(\eta,\bar\eta)$ have a chance of contributing to $Z_N$.

From Section \ref{subsec: large N}, we know that the BPS spectrum associated with global $\AdS_3$, prior to spectral flow, is given by the projection of the Hilbert space $\cH_{\rm KK}$ of Kaluza-Klein states \cite{deBoer:1998kjm} to the chiral primary sector:
\be\label{eq: unflowed}
\Zhat_{1}^{(0,0)} = \lim_{q,\qb \to 0} \Tr_{\cH_{\rm KK}} \left( q^{L_0 - J_0} \bar{q}^{\bar{L}_0 - \bar{J}_0} y^{2 J_0} \yb^{2 \bar{J}_0} \right) \,,
\ee
which is equal to the $N\rightarrow \infty$ limit $Z_\infty$ of the chiral primary partition function \eqref{eq: chiral primary part fn recall} of the CFT.
Now suppose we flow our background by $(\eta,\bar\eta)$. 
The BPS spectrum associated with spectrally-flowed $\AdS_3$ will also be given by the projection of the Hilbert space $\cH_{\rm KK}$ of Kaluza-Klein states to the chiral primary sector, but where the states are now weighted by the spectrally-flowed generators of the $\cN = (4,4)$ algebra:
\be \label{eq: spectral flow part fn}
\lim_{q,\qb \to 0} \, q^{\frac{\eta(\eta-2)}{16 G_N}} \qb^{\frac{\bar\eta(\bar\eta-2)}{16 G_N}} y^{\frac{\eta}{4 G_N}} \yb^{\frac{\bar\eta}{4 G_N}} \, \Tr_{\cH_{\rm KK}} \left( q^{L_0 - (1 - \eta) J_0} \bar{q}^{\bar{L}_0 - (1 - \bar\eta) \bar{J}_0} y^{2 J_0} \yb^{2 \bar{J}_0} \right) \,,
\ee
where the gravitational coupling is $G_N^{(3)} = 1/4N$ by \cite{Brown:1986nw}.\footnote{This is not to be confused with the spectral flow of the boundary CFT.}
The limit $q,\bar{q}\rightarrow 0$ only exists provided that the spectral flow parameters take the values
\be \label{eq: BPS spectral flows}
(\eta,\bar\eta) = (0,0), \ (2,0), \ (0,2), \ (2,2)\,,
\ee
since otherwise the negative powers of $q,\bar{q}$ inside the trace can overwhelm the powers from the prefactor and hence the quantity in \eqref{eq: spectral flow part fn} would diverge in the limit $q,\bar{q}\rightarrow 0$.
Therefore, only these spectral flows of global $\AdS_3$ correspond to allowed contributions to the chiral primary partition function $Z_N$. 
We will refer to flows by the values \eqref{eq: BPS spectral flows} as \textit{BPS spectral flows}.

Observe also the following: Taking the limit $q,\qb \to 0$ in \eqref{eq: spectral flow part fn} after spectral flow amounts to projecting the space $\cH_{\rm KK}$ of Kaluza-Klein states on to a subspace of states satisfying a different charge constraint. For example, consider the partition function of chiral primary excitations of global $\AdS_3$ flowed by $(\eta,\bar\eta) = (2,0)$. The states in $\cH_{\rm KK}$ are now counted with weights
\be \label{eq: 20 part fn}
\lim_{q,\qb \to 0} \, y^{2 N} \, \Tr_{\cH_{\rm KK}} \left( q^{L_0 + J_0} \bar{q}^{\bar{L}_0 - \bar{J}_0} y^{2 J_0} \yb^{2 \bar{J}_0} \right).
\ee
The projection $q,\qb \to 0$ of the flowed background now enforces the condition
\be
L_0 + J_0 = 0
\ee
for \textit{anti-chiral} primary states on the left and the chiral primary condition $\Lb_0 - \Jb_0 = 0$ on the right, \textit{in terms of the flowed generators}. The expression \eqref{eq: 20 part fn} nevertheless represents a valid contribution to $Z_N$, because the states counted in \eqref{eq: 20 part fn} are left-right chiral primaries with respect to the original, unflowed generators. We will indeed find that the one-loop chiral primary partition function of $\AdS_3$ flowed by $(\eta,\bar\eta) = (2,0)$ contributes to the full answer $Z_N$ within a large region in the space of fugacities $y,\yb$ (see Section \ref{sec: finite N saddles}).

The partition function \eqref{eq: 20 part fn} of our Chern-Simons supergravity on $\AdS_3$ flowed by $(\eta,\bar\eta) = (2,0)$ evaluates to the quantity
\be
\Zhat_{1}^{(2,0)} = y^{2N} \Zhat_{1}^{(0,0)}(y^{-1},\bar{y}),
\ee
coinciding with the residue at the pole $(2,0)_1^0$ of the grand-canonical partition function $\cZ(p;y,\yb)$. Similarly, we find that flowing by $(\eta,\bar\eta) = (0,2)$ and $(2,2)$ yield
\ie
\Zhat_{1}^{(0,2)} &= \bar{y}^{2N} \Zhat_{1}^{(0,0)}(y,\bar{y}^{-1}) \\
\Zhat_{1}^{(2,2)} &= y^{2N} \bar{y}^{2N} \Zhat_{1}^{(0,0)}(y^{-1},\bar{y}^{-1}).
\fe
We recognize these expressions as the residues at the poles $(0,2)_1^0$ and $(2,2)_1^0$ of the grand-canonical partition function $\cZ$.

Therefore, we identify the residues $\Zhat_{1}^{(2,0)}$, $\Zhat_{1}^{(0,2)}$, and $\Zhat_{1}^{(2,2)}$ as the partition functions of Kaluza-Klein states on BPS spectral flows of $\adsM$ by $(\eta,\bar\eta) = (2,0),(0,2)$, and $(2,2)$.

\subsection{States dual to spectral flows of $\adsS$} \label{subsec: states dual to flows}

Which states in the symmetric orbifold $\symM$ are holographically dual to the BPS spectral flows of $\adsM$?

The unflowed, i.e. $(\eta,\bar\eta) = (0,0)$, global $\adsM$ solution of IIB string theory is dual to the vacuum state $| 1 \ra^{\otimes N}$ in the NS sector of the symmetric orbifold $\symM$ \cite{Maldacena:1998bw}. Each NS vacuum $|1 \ra = |1\ra^{\rm BPS}$ is a left-right chiral primary state with $h = \hb = j = \jb = 0$.

We can ask whether the BPS spectral flows of $\adsM$ can be identified with operations on $| 1 \ra^{\otimes N}$. We recall the following fact about $\cN = (4,4)$ symmetric orbifold SCFTs that we noted in Section \ref{subsec: chiral spectrum}. Let $| w \ra^{\rm BPS}$ be the $w$-cycle BPS state with 
\be
h = j = \frac{w-1}{2}, \quad \hb = \jb = \frac{w-1}{2}.
\ee
Given a state $| w \ra^{\rm BPS}$ in $\symM$, it is possible to construct other left-right chiral primary states
\ie \label{eq: other 1-cycles}
| w_{+-} \ra^{\rm BPS} &:= J_{-1}^+ | w \ra^{\rm BPS} \\
| w_{-+} \ra^{\rm BPS} &:= \Jb_{-1}^+ | w \ra^{\rm BPS} \\
| w_{++} \ra^{\rm BPS} &:= J_{-1}^+ \Jb_{-1}^+ | w \ra^{\rm BPS}
\fe
with the quantum numbers
\be
| w_{\mp \mp} \ra^{\rm BPS} : \quad h = j = \frac{w \mp 1}{2}, \quad \hb = \jb = \frac{w \mp 1}{2}
\ee
We have used the chiral primary modes $J_{-1}^+, \Jb_{-1}^+$ of the left and right R-currents $J^{+}(z)$. In this notation, we have $| w \ra^{\rm BPS} = | w_{--} \ra^{\rm BPS}$.

By acting with the R-current modes $J_{-1}^{(i)+}, \Jb_{-1}^{(i)+}$ on every copy of the seed theory in the NS vacuum $| 1 \ra^{\otimes N}$ of $\symM$, we can build a quartet
\be \label{eq: quartet BPS vacua}
 \left\{ \ | 1_{--} \ra^{\otimes N}, \ | 1_{+-} \ra^{\otimes N}, \ | 1_{-+} \ra^{\otimes N}, \ | 1_{++} \ra^{\otimes N} \right\}
\ee
of chiral primary ``vacuum'' states with the quantum numbers
\ie \label{eq: pm vacua}
| 1_{+-} \ra^{\otimes N} \ &: \quad h = j = N, \quad \hb = \jb = 0 \\
| 1_{-+} \ra^{\otimes N} \ &: \quad h = j = 0, \quad \ \hb = \jb = N \\
| 1_{++} \ra^{\otimes N} \ &: \quad h = j = N, \quad \hb = \jb = N.
\fe
These chiral primary states have charges \eqref{eq: flow ads charges} that agree with those of the BPS spectral flows of $\adsM$ backgrounds. 

Based on the agreement between the classical charges, it may appear natural to identify the chiral primary states $| 1_{\mp \mp} \ra^{\otimes N}$ with the BPS spectral flows of $\adsS$. While this identification may be appropriate in the strict large $N$ limit, identifying $| 1_{+-} \ra^{\otimes N}$, $| 1_{-+} \ra^{\otimes N}$, $| 1_{++} \ra^{\otimes N}$ with the backgrounds flowed by $(\eta,\bar\eta) = (2,0), (0,2), (2,2)$ is delicate at finite $N$. We will describe the issue and propose an alternative perspective valid at finite $N$.

To illustrate our point in a simple context, consider the $\yb = 0$ limit of the one-loop BPS partition function on $\adsT$ flowed by $(\eta,\bar\eta) = (2,0)$:
\be \label{eq: counter 1}
\Zhat_k^{(2,0)}(y,0)  =  y^{2N} \frac{(1+ \varepsilon y^{-1})^2}{1-y^{-2}}  =  y^{2N} \left( 1 + \frac{2 \varepsilon}{y} + \frac{2}{y^2} + \frac{2 \varepsilon}{y^3} + \cdots \right) \,.
\ee
The limit $\yb=0$ projects $\Zhat_k^{(2,0)}$ onto the sector with zero right-moving R-charge $\jb = 0$. The classical identification would suggest that \eqref{eq: counter 1} should match the spectrum of $S_N$-invariant excitations on the state
\be
|1_{+-}\ra^{\otimes N} = \left( \sum_{i=1}^N \psi_{-\frac{1}{2}}^{(i)+-} \right)^N \left( \sum_{i=1}^N \psi_{-\frac{1}{2}}^{(i)++} \right)^N |1_{--}\ra^{\otimes N} \,,
\ee
where the state was written in terms of $S_N$-invariant combinations of $\Tfour$ oscillators using that $J_{-1}^{(i)+} = \psi_{-\frac{1}{2}}^{(i)++} \psi_{-\frac{1}{2}}^{(i)+-}$. The only $S_N$-invariant chiral primary ``excitations'' of $|1_{+-}\ra^{\otimes N}$ with $\jb=0$ are given by the removal of copies of $\left( \sum_{i=1}^N \psi_{-\frac{1}{2}}^{(i)+\mp} \right)$ from $|1_{+-}\ra^{\otimes N}$. These excitations have the spectrum
\be \label{eq: counterspectrum}
y^{2N} \left( 1 + \frac{2 \varepsilon}{y} + \frac{2}{y^2} + \frac{2 \varepsilon}{y^3} + \cdots + \frac{1}{y^{2N}} \right),
\ee
and the perturbative part of this spectrum, i.e.\ the terms inside the bracket, tends to that in \eqref{eq: counter 1} in the large $N$ limit.\footnote{The difference between \eqref{eq: counter 1} and \eqref{eq: counterspectrum}, including the overall $y^{2N}$, is non-zero and independent of $N$.} The obvious difference between \eqref{eq: counterspectrum} and the one-loop partition function \eqref{eq: counter 1} is that, at finite $N$, the latter spectrum is not bounded below and its series expression diverges for $|y|<1$.

These properties of \eqref{eq: counter 1} have an interpretation in the gravitational path integral on the spectrally-flowed $\adsT$ background whose periodicity is twisted by the chemical potential for $J_0$: there exist fluctuation modes on this background whose mass-squared is negative for $|y|<1$. The contours for these modes must be Wick-rotated for the path integral to be well-defined. The spectrum computed with the rotated contours would give
\be \label{eq: counter 2}
\Zhat_k^{(2,0)}(y,0) =  - y^{2N} \frac{(1+ \varepsilon y)^2}{1-y^{2}} = - y^{2N} \left( 1 + 2\varepsilon y + 2 y^2 + 2 \varepsilon y^3 + \cdots \right),
\ee
the analytic continuation of \eqref{eq: counter 1} (where we have used $\varepsilon^2=1$).

The spectrum \eqref{eq: counter 2} computed with the rotated contours suggests a more direct CFT description, valid at finite $N$, for the supersymmetric partition functions \eqref{eq: three other} of flowed backgrounds. We claim that the path integral quantization of BPS fluctuations of spectrally-flowed $\adsM$ geometries, defined with rotated contours for the negative modes, produces bulk states that are holographically dual to (a subset of) the chiral primary states of $\symM$ that become null at a finite value of $N$.\footnote{The fuller proposal involves $\orbifold$ orbifolds and their spectral flows and is discussed in Section \ref{sec: states dual to orbifolds}.}

Let us see this statement in action. In the limit $\yb=0$, it can be shown (see Section \ref{sec: finite N saddles}) that $Z_N$ is expressed exactly as the sum
\ie \label{eq: dual finite N states example}
Z_N(y,0) &= \Zhat_1^{(0,0)}(y,0) + \Zhat_1^{(2,0)}(y,0) \\
&= 1 + 2 \varepsilon y + 2 y^2 + 2 \varepsilon y^3 + \cdots + 2 \varepsilon y^{2N-1} + y^{2 N}.
\fe
of contributions from the unflowed background
\be \label{eq: unflowed counter}
\Zhat_1^{(0,0)}(y,0) = \frac{(1+ \varepsilon y)^2}{1-y^2} = 1 + 2\varepsilon y + 2 y^2 + 2\varepsilon y^3 + \cdots
\ee
and from the left-flowed background $\Zhat_1^{(2,0)}(y,0)$ \eqref{eq: counter 2}. The terms $\Zhat_1^{(0,0)}$ and $\Zhat_1^{(2,0)}$ are residues at the only two poles of the grand-canonical partition function $\cZ_{\Tfour}$ that survive this limit. The states counted in $Z_N(y,0)$ are of the form
\be
\left( \sum_{i=1}^N \psi_{-\frac{1}{2}}^{(i)+-} \psi_{-\frac{1}{2}}^{(i)++} \right)^n \left( \sum_{i=1}^N \psi_{-\frac{1}{2}}^{(i)+-} \right)^m \left( \sum_{i=1}^N \psi_{-\frac{1}{2}}^{(i)++} \right)^\ell |1_{--}\ra^{\otimes N},
\ee
where those with large enough powers $n,m,\ell$ vanish due to the fact that there are only $N$ fermion modes of two flavors. There are no twisted sector excitations since any twist operator in the symmetric orbifold has $\hb>0$.

A state-counting interpretation for the partition functions $\Zhat_1^{(0,0)}$ and $\Zhat_1^{(2,0)}$ is as follows. The BPS Hilbert space $\cH_N$ of $\symT$ at finite $N$ can be thought of as the BPS Hilbert space $\cH_\infty$ at large $N$ equipped the finite $N$ constraints. Denote the $S_N$-invariant combinations of chiral primary oscillators in the ``large $N$'' symmetric orbifold as
\be
\left| \psi^- \psi^+ \right| = \sum_{i=1}^N \psi_{-\frac{1}{2}}^{(i)+-} \psi_{-\frac{1}{2}}^{(i)++}, \qquad \left| \psi^A \right| = \sum_{i=1}^N \psi_{-\frac{1}{2}}^{(i)+A}.
\ee
The states counted by $\Zhat_1^{(0,0)}(y,0)$ are the operators
\ie
1 :& \qquad 1 \\
2 \varepsilon y^{2n+1} :& \qquad \left| \psi^- \psi^+ \right|^n \, \left| \psi^- \right| , \quad \left| \psi^- \psi^+ \right|^n \, \left| \psi^+ \right| \\
2 y^{2n+2} :& \qquad \left| \psi^- \psi^+ \right|^{n+1}, \quad \left| \psi^- \psi^+ \right|^{n}\, \left| \psi^- \right| \, \left| \psi^+ \right|
\fe
acting on $|1_{--}\ra^{\otimes N}$, where $n=0,1,2,\cdots$. As \eqref{eq: unflowed counter} indicates, these states overcount the finite $N$ answer.

It is not hard to recognize that $\Zhat_1^{(2,0)}$, expressed as a power series \eqref{eq: counter 2} in the domain $|y|<1$, is the partition function over states that are null for a value of $N$. The null states can be written as
\ie
y^{2N} :& \qquad \left| \psi^- \psi^+ \right|^{N} - \left| \psi^- \psi^+ \right|^{N-1} \, \left| \psi^- \right| \, \left| \psi^+ \right| \\
2\varepsilon y^{2N+2n+1} :& \qquad \left| \psi^- \psi^+ \right|^{N+n} \, \left| \psi^- \right| , \quad \left| \psi^- \psi^+ \right|^{N+n} \, \left| \psi^+ \right| \\
2 y^{2N+2n+2} :& \qquad \left| \psi^- \psi^+ \right|^{N+n+1}, \quad \left| \psi^- \psi^+ \right|^{N+n}\, \left| \psi^- \right| \, \left| \psi^+ \right|
\fe
acting on $|1_{--}\ra^{\otimes N}$. The spectrum of these null states must be subtracted from $\Zhat_1^{(0,0)}$. This is the CFT explanation for the overall minus sign \eqref{eq: counter 2} in $\Zhat_1^{(2,0)}$. In the gravitational path integral at one-loop, we expect the minus sign of $\Zhat_1^{(2,0)}$ to arise from the aforementioned contour rotations for a pair of modes that have negative mass-squared when $|y|<1$. 

We thus propose the following state-counting interpretation for the partition functions of spectrally-flowed $\adsS$ defined with rotated contours for the negative modes: The bulk states counted by the partition functions of spectrally-flowed $\adsM$ are not dual to any physical states of $\symM$; rather, their microcanonical contribution to the full spectrum indicates that they are holographicaly dual to the chiral primary states of $\symM$ that become null at a finite value of $N$.

\section{$\orbifold$ orbifolds} \label{sec: orbifolds}

In this section, we show that the grand-canonical residues at higher $k$ coincide with the supersymmetric partition functions of strings on $\orbifoldM$ backgrounds and their asymmetric generalizations, under the assumption of a certain Gauss constraint. This constraint on the multiparticle spectrum of $\orbifoldM$ backgrounds appears not to be necessary from the worldsheet treatment of orbifolds, but we find that it is nevertheless required in the holographic computation of the BPS partition function $Z_N$ of $\symM$.

\subsection{Geometries labelled by higher $k$} \label{subsec: higher k}

We still have left the four infinite sequences of simple poles at higher $k>1$
\ie
(0,0)_k^m:& \quad p \, = \, e^{2 \pi i \frac{m}{k}} y^{-\left(1 - \frac{1}{k}\right)} \bar{y}^{-\left(1 - \frac{1}{k}\right)} \\
(2,0)_k^m:& \quad p \, = \, e^{2 \pi i \frac{m}{k}} y^{-\left(1 + \frac{1}{k}\right)} \bar{y}^{-\left(1 - \frac{1}{k}\right)} \\
(0,2)_k^m:& \quad p \, = \, e^{2 \pi i \frac{m}{k}} y^{-\left(1 - \frac{1}{k}\right)} \bar{y}^{-\left(1 + \frac{1}{k}\right)} \\
(2,2)_k^m:& \quad p \, = \, e^{2 \pi i \frac{m}{k}} y^{-\left(1 + \frac{1}{k}\right)} \bar{y}^{-\left(1 + \frac{1}{k}\right)}
\fe
on top of the spectrally-flowed $\adsM$ backgrounds. We focus on the case $(0,0)$ in the next few subsections and then generalize the discussion to the other three cases in Section \ref{subsec: orbifold spectral flows}.

Consider the sum (over $m$) of grand-canonical residues at the poles $(0,0)_k^m$ with fixed $k$. The result can be written as
\be \label{eq: cc orbifold residue}
\Zhat_{k}^{(0,0)}(y,\bar{y}) = \frac{1}{k} \sum_{m=0}^{k-1} D_{k}^{(0,0)} \! \left( e^{i \pi m (1 + \delta)} y,e^{i \pi m(1 - \delta)} \yb \right)
\ee
where $\delta$ is any odd integer. 
For the $\Kthree$ case, this function is
\begin{multline} \label{eq: cc orbifold K3 detail}
D_{k}^{(0,0)}(y,\yb) = y^{N\left(1 - \frac{1}{k}\right)} \bar{y}^{N\left(1 - \frac{1}{k}\right)} \prod_{\substack{n=1 \\ n \neq k}}^\infty \frac{1}{(1- y^{\frac{n}{k}-1} \bar{y}^{\frac{n}{k}-1})} \\
\times \prod_{n=1}^\infty \frac{1}{(1-y^{\frac{n}{k}-1} \bar{y}^{\frac{n}{k}+1})(1-y^{\frac{n}{k}+1} \bar{y}^{\frac{n}{k}-1})(1-y^{\frac{n}{k}+1} \bar{y}^{\frac{n}{k}+1})(1-y^{\frac{n}{k}} \bar{y}^{\frac{n}{k}})^{20}}
\end{multline}
and for the $\Tfour$ case, it is
\begin{multline} \label{eq: cc orbifold T4 detail}
D_{k}^{(0,0)}(y,\yb) = y^{N\left(1 - \frac{1}{k}\right)} \bar{y}^{N\left(1 - \frac{1}{k}\right)} \prod_{\substack{n=1 \\ n \neq k}}^\infty \frac{1}{(1- y^{\frac{n}{k}-1} \bar{y}^{\frac{n}{k}-1})} \\
\times \prod_{n=1}^\infty \frac{(1 + \varepsilon y^{\frac{n}{k}-1} \bar{y}^{\frac{n}{k}})^2 (1 + \varepsilon y^{\frac{n}{k}} \bar{y}^{\frac{n}{k}-1})^2 (1 + \varepsilon y^{\frac{n}{k}+1} \bar{y}^{\frac{n}{k}})^2 (1 + \varepsilon y^{\frac{n}{k}} \bar{y}^{\frac{n}{k}+1})^2}{(1-y^{\frac{n}{k}-1} \bar{y}^{\frac{n}{k}+1})(1-y^{\frac{n}{k}+1} \bar{y}^{\frac{n}{k}-1})(1-y^{\frac{n}{k}+1} \bar{y}^{\frac{n}{k}+1})(1-y^{\frac{n}{k}} \bar{y}^{\frac{n}{k}})^{4}}.
\end{multline}
The quantity $D_{k}^{(0,0)} \! \left( e^{i \pi m (1 + \delta)} y,e^{i \pi m(1 - \delta)} \yb \right)$ is independent of $\delta$ for any choice of odd integer $\delta$. We will find in this section that \eqref{eq: cc orbifold residue} is the supersymmetric one-loop partition function of IIB strings on $\orbifoldM$. See Appendix \ref{app: relevant formulas} for explicit formulas.

\noindent \textit{Fractional charges}

The infinite products of \eqref{eq: cc orbifold residue}, which we interpret as the one-loop determinant around the $\orbifold$ background, contain fractional powers of $y,\yb$ in units of $1/k$. This suggests that states obtained by quantizing the fluctuations of $\orbifoldM$ possess R-charges $j,\jb$ that are quantized in units of $1/2k$.\footnote{Note that this feature is \textit{not} one which is exhibited by the fractional modes in the Hilbert space of the dual symmetric orbifold CFT. The fractional modes in $\symM$ have conformal dimensions $h,\hb$ that are quantized in units of $\frac{1}{2\cdot 1}, \frac{1}{2\cdot 2}, \cdots, \frac{1}{2 \cdot N}$ but have charges $j,\jb$ that are half-integer quantized.} We will see in the next subsection that this is indeed the case. The sum over the $\bZ_k$-images $m=0,1,2,\cdots,k-1$, interpreted as the Gauss law constraint, projects the spectrum onto that involving only states of half-integer $j,\jb$.

\noindent \textit{Accumulation point and negative modes}

The powers $y^{N\left(1 - \frac{1}{k}\right)} \bar{y}^{N\left(1 - \frac{1}{k}\right)}$, accounting for the classical R-charges of the $\orbifold$ solution, have an accumulation point as $k \to \infty$. This would normally pose a problem for the convergence of the sum over saddles.\footnote{See \cite{Stanford:2025llj} for recent discussion in a closely-related setting.}

For the case of $\Kthree$, what saves the day is the fact that the ``one-loop determinant'' contains bosonic negative modes $n<k$. Let us examine the terms in \eqref{eq: cc orbifold K3 detail} corresponding to the contribution from modes that are negative when $|y|,|\yb|<1$:
\be
\frac{1}{\prod_{n=1}^{k-1} (1 - y^{\frac{n}{k}-1} \yb^{\frac{n}{k}-1})(1-y^{\frac{n}{k}-1} \bar{y}^{\frac{n}{k}+1})(1-y^{\frac{n}{k}+1} \bar{y}^{\frac{n}{k}-1})}.
\ee
For simplicity, we restrict to the domain of fugacities where $|y|=|\yb|<1$.\footnote{The behavior of these partition functions in more general fugacity domains $y,\yb$ is quite rich. The general case is discussed in Section \ref{sec: finite N saddles}.} Then the latter two denominators no longer correspond to negative mode contributions. However, the first term is most naturally expressed in the form
\be \label{eq: analytic cont cc}
\frac{1}{\prod_{n=1}^{k-1} (1 - y^{\frac{n}{k}-1} \yb^{\frac{n}{k}-1})} = \frac{ (-1)^{k-1} y^{\frac{1}{2}(k-1)} \yb^{\frac{1}{2}(k-1)}}{\prod_{n=1}^{k-1} (1 - y^{1-\frac{n}{k}} \yb^{1-\frac{n}{k}})}.
\ee
We see that the effect of bosonic negative modes in the one-loop determinant of $\orbifoldK$ is to shift the ground state energy and R-charge of $\orbifold$ from the classical value $\frac{N}{2}\left( 1- \frac{1}{k} \right)$ to the quantum-corrected value
\be
J_0 = \Jb_0 = \frac{N}{2}\left( 1- \frac{1}{k} \right) + \frac{1}{4}(k-1).
\ee
The one-loop correction to the ground state charge ``resolves'' the accumulation point: it ensures that the overall charge weights
\be
y^{N \left( 1 - \frac{1}{k} \right) + \frac{1}{2}(k-1)} \, \yb^{N \left( 1 - \frac{1}{k} \right) + \frac{1}{2}(k-1)}
\ee
carried by $\orbifold$ do not accumulate as $k \to \infty$ for any fixed $N$.

The presence of bosonic negative modes in the determinant has another important consequence. As seen in \eqref{eq: analytic cont cc}, geometries labelled by higher $k$ will contribute to the full partition function with alternating overall signs $(-1)^{k-1}$. In a gravitational path integral computation at one-loop, we expect these signs to be produced by a contour rotation for an even number of bosonic negative modes on $\orbifoldM$, in analogy to that observed in the context of giant graviton brane partition functions in $\AdS_5 \times {\rm S}^5$ \cite{Lee:2024hef}.

The case of $\Tfour$ is more subtle. Here, the quantum correction to the ground state energy/charge from bosonic negative modes cancels with that from fermionic negative modes. Hence, the accumulation point problem would appear to persist. Nonetheless, in Section \ref{sec: finite N saddles}, we will show that a regularization scheme renders the sum over the supersymmetric partition functions of $\orbifoldT$ well-defined. The result of the regularized sum exactly reproduces the spectrum of chiral primary states in $\symT$ at finite $N$.

\subsection{Classical aspects} \label{subsec: classical aspects}

We now discuss geometric properties of the $\orbifold$ solutions. We will show that these geometries have asymptotically $\AdS_3$ boundary conditions and that they possess classical charges that are in agreement with those in the residues \eqref{eq: cc orbifold residue}.

The metric of $\orbifold$ in the string frame is
\be \label{eq: orbifold metric}
ds^2 = Q \left[-(r^2 + 1) dt^2 + \frac{dr^2}{r^2 + 1} + r^2 d\phi^2 + d\theta^2 + \cos^2\theta d\psi^2 + \sin^2\theta d\chi^2 \right],
\ee
where $Q=\sqrt{Q_1 Q_5}$. The coordinates are identified as
\be \label{eq: orbifold action}
(\phi,\psi,\chi) \sim (\phi + \frac{2\pi}{k}, \psi - \frac{2\pi}{k},\chi) \sim (\phi, \psi + 2\pi, \chi) \sim (\phi,\psi,\chi + 2\pi),
\ee
and the fixed locus $\bR \times S^1$ of the orbifold is located at $r=0$ and $\theta=\frac{\pi}{2}$. The simultaneous orbifold of $AdS_3$ and $S^3$ by the cyclic group $\bZ_k$ is compatible with the supersymmetries preserved by the left-right chiral primary sector. In terms of the coordinates \eqref{eq: orbifold metric}, the energies and charges are given by the Killing vectors
\ie\label{eq:Killingvectors}
L_0 &= \frac{1}{2}(H + P_\phi) = \frac{i}{2} (\pa_t - \pa_\phi) \\
\Lb_0 &= \frac{1}{2}(H - P_\phi) = \frac{i}{2} (\pa_t + \pa_\phi) \\
J_0 &= \frac{1}{2}(-J_{\chi} + J_{\psi}) = \frac{i}{2} (\pa_{\chi} - \pa_{\psi}) \\
\Jb_0 &= -\frac{1}{2}(J_{\chi} + J_{\psi}) = \frac{i}{2} (\pa_{\chi} + \pa_{\psi}).
\fe
The $\bZ_k$ group acting as \eqref{eq: orbifold action} is generated by
\begin{equation}\label{eq:Zkaction}
g = e^{\frac{2\pi i}{k} (L_0 - \Lb_0)} e^{-\frac{2\pi i}{k}(J_0 - \Jb_0)} = e^{\frac{2\pi i}{k} (L_0 - J_0)} \otimes e^{-\frac{2\pi i}{k}(\Lb_0 - \Jb_0)},
\end{equation}
so it is apparent that left-right chiral primary states $|\varphi\ra$ are preserved under
\be
g^m |\varphi\ra = |\varphi\ra, \qquad g^m \in \bZ_k
\ee
where $m=0,1,\cdots, k-1$.

It is not immediately obvious that the metric \eqref{eq: orbifold metric} with identifications \eqref{eq: orbifold action} satisfies the boundary conditions required for inclusion in an $\AdS_3$ gravitational path integral. Let us show that the metric can be brought into a form that is manifestly asymptotically $\AdS_3$. It turns out that there are infinitely many distinct transformations that relate $\orbifold$ to asymptotically-$\AdS_3$ metrics with conical defects of deficit angle $2\pi (1-\frac{1}{k})$. Upon reduction to three-dimensions, these spaces differ only by discrete choices for the asymptotic values of the gauge fields. Hence we need to determine which transformations correspond to solutions satisfying the left- and right-moving chiral primary conditions.

Consider the coordinates
\ieg \label{eq: coordinate transformation cc}
\tilde{t} = k t, \qquad \tilde{r} = \frac{1}{k} r, \qquad \tilde{\phi} = k \phi, \\
\tilde{\psi} = \psi - (k \nu - 1) \phi, \qquad \tilde{\chi} = \chi - (k \nu - 1) t,
\feg
labelled by integer $\nu$.\footnote{There are more general transformations that make the coordinates $2\pi$-periodic. Requiring the defect solutions to be supersymmetric constrains the transformations to be of the written type \cite{Balasubramanian:2000rt}.} The resulting metrics
\begin{multline}
ds^2 = Q \bigg[- \left( \tilde{r}^2 + \frac{1}{k^2} \right) d\tilde{t}^2 + \frac{1}{\tilde{r}^2 + \frac{1}{k^2}} d\tilde{r}^2 + \tilde{r}^2 d\tilde{\phi}^2 \\
+ d\tilde{\theta}^2 + \cos^2\tilde{\theta} \left( d\tilde{\psi} + \left( \nu - \frac{1}{k} \right) d\tilde{\phi} \right)^2 + \sin^2\tilde{\theta} \left( d\tilde{\chi} + \left( \nu - \frac{1}{k}\right) d\tilde{t} \right)^2 \bigg],
\end{multline}
have the identifications
\be
(\tilde{\phi},\tilde{\psi},\tilde{\chi}) \sim (\tilde{\phi} + 2\pi, \tilde{\psi} - 2\pi \nu,\tilde{\chi}) \sim (\tilde{\phi}, \tilde{\psi} + 2\pi, \tilde{\chi}) \sim (\tilde{\phi},\tilde{\psi},\tilde{\chi} + 2\pi).
\ee
The Kaluza-Klein reduction on $S^3$ gives rise to the three-dimensional metric \cite{Balasubramanian:2000rt,Maldacena:2000dr}
\be \label{eq: 3d metric after KK}
ds^2 = Q \left[- \left( \tilde{r}^2 + \frac{1}{k^2} \right) d\tilde{t}^2 + \frac{1}{\tilde{r}^2 + \frac{1}{k^2}} d\tilde{r}^2 + \tilde{r}^2 d\tilde{\phi}^2 \right]
\ee
and the $SU(2)_L \times SU(2)_R$ gauge fields
\be
A = \left( \nu - \frac{1}{k} \right) \sigma^3 d\tilde{\phi}, \qquad \bar{A} = \left( \nu - \frac{1}{k} \right) \sigma^3 d\tilde{\phi}.
\ee

We can compute the charges associated with the solutions obtained from the KK reduction. Following \cite{Balasubramanian:2000rt}, 3d gravity solutions of the form
\ieg
ds^2 = - \left( r^2 - M_3 + \frac{16 G_3^2 J_3^2}{r^2} \right) dt^2 + \left(r^2 - M_3 + \frac{16 G_3^2 J_3^2}{r^2} \right)^{-1} dr^2 + r^2 \left(d\phi - \frac{4 G_3 J_3}{r^2} dt \right)^2 \\
A = A_\phi^3 \, \sigma^3 d\phi , \quad \bar{A} = \bar{A}_\phi^3 \, \sigma^3 d\phi
\feg
possess the quantum numbers
\ieg \label{eq: classical charges relation}
L_0 = \frac{c}{24} M_3 + \frac{1}{2} J_3 + \frac{c}{24} (A_\phi^3)^2 + \frac{c}{24} \\
\Lb_0 = \frac{c}{24} M_3 - \frac{1}{2} J_3 + \frac{c}{24} (\bar{A}_\phi^3)^2 + \frac{c}{24} \\
J_0 = \frac{c}{12} A_\phi^3, \qquad \Jb_0 = \frac{c}{12} \bar{A}_\phi^3
\feg
of the dual CFT. We used $c = \frac{3}{2 G_3}$ and set the AdS radius to 1. Our solutions have $M_3 = -\frac{1}{k^2}$ and $J_3 = 0$, so we find
\ie
L_0 = \Lb_0 &= \frac{N}{4k} \left( k \nu^2 - 2 \nu + k \right) \\
J_0 = \Jb_0 &= \frac{N}{2} \left( \nu - \frac{1}{k} \right)
\fe
These expressions suggest that we can interpret $\nu$ as a spectral flow parameter acting simultaneously on the left- and right-movers. The solution $\nu=0$ corresponds to the Ramond ground state with $L_0 - \frac{c}{24} = \Lb_0 - \frac{c}{24} = 0$ and R-charges $J_0 = \Jb_0 = - \frac{N}{2 k}$. However, for our purposes, we need solutions that satisfy the chiral primary conditions in the NS sector.

It is easy to see that only the choice $\nu = 1$ is consistent with the left and right chiral primary conditions $L_0 - J_0 = \Lb_0 - \Jb_0 =0$ in the NS sector, where
\be \label{eq: orbifold classical charges}
h = \hb = j = \jb = \frac{N}{2} \left( 1 - \frac{1}{k} \right).
\ee
Counting these contributions with weights $y^{2 J_0} \yb^{2 \Jb_0}$ as before, these charges reproduce precisely the ``on-shell'' piece $y^{N \left(1 - \frac{1}{k}\right)} \yb^{N \left(1 - \frac{1}{k}\right)}$ of the $(0,0)$ residues \eqref{eq: cc orbifold residue} that we identified with chiral primary partition functions of IIB strings on $\orbifoldM$.

The case $k=1$ brings us back to the global $\AdS_3$ solution with the $SU(2) \times SU(2)$ gauge fields turned off in Section \ref{sec: large N and spectral flows}.

\subsection{Strings on $\orbifold$ orbifolds} 
\label{subsec: strings on orbifolds}

In this subsection, we show that the spectrum $D_k^{(0,0)}$ in \eqref{eq: cc orbifold T4 detail}, which enters into the residue $\Zhat_{k}^{(0,0)}$, is reproduced in terms of the worldsheet partition function on $\orbifoldT$ in the chiral primary sector.

The perturbative spectrum on $\orbifoldT$ backgrounds has been computed from the worldsheet perspective in several places in the literature \cite{Martinec:2001cf,Martinec:2002xq,Gaberdiel:2023dxt}. The full spectrum of $\orbifoldT$ with one unit $\mathsf{k}=1$ of pure NS-NS flux was computed in \cite{Gaberdiel:2023dxt}.

What is new in our work is a proposal for how to sum the perturbative BPS spectra on various $\orbifold$ backgrounds into a result which reproduces the finite $N$ chiral primary partition function $Z_N$ of $\symM$. In particular, we find in Section \ref{subsec: gauss law} that, to reproduce $\Zhat_{k}^{(0,0)}$ from the string spectrum on $\orbifold$, one has to impose an additional Gauss constraint on the fractionally-charged spacetime spectrum. This corresponds to an extra $\bZ_k$-projection of the spectrum that is not required in the worldsheet treatment of orbifold backgrounds. This constraint is required also to reproduce $\Zhat_{k}^{(2,0)}, \Zhat_{k}^{(0,2)}, \Zhat_{k}^{(2,2)}$ from strings on asymmetric orbifolds in Section \ref{subsec: orbifold spectral flows}.

The partition function of chiral primary states in AdS$_3$/CFT$_2$ is independent of the units $\mathsf{k}$ of NS-NS flux. It is thus reasonable to expect that the spectrum of chiral primary excitations on $\orbifold$ is invariant under changes in $\mathsf{k}$. This allows us to use the results of the recent work \cite{Gaberdiel:2023dxt}, which computed the tensionless string spectrum on $\orbifoldT$ in the hybrid formalism based on a free-field realization of the current algebra $\mathfrak{psu}(1,1|2)_{\mathsf{k}=1}$. We review their computation of the spectrum on $\orbifold$ in the current subsection, focusing specifically on the chiral primary sector. While worldsheet methods are used in our work to compute the orbifold spectrum, we expect that the same BPS spectrum can be found via a direct Kaluza-Klein decomposition of IIB supergravity on $\orbifoldM$.

\subsubsection*{Fractional worldsheet spectral flow}

We now describe the spectrum of the tensionless string on the $\orbifold$ orbifold, based on the review of tensionless strings on $\adsT$ in Section \ref{ssec:WorldSheet}.\footnote{Strings on $\orbifold$ with $Q_5\geq 2$ can be studied in the RNS formalism, see \cite{Martinec:2001cf,Martinec:2002xq,Martinec:2017ztd,Martinec:2018nco,Martinec:2019wzw,Martinec:2020gkv,Bufalini:2021ndn,Martinec:2022okx,Martinec:2023zha}.}

Let us first look at string in the orbifold geometry $\orbifoldM$. The $\orbifold$ part is given by the original metric \eqref{eq: orbifold metric} 
but with the identification 
\begin{equation}
(\phi,\psi,\chi) \sim (\phi + \frac{2\pi}{k}, \psi - \frac{2\pi}{k},\chi) 
\sim (\phi+ 2\pi, \psi , \chi)
\sim (\phi, \psi + 2\pi, \chi) \sim (\phi,\psi,\chi + 2\pi).
\end{equation}
Therefore the boundary conditions of relevent fields in the worldsheet CFT that close up to the orbifold action are
\begin{equation}\label{eq:BoundaryCondition}
\begin{aligned}
t(\tau, \sigma+2\pi)&= t(\tau, \sigma)
\,, \qquad 
\phi(\tau, \sigma+2\pi)= \phi(\tau, \sigma)+\frac{2\pi n}{k} 
\,, \qquad \\
\chi(\tau, \sigma+2\pi)&= \chi(\tau, \sigma)
\,, \qquad 
\psi(\tau, \sigma+2\pi)= \psi(\tau, \sigma)-\frac{2\pi n}{k} \,,
\end{aligned}
\end{equation}
where 
\begin{equation}
n=1,2,\dots, k
\end{equation}
and $(\tau, \sigma)$ are the closed string worldsheet coordinate. For $n=1,2,\dots, k-1$, these are new closed string solutions that do not satisfy the original close string boundary condition; they correspond to the $n$-th twisted sector. In the $n$-th twisted sector, the endpoints of the string are identified up to the action of $g^n \in \bZ_k$ where $g = e^{\frac{2\pi i}{k} (\cJ_0^3 - \cK_0^3)} \otimes e^{-\frac{2\pi i}{k} (\bar\cJ_0^3 - \bar\cK_0^3)}$ in terms of worldsheet charges.

Recall that the spectrum of the tensionless string worldsheet CFT is generated by a sum over spectral flows from the unflowed sector by \eqref{eq:SFonworldsheet}. Comparing \eqref{eq:BoundaryCondition} with \eqref{eq:BCoriginal} (and the analoguous boundary condition for $S^3$), one can see that to describe the worldsheet CFT of tensionless string in the orbifold geometry $\orbifoldM$, one should choose the worldsheet spectral flow parameter to be
\begin{equation} \label{eq: fractional flow}
w=\frac{n}{k}, \qquad n\in \bZ_{>0}.
\end{equation}
Compared with the original worldsheet CFT, the orbifolded theory contains both an untwisted sector and the new twisted sectors. 
The untwisted sector corresponds to 
\begin{equation}
w\in \bZ_{>0}    \qquad \qquad \textrm{(untwisted sector)},
\end{equation}
as in the original worldsheet CFT. The twisted sectors are given by  
\begin{equation}
w=\frac{m}{k} + \bZ_{\geq 0}, \qquad m=1,2,\dots, k-1 \qquad \textrm{(twisted sectors)}.  
\end{equation}
In general not all of the twisted sector states will survive the orbifold invariance condition, i.e. the condition that the twisted sector state is invariant under the action of $g \in \bZ_k$. For a state to be invariant under $g$, its spacetime conformal dimensions and charges must satisfy
\be
(h-j) - (\hb - \jb) \ \in \ k \cdot \bZ.
\ee
However, it is apparent that the left-right chiral primary states automatically satisfy this condition. So, as long as we restrict ourselves to the chiral primary sector, all states obtained by the fractional spectral flow \eqref{eq: fractional flow} survive the $\bZ_k$-orbifold projection.\footnote{This $\bZ_k$-orbifold projection should be distinguished from the \textit{extra} $\bZ_k$-projection to be discussed in the next subsection. As we will see, not all of the chiral primary states in the twisted sector will survive the extra projection.}

\subsubsection*{BPS spectrum on $\orbifold$}
Let us now compute the spectrum of the orbifold worldsheet CFT.

Within each value of the $w=\frac{n}{k}$ spectrally-flowed sector of the worldsheet CFT on $\orbifoldT$, the left-moving chiral primary states form a quartet:
\begin{equation}\label{eq:BPSquartetWS1}
\begin{array}{ccc}
& h=j=\frac{w+1}{2} & \\
h=j=\frac{w}{2} & & h=j=\frac{w}{2}  \\
& h=j=\frac{w-1}{2}&
\end{array}     
\end{equation}
and similarly for the right-moving sector.
Prior to the projection onto $\bZ_k$-invariant states, the worldsheet spectrum is given by summing over all the spectrally-flow sector with 
\begin{equation}
\mathsf{z}^{\rm WS}_k(q) = \sum_{n=1}^\infty \mathsf{z}_{\frac{n}{k}}^{\rm WS}(q)\,.   
\end{equation}
Unlike in the RNS formalism, there is no need to sum over spin-structures in the hybrid formalism.

The partition function and index ($\varepsilon=\pm1$) of the worldsheet theory on $\orbifoldT$ specialized to the chiral primary sector is
\begin{equation}\label{eq:WSindexZk}
\mathsf{z}^{\rm WS}_{k}(y,\yb) = \sum_{n=1}^\infty \left| y^{\frac{n}{k}-1}(1 + 2 \varepsilon y + y^2) \right|^2 -1.
\end{equation}
To compute the multiparticle spectrum, it is convenient to write the bosonic and fermionic part of the partition function separately:
\begin{equation}\label{eq:WSpartitionfunctionZk}
\begin{aligned}
\mathsf{z}_{k,B}^{\rm WS}(y,\yb) 
&= \sum_{n=1}^\infty  \left[y^{\frac{n}{k}-1} \yb^{\frac{n}{k}-1}(1 + y^2 + \yb^2 + y^2 \yb^2 + 4 y \yb) \right]  - 1   \\
\mathsf{z}_{k,F}^{\rm WS}(y,\yb) &= \sum_{n=1}^\infty y^{\frac{n}{k}-1} \yb^{\frac{n}{k}-1}(2 y + 2 \yb + 2 y \yb^2 + 2 y^2 \yb) 
\end{aligned}    
\end{equation}
An analogous computation can be done in the case of $\orbifoldK$ by considering fractional spectral flows of \eqref{eq: K3 worldsheet spectrum}.

\subsubsection*{The multiparticle spectrum}

To find the spacetime partition function on the $\orbifold$ background, we need to consider multiparticle states built from single-string states on the worldsheet. 
The spacetime partition function (resp. index) can be found from the worldsheet partition function (resp. index) as follows. 

For the partition function, the bosonic part $\mathsf{z}^{\textrm{WS}}_{\textrm{B}}(q)$ is multiparticled as
\begin{equation}\label{eq:PEB}
\PE[f(q)] = \exp \left( \sum_{m=1}^\infty \frac{1}{m} f(q^m) \right),
\end{equation}
while the fermionic part $\mathsf{z}^{\textrm{WS}}_{\textrm{F}}(q)$ is multiparticled as
\begin{equation}\label{eq:PEF}
\widetilde{\PE}[f(q)] = \exp \left( \sum_{m=1}^\infty \frac{(-1)^{m+1}}{m} f(q^m) \right).
\end{equation}
The multiparticled index is obtained by applying $\PE$ to
\begin{equation}
\mathsf{z}^{\textrm{WS}}_{\textrm{ind}}(q)=\mathsf{z}^{\textrm{WS}}_{\textrm{B}}(q)-\mathsf{z}^{\textrm{WS}}_{\textrm{F}}(q).
\end{equation}
The variable $q$ here denotes all the fugacities collectively.

Let us multiparticle the single-string spectrum on $\orbifoldT$ (see \eqref{eq:WSindexZk} or \eqref{eq:WSpartitionfunctionZk}) via the plethystic exponential, after including the classical contribution \eqref{eq: orbifold classical charges}. We find that the partition function
\be
y^{N\left(1 - \frac{1}{k}\right)} \yb^{N\left(1 - \frac{1}{k}\right)} \, \PE[\mathsf{z}^{\rm WS}_{k,B}(y,\yb)] \, \widetilde{\PE}[\mathsf{z}^{\rm WS}_{k,F}(y,\yb)]
\ee
and the index
\be
y^{N\left(1 - \frac{1}{k}\right)} \yb^{N\left(1 - \frac{1}{k}\right)} \, \PE[\mathsf{z}^{\rm WS}_{k}(y,\yb)]
\ee
gives, respectively, the spacetime spectrum
\begin{multline} \label{eq: orbifold multiparticle cc}
D_{k}^{(0,0)}(y,\yb) = y^{N\left(1 - \frac{1}{k}\right)} \bar{y}^{N\left(1 - \frac{1}{k}\right)} \prod_{\substack{n=1 \\ n \neq k}}^\infty \frac{1}{(1- y^{\frac{n}{k}-1} \bar{y}^{\frac{n}{k}-1})} \\
\times \prod_{n=1}^\infty \frac{(1 + \varepsilon y^{\frac{n}{k}-1} \bar{y}^{\frac{n}{k}})^2 (1 + \varepsilon y^{\frac{n}{k}} \bar{y}^{\frac{n}{k}-1})^2 (1 + \varepsilon y^{\frac{n}{k}+1} \bar{y}^{\frac{n}{k}})^2 (1 + \varepsilon y^{\frac{n}{k}} \bar{y}^{\frac{n}{k}+1})^2}{(1-y^{\frac{n}{k}-1} \bar{y}^{\frac{n}{k}+1})(1-y^{\frac{n}{k}+1} \bar{y}^{\frac{n}{k}-1})(1-y^{\frac{n}{k}+1} \bar{y}^{\frac{n}{k}+1})(1-y^{\frac{n}{k}} \bar{y}^{\frac{n}{k}})^{4}}.
\end{multline}
The one-loop supersymmetric partition function of $\orbifoldT$ agrees precisely with the expression $D_{k}^{(0,0)}$ in \eqref{eq: cc orbifold T4 detail} appearing in the residue $\Zhat_k^{(0,0)}$ of the grand-canonical partition function $\cZ(p;y,\yb)$ of $\symT$.

\subsection{Gauss law} \label{subsec: gauss law}

While \eqref{eq: orbifold multiparticle cc} is a perfectly-valid partition function $D_{k}^{(0,0)}(y,\yb)$ associated to $\orbifoldT$ from the perspective of the worldsheet, our residue
\be \label{eq: residue gauss law}
\Zhat_{k}^{(0,0)}(y,\bar{y}) = \frac{1}{k} \sum_{m=0}^{k-1} D_{k}^{(0,0)} \! \left( e^{i \pi m (1 + \delta)} y,e^{i \pi m(1 - \delta)} \yb \right)
\ee
suggests that this is not the end of the story.\footnote{Recall $\delta$ is any odd integer. $\Zhat_{k}^{(0,0)}$ is independent of $\delta$ for any choice of odd integer $\delta$.} Namely, the residue consists of a sum over actions on the fractionally-charged BPS spectrum on $\orbifoldM$.

To understand what the sum is doing, it is useful to endow a Hilbert space interpretation for the residue \eqref{eq: residue gauss law}. Let
\be
\cH_k = \cH_k^{\rm cl} \otimes \cH_k^{\rm pert}
\ee
be the Hilbert space associated to the chiral primary sector of $\orbifoldM$ and its perturbative excitations, such that we can write
\be
\Tr_{\cH_k} \! \left( y^{2 J_0} \yb^{2 \Jb_0} \right) = D_k^{(0,0)}(y,\yb).
\ee
$\cH_k^{\rm cl}$ is a one-dimensional Hilbert space consisting of a state with the charges $j = \jb = \frac{N}{2}\left( 1- \frac{1}{k} \right)$ of the classical $\orbifold$ solution. $\cH_k^{\rm pert}$ is the multi-string Hilbert space of chiral primary excitations on $\orbifoldM$ that we computed in Section \ref{subsec: strings on orbifolds}. Then the residue $\Zhat_{k}^{(0,0)}$ can be written in terms of a sum over traces over $\cH_k$:
\ie \label{eq: cc orbifold match}
\Zhat_{k}^{(0,0)}(y,\bar{y}) \, &= \, \frac{1}{k} \sum_{m=0}^{k-1} \Tr_{\cH_k} \! \left(e^{2\pi i m \left[ (1 + \delta) J_0 + (1 - \delta) \Jb_0 \right]} \, y^{2 J_0} \yb^{2 \Jb_0} \right) \\
&= \, \Tr_{\cH_k} \! \left( P_k \, y^{2 J_0} \yb^{2 \Jb_0} \right)
\fe
where we defined the projection operator
\be
P_k = \frac{1}{k} \sum_{m=0}^{k-1} \Omega^m, \qquad \Omega = e^{2\pi i \left[ (1 + \delta) J_0 + (1 - \delta) \Jb_0 \right]} \in \bZ_k.
\ee
The group $\bZ_k$ is generated by $\Omega$ since $\cH_k$ consists of states with fractional R-charges. Therefore, we find that the sum over $\bZ_k$ images in $P_k$ projects $\cH_k$ onto the subspace of $\bZ_k$-invariant states with half-integer $j,\jb$. This amounts to a Gauss law constraint on the spacetime spectrum of $\orbifoldM$.

Here, it is important that the projection operator $P_k$ acts on the tensor product $\cH_k = \cH_k^{\rm cl} \otimes \cH_k^{\rm pert}$ rather than on only the perturbative Hilbert space $\cH_k^{\rm pert}$. The weight $y^{N\left(1 - \frac{1}{k}\right)} \bar{y}^{N\left(1 - \frac{1}{k}\right)}$ of the classical contribution has fractional powers if $N$ is not divisible by $k$. If $\bZ_k$ acted only on $\cH_k^{\rm pert}$, the partition function would count states with R-charges $j,\jb$ that are half-integer spaced but that are not in general half-integers. In order for $\Zhat_k^{(0,0)}$ to contain only the states with half-integer R-charges, the projection onto $\bZ_k$-invariant states must act on $\cH_k$.

Interestingly, this $\bZ_k$ action enforcing Gauss law is different from that used to define the $\orbifold$ orbifold. Recall that the latter $\bZ_k$ acting on states in $\adsS$ was generated by $g = e^{\frac{2\pi i}{k} (L_0 - J_0)} \otimes e^{-\frac{2\pi i}{k}(\Lb_0 - \Jb_0)}$ in terms of spacetime charges. (The factor $1/k$ in the exponent is needed here because the excitations on $\adsS$ do not have fractional conformal dimensions and charges.) It is clear that a projection $\frac{1}{k}\sum_{m=0}^{k-1} g^m$ of the spectrum with respect to $g$ acts trivially on all chiral primary excitations of $\orbifold$ regardless of whether they are fractionally-charged. In contrast, the Gauss law projection $P_k$ acts non-trivially on states with fractional charges.

We thus find that the insertion of $P_k$ corresponds to an \textit{extra} $\bZ_k$-projection, acting on the multiparticle spectrum of $\orbifoldM$, that is not required in the worldsheet treatment of orbifolds but is nevertheless required in the holographic computation of the partition function $Z_N$ of $\symM$. Assuming that this Gauss constraint is an ingredient in the bulk prescription for computing the contribution of the $\orbifold$ orbifold geometry to the full BPS partition function $Z_N$, we reproduce the residue $\Zhat_k^{(0,0)}(y,\yb)$ from the trace over the Hilbert space $\cH_k$ of $\orbifoldM$ and its chiral primary excitations in \eqref{eq: cc orbifold match}.

It would be important to investigate the gravitational path integral origin of the presence of what appears to be a $\bZ_k$-family of $\orbifold$ saddles that implement Gauss law.\footnote{One possibility is that, in a Euclidean path integral computation, each of the $k$ terms in $\Zhat_k^{(0,0)}$ may arise as a saddle that are related to the others under a $\bZ_k$ action. Though we do not explore this perspective in the current work, this would be an intriguing physical explanation for the Gauss law constraint.}

\subsection{Spectral flows and asymmetric orbifolds} \label{subsec: orbifold spectral flows}

Thus far, we found that the one-loop BPS partition functions of $\orbifoldM$ are encoded in the residues of the $(0,0)_k$ poles. Here, we show that a more general class of $\orbifold$ orbifolds are necessary to provide a physical interpretation for the other three sequences of poles $(2,0)_k$, $(0,2)_k$, and $(2,2)_k$.

As was done in the $(0,0)$ case, we consider a sum (over $m$) of grand-canonical residues at the poles $(2,0)_k^m$, $(0,2)_k^m$, and $(2,2)_k^m$ with fixed $k$:
\ieg \label{eq: abstract orbifold residue}
\Zhat_{k}^{(0/2,0/2)}(y,\bar{y}) = \frac{1}{k} \sum_{m=0}^{k-1} D_{k}^{(0/2,0/2)} \! \left( e^{\pm i \pi m (1 + \delta)} y,e^{\pm i \pi m(1 - \delta)} \yb \right) \\
D_{k}^{(0/2,0/2)}(y,\yb) = y^{N\left(1 \mp \frac{1}{k}\right)} \bar{y}^{N\left(1 \mp \frac{1}{k}\right)} \prod_{\substack{n=1 \\ n \neq k}}^\infty \frac{1}{ 1- y^{\pm \left(\frac{n}{k}- 1 \right)} \bar{y}^{\pm \left(\frac{n}{k}- 1 \right)} } \times \prod_{n=1}^\infty \, ({\rm others})
\feg
for both $\symT$ and $\symK$. We refer the readers to Appendix \ref{app: relevant formulas} for explicit formulas of the residues. As before, the result is independent of $\delta$ for any choice of odd integer $\delta$. We will identify \eqref{eq: abstract orbifold residue} with the one-loop supersymmetric partition functions of IIB strings on ``asymmetric'' $\orbifoldM$ orbifold backgrounds.

Following \cite{Gaberdiel:2023dxt}, let us define a two-parameter family of asymmetric $\bZ_k$-orbifolds, given by the identification on the original metric \eqref{eq: orbifold metric} 
\ie
(\phi,\psi,\chi)& \sim \left( \phi+\frac{2\pi}{k}, \psi - \frac{2\pi(s+\bar s +1)}{k}, \chi + \frac{2\pi (s-\bar s)}{k} \right)\\
& \sim (\phi, \psi + 2\pi, \chi)\sim  (\phi,\psi,\chi + 2\pi)
\fe
where $s,\bar{s}$ are integers defined modulo $k$. The group $\bZ_k$ is now generated in terms of spacetime charges by
\be
g = e^{\frac{2\pi i}{k} (L_0 - (2s+1) J_0)} \otimes e^{-\frac{2\pi i}{k} (\Lb_0 - (2\bar{s}+1) \Jb_0)}.
\ee
The case $s=\bar{s} = 0$ corresponds to the $(0,0)_k$ orbifold we studied earlier.

What are the classical dimensions and charges of the asymmetric $\orbifold$ orbifold solutions, and which of these solutions should we associate with the $(2,0)_k$, $(0,2)_k$, and $(2,2)_k$ residues? To bring the metric in a form that is asymptotically $\AdS_3$, we use the coordinates
\ieg
\tilde{t} = k t, \qquad \tilde{r} = \frac{1}{k} r, \qquad \tilde{\phi} = k \phi, \\
\tilde{\psi} = \psi - \left( k - (s+\bar{s}+1) \right) \phi - (s-\bar{s}) t \\
\tilde{\chi} = \chi - (s-\bar{s}) \phi - \left( k - (s+\bar{s}+1) \right) t.
\feg
These transformations are a slight generalization of the earlier ones \eqref{eq: coordinate transformation cc} with $\nu = 1$. The resulting metrics
\ie
ds^2 = Q \bigg[ - &\left( \tilde{r}^2 + \frac{1}{k^2} \right) d\tilde{t}^2 + \frac{1}{\tilde{r}^2 + \frac{1}{k^2}} d\tilde{r}^2 + \tilde{r}^2 d\tilde{\phi}^2 \\
+ d\tilde{\theta}^2 &+ \cos^2\tilde{\theta} \left( d\tilde{\psi} + \left( 1 - \frac{s+\bar{s}+1}{k} \right) d\tilde{\phi} + \left( \frac{s-\bar{s}}{k} \right) d\tilde{t} \right)^2 \\
&+ \sin^2\tilde{\theta} \left( d\tilde{\chi} + \left( \frac{s-\bar{s}}{k} \right) d\tilde{t} + \left( 1 - \frac{s+\bar{s}+1}{k}\right) d\tilde{t} \right)^2 \bigg]
\fe
have the identifications
\be
(\tilde{\phi},\tilde{\psi},\tilde{\chi}) \sim (\tilde{\phi} + 2\pi, \tilde{\psi} - 2\pi,\tilde{\chi}) \sim (\tilde{\phi}, \tilde{\psi} + 2\pi, \tilde{\chi}) \sim (\tilde{\phi},\tilde{\psi},\tilde{\chi} + 2\pi).
\ee
The Kaluza-Klein reduction on $S^3$ yields the three-dimensional metric \eqref{eq: 3d metric after KK} and the $SU(2)_L \times SU(2)_R$ gauge fields
\be
A = \left( 1 - \frac{2s+1}{k} \right) \sigma^3 d\tilde{\phi}, \qquad \bar{A} = \left( 1 - \frac{2\bar{s}+1}{k} \right) \sigma^3 d\tilde{\phi}.
\ee
Using the relation \eqref{eq: classical charges relation} between the classical solution and charges, we find that these solutions carry the conformal dimensions and charges
\ieg \label{eq: classical charges asymmetric}
L_0 = \frac{N}{2} \left( 1 - \frac{2s + 1}{k} \right) + \frac{N s (s+1)}{k^2} \\
\Lb_0 = \frac{N}{2} \left( 1 - \frac{2\bar{s} + 1}{k} \right) + \frac{N \bar{s} (\bar{s}+1)}{k^2} \\
J_0 = \frac{N}{2} \left( 1 - \frac{2s + 1}{k} \right) \\
\Jb_0 = \frac{N}{2} \left( 1 - \frac{2\bar{s} + 1}{k} \right).
\feg
Only the values $(s,\bar{s}) = (0,0), (-1,0), (0,-1)$, and $(-1,-1)$ correspond to chiral primary solutions, and the latter three carry charges that are in agreement with the classical weights
\be
y^{N \left( 1 + \frac{1}{k} \right)} \yb^{N \left( 1 - \frac{1}{k} \right)}, \quad y^{N \left( 1 - \frac{1}{k} \right)} \yb^{N \left( 1 + \frac{1}{k} \right)}, \quad y^{N \left( 1 + \frac{1}{k} \right)} \yb^{N \left( 1 + \frac{1}{k} \right)}
\ee
appearing in the grand-canonical residues $\Zhat_k^{(2,0)}$, $\Zhat_k^{(0,2)}$, and $\Zhat_k^{(2,2)}$, respectively. Thus, we can identify the grand-canonical poles $(2,0)_k$, $(0,2)_k$, and $(2,2)_k$ with the asymmetric $\orbifold$ orbifolds labelled by $(s,\bar{s}) = (-1,0), (0,-1)$, and $(-1,-1)$.

The dependence of the quantum numbers \eqref{eq: classical charges asymmetric} on $(s,\bar{s})$ suggests that we can view these asymmetric orbifold solutions as arising from spectral flows of the original $\orbifold$ solution either by (1) the fractional values $\eta = -\frac{2 s}{k}, \bar\eta = -\frac{2 \bar{s}}{k}$ or, alternatively, (2) the integer values $\eta = -2 s, \bar\eta = -2 \bar{s}$ but with an effective central charge $c_{\rm eff} = c/k$. For our purposes, the latter perspective will be more useful. This enlarged family of spectral flows for $\orbifold$ contains, at $k=1$, the BPS spectral flows of $\adsS$. A way to understand why the spectral flow here acts with the effective value $c_{\rm eff} = c/k$ is that the boundary dual of the classical $\orbifold$ solution is the chiral primary state $(| k_{--} \ra^{\rm BPS})^{\otimes N/k}$ in the symmetric orbifold (see Section \ref{sec: states dual to orbifolds}). Due to the number of cycles, the effective theory of excitations on $(| k_{--} \ra^{\rm BPS})^{\otimes N/k}$ has an effective central charge that is reduced by a factor of $k$ relative to that on the NS vacuum $| 1_{--} \ra^{\otimes N}$.

Before computing the one-loop spectrum on asymmetric orbifolds, we need to find out what the chiral primary condition on the perturbative excitations of the original $\orbifold$ orbifold (associated to $(0,0)_k$ poles) gets mapped to under a spectral flow to asymmetric orbifolds. Let us write the BPS partition function of the original $\orbifold$ as
\be
\lim_{q,\qb \to 0} \, y^{N\left(1-\frac{1}{k}\right)} \yb^{N\left(1-\frac{1}{k}\right)} \Tr_{\cH_k^{\rm pert}} \left( q^{l_0 - j_0} \qb^{\lb_0 - \jb_0} y^{2 j_0} y^{2 \jb_0} \right),
\ee
prior to imposing Gauss law. We separated the charges as
\ie
L_0 &= L_0^{\rm cl} + l_0 \\
J_0 &= J_0^{\rm cl} + j_0
\fe
into parts acting on $\cH_k^{\rm cl}$ and $\cH_k^{\rm pert}$, respectively, to emphasize that the spectral flow operation with $c_{\rm eff}$ will act on the generators $l_0, j_0$ for the charges of perturbative excitations on $\orbifold$. Similar statements apply to the right-moving sector. The limit $q,\qb \to 0$ projects $\cH_k^{\rm pert}$ to the left- and right-chiral primary states $l_0 - j_0 = \lb_0 - \jb_0 =0$.

Now consider the spectral flow of $\orbifold$ by $(\eta,\bar\eta) = (- 2 s, -2 \bar{s})$ with the effective central charge $c_{\rm eff} = c/k$. The partition function of the flowed $\orbifold$ is
\be
\lim_{q,\qb \to 0} \, q^{\frac{N s(s+1)}{k}} \qb^{\frac{N \bar{s}(\bar{s}+1)}{k}} y^{N\left(1-\frac{2s+1}{k}\right)} \yb^{N\left(1-\frac{2\bar{s}+1}{k}\right)} \, \Tr_{\cH_k^{\rm pert}} \left( q^{l_0 - (2s+1) j_0} \qb^{\lb_0 - (2\bar{s}+1) \jb_0} y^{2 j_0} y^{2 \jb_0} \right).
\ee
The result is non-vanishing under the projection $q,\qb \to 0$ only if
\be
(s,\bar{s}) = (0,0), \,(-1,0),\, (0,-1),\, (-1,-1),
\ee
so only these classes of asymmetric orbifolds are valid contributions to $Z_N$. Furthermore, note that taking $s=-1$ would correspond to imposing the \textit{anti-chiral} primary condition
\be
l_0 + j_0 = 0
\ee
on the left, and similarly for $\bar{s}=-1$ on the right, in terms of the flowed generators. This generalizes our findings in Section \ref{subsec: bulk spectral flows} regarding the need to consider left and/or right anti-chiral primary excitations with respect to the flowed generators on spectrally-flowed $\adsS$.

Let us now compute the spectrum of BPS excitations on asymmetric $\orbifoldT$ orbifolds, using again the worldsheet analysis in the tensionless limit \cite{Gaberdiel:2023dxt}. Now the boundary conditions of fields that close up to the worldsheet orbifold action
\be
g = e^{\frac{2 \pi i}{k} (\cJ_0^3 - (2s+1)\cK_0^3)} \otimes e^{-\frac{2 \pi i}{k} (\bar{\cJ}_0^3 - (2\bar{s}+1)\bar{\cK}_0^3)}
\ee
are
\ie \label{eq: asymmetric BC} 
t(\tau, \sigma+2\pi)&= t(\tau, \sigma)\\
\phi(\tau, \sigma+2\pi)&= \phi(\tau, \sigma)+\frac{2\pi n}{k} \\
\chi(\tau, \sigma+2\pi)&= \chi(\tau, \sigma)+\frac{2\pi n (s-\bar{s})}{k} \\
\psi(\tau, \sigma+2\pi)&= \psi(\tau, \sigma)-\frac{2\pi n (s+\bar{s}+1)}{k}.
\fe
The boundary condition of $\phi(\tau,\sigma)$ is unchanged from that in \eqref{eq:BoundaryCondition}, so, by comparison to \eqref{eq:BCoriginal}, the worldsheet spectral flow $w$ will continue to take the values
\be
w = \frac{n}{k}, \quad n= 1,2,3,\cdots.
\ee
The orbifold invariance condition for asymmetric orbifolds is
\be \label{eq: asymmetric invariance}
(h - (2s+1) j) - (\hb - (2\bar{s}+1) \jb) \ \in \ k \cdot \bZ.
\ee
As in the $(0,0)$ case, we will be restricting to only the (anti-)chiral primary excitations on asymmetric orbifolds. They automatically satisfy \eqref{eq: asymmetric invariance}, so this condition will not impose any further constraints on our spectrum.

We now turn to the spectrum of the asymmetric orbifold worldsheet CFT. For $s=0$, the relevant left-moving states are chiral primaries. They form the quartet
\be
\begin{array}{ccc}
& h=j=\frac{w+1}{2} & \\
h=j=\frac{w}{2} & & h=j=\frac{w}{2}  \\
& h=j=\frac{w-1}{2}&
\end{array}     
\ee
as before. A new feature of asymmetric orbifolds that is that, for $s=-1$, the relevant left-moving states are anti-chiral primaries. They form the quartet
\be
\begin{array}{ccc}
& -h=j=-\frac{w-1}{2} & \\
-h=j=-\frac{w}{2} & & -h=j=-\frac{w}{2}  \\
& -h=j=-\frac{w+1}{2}.&
\end{array}     
\ee
The right-moving spectrum for $\bar{s}=0,-1$ is analogous. Summing over the spectrally-flow sectors, the BPS partition function and index of the worldsheet theory on asymmetric $\orbifoldT$ orbifolds are
\be
\mathsf{z}^{\rm WS}_{k}(y,\yb) = \sum_{n=1}^\infty \left[ y^{(2s+1)\frac{n}{k}} \yb^{(2\bar{s}+1)\frac{n}{k}} \left| y^{-1} (1 + 2 \varepsilon y + y^2) \right|^2 \right] -1
\ee
with bosonic and fermionic parts
\ie
\mathsf{z}_{k,B}^{\rm WS}(y,\yb) 
&= \sum_{n=1}^\infty  \left[y^{(2s+1)\frac{n}{k}-1} \yb^{(2\bar{s}+1)\frac{n}{k}-1}(1 + y^2 + \yb^2 + y^2 \yb^2 + 4 y \yb) \right]  - 1   \\
\mathsf{z}_{k,F}^{\rm WS}(y,\yb) &= \sum_{n=1}^\infty y^{(2s+1)\frac{n}{k}-1} \yb^{(2\bar{s}+1)\frac{n}{k}-1}(2 y + 2 \yb + 2 y \yb^2 + 2 y^2 \yb).
\fe

Let us compute the spacetime spectrum from the worldsheet spectrum on asymmetric $\orbifoldT$. After including the classical contribution, we find the one-loop supersymmetric partition function
\be
y^{N\left(1 - \frac{1}{k}\right)} \yb^{N\left(1 - \frac{1}{k}\right)} \, \PE[\mathsf{z}^{\rm WS}_{k,B}(y,\yb)] \, \widetilde{\PE}[\mathsf{z}^{\rm WS}_{k,F}(y,\yb)]
\ee
and the index
\be
y^{N\left(1 - \frac{1}{k}\right)} \yb^{N\left(1 - \frac{1}{k}\right)} \, \PE[\mathsf{z}^{\rm WS}_{k}(y,\yb)]
\ee
of asymmetric $\orbifoldT$. They agree with the expressions
\be
D_{k}^{(0/2,0/2)}(y,\yb) = y^{N\left(1 \mp \frac{1}{k}\right)} \bar{y}^{N\left(1 \mp \frac{1}{k}\right)} \prod_{\substack{n=1 \\ n \neq k}}^\infty \frac{1}{ 1- y^{\pm \left(\frac{n}{k}- 1 \right)} \bar{y}^{\pm \left(\frac{n}{k}- 1 \right)} } \times \prod_{n=1}^\infty \, ({\rm others})
\ee
appearing in the residues $\Zhat_k^{(0/2,0/2)}$ of the grand canonical partition function $\cZ(p;y,\yb)$ of $\symT$. 

As before, our residue
\be
\Zhat_{k}^{(0/2,0/2)}(y,\bar{y}) = \frac{1}{k} \sum_{m=0}^{k-1} D_{k}^{(0/2,0/2)} \! \left( e^{\pm i \pi m (1 + \delta) } y,e^{\pm i \pi m (1 - \delta)} \yb \right)
\ee
suggests that the spectrum on asymmetric $\orbifold$ orbifolds are subject to the Gauss law. The projection operator implementing Gauss law now takes the form
\be
P_k = \frac{1}{k} \sum_{m=0}^{k-1} \Omega^m, \qquad \Omega = e^{2\pi i m \left[ (2s+1)(1 + \delta) J_0 + (2\bar{s}+1) (1 - \delta) \Jb_0 \right]} \in \bZ_k,
\ee
where only the values $s,\bar{s} = 0,-1$ are considered. The operator $P_k$ projects the spectrum to states with half-integer R-charges $j,\jb \in \frac{1}{2}\bZ$. Again, we make the crucial assumption that this Gauss constraint is an ingredient in the bulk prescription for computing the contribution of the asymmetric $\orbifold$ orbifolds to the full BPS partition function $Z_N$. Under this assumption, we reproduce the residues $\Zhat_k^{(0/2,0/2)}(y,\yb)$ in \eqref{eq: abstract orbifold residue} from
\be
\Zhat_k^{(0/2,0/2)}(y,\yb) = \Tr_{\cH_k} \left( P_k \, y^{2 J_0} \yb^{2 \Jb_0} \right),
\ee
the trace over the Hilbert space $\cH_k$ of the IIB theory on asymmetric $\orbifoldM$ backgrounds and their (anti-)chiral primary excitations.

\section{States dual to $\orbifold$ and spectral flows} \label{sec: states dual to orbifolds}

In \cite{Martinec:2001cf,Martinec:2002xq,Gaberdiel:2023dxt}, it was proposed that the holographic dual of the $\orbifold$ geometry is the non-perturbative ``vacuum'' state
\be
(| k_{--} \ra^{\rm BPS})^{\otimes N/k}: \ h = j = \hb = \jb = \frac{N}{2}\left(1 - \frac{1}{k}\right)
\ee
in the NS sector of $\symM$, consisting of $N/k$-many cycles $| k_{--} \ra^{\rm BPS}$ of length $k$.\footnote{In these works, it was assumed that $N$ is divisible by $k$. We relax this assumption in our work; $N = 1/4 G_N^{(3)}$ is an abstract parameter from the perspective of a gravity computation. We find that saddles for which $N/k$ is not an integer will be essential to reproduce $Z_N$.} Let us refer to this state as the $k$-wound vacuum. A minor generalization to asymmetric $\orbifold$ suggests that we may identify those labelled by $(s,\bar{s}) = (-1,0), (0,-1), (-1,-1)$ with the states
\ie
(| k_{+-} \ra^{\rm BPS})^{\otimes N/k} &: \ h = j = \frac{N}{2}\left( 1 + \frac{1}{k} \right), \quad \hb = \jb = \frac{N}{2}\left( 1 - \frac{1}{k} \right) \\
(| k_{-+} \ra^{\rm BPS})^{\otimes N/k} &: \ h = j = \frac{N}{2}\left( 1 - \frac{1}{k} \right), \quad \hb = \jb = \frac{N}{2}\left( 1 + \frac{1}{k} \right) \\
(| k_{++} \ra^{\rm BPS})^{\otimes N/k} &: \ h = j = \frac{N}{2}\left( 1 + \frac{1}{k} \right), \quad \hb = \jb = \frac{N}{2}\left( 1 + \frac{1}{k} \right),
\fe
respectively. These states have energies of order $\sim N$ above the true vacuum $|1_{--}\ra^{\otimes N}$ in the NS sector of the symmetric orbifold. The worldsheet spectrum on $\orbifold$ orbifolds was found \cite{Gaberdiel:2023dxt} to be in agreement with the spectrum of perturbative CFT excitations of the $k$-wound vacuum.\footnote{The $\bZ_k$-twisted sector of ``excitations'' on the $k$-wound vacuum in the CFT consists of $k$-cycles that fractionate into smaller cycles carrying fractional charges: $(k) \to (1)^{n_1} (2)^{n_2} \cdots (k)^{n_k}$ with $\sum_{j=1}^k j n_j = k$. These ``excitations'' have negative conformal dimension and charges, so the $k$-wound vacuum is unstable for $k > 1$.} As we observed earlier, a crucial fact about the $\bZ_k$-twisted sector of the spectrum $\Zhat_k^{(0,0)}$ on $\orbifold$ orbifolds is that it contains modes whose mass-squared is negative in the fugacity domain $|y|,|\yb|<1$. For the path integral over these modes to be well-defined, the contours for these modes would need to be Wick rotated to the imaginary axes.

What is the CFT interpretation of the $\orbifold$ spectrum computed with rotated contours for these negative modes? We propose that a path integral quantization of BPS fluctuations of $\orbifold$ geometries and their spectral flows, defined with rotated contours for the negative modes, produces bulk states that are holographically dual to the chiral primary states of $\symM$ that become null at a finite value of $N$. This is directly analogous to what we found for spectrally-flowed $\adsS$ in Section \ref{subsec: states dual to flows}.

Let us demonstrate our claim by considering the contribution of $\orbifoldT$ to the chiral primary spectrum of $\symT$ at $N=1$
\be \label{eq: n1 result state counting}
Z_{N=1} = 1 + (4 \varepsilon) y + 6 y^2 + (4 \varepsilon) y^3 + y^4,
\ee
where we set $\yb = y$ for simplicity. $\varepsilon=\pm 1$ for the partition function and index, respectively. Due to subtleties related to fermionic negative modes in the $\Tfour$ case (see Section \ref{subsec: deriving residue formula}), we will keep $\varepsilon$ abtract and use it to regularize the microcanonical contributions of $\orbifold$ orbifolds.

In the domain $|y|=|\yb|<1$ of fugacity space, it turns out that only the $(0,0)$ set of orbifold geometries contribute to the chiral primary spectrum:
\be \label{eq: saddle sum state counting}
Z_N(y,\yb) = \sum_{k=1}^\infty \Zhat_k^{(0,0)}(y,\yb), \qquad |y|=|\yb|<1.
\ee
We will explain why this is the case in Section \ref{sec: finite N saddles}. Assuming for the time being that this formula holds, we can study the state-counting interpretation of the supersymmetric partition functions on $\orbifoldT$:
\ie
&\widehat{Z}_{k}^{(0,0)}(y,\bar{y}) = \sum_{m=0}^{k-1}  \frac{1}{k} e^{-2 \pi i N \frac{m}{k}} y^{N\left(1 - \frac{1}{k}\right)} \bar{y}^{N\left(1 - \frac{1}{k}\right)} \prod_{\substack{n=1 \\ n \neq k}}^\infty \frac{1}{(1-e^{2 \pi i n \frac{m}{k}} y^{\frac{n}{k}-1} \bar{y}^{\frac{n}{k}-1})} \\
&\times \prod_{n=1}^\infty \frac{(1 + \varepsilon e^{2 \pi i n \frac{m}{k}} y^{\frac{n}{k}-1} \bar{y}^{\frac{n}{k}})^2 (1 + \varepsilon e^{2 \pi i n \frac{m}{k}} y^{\frac{n}{k}} \bar{y}^{\frac{n}{k}-1})^2 (1 + \varepsilon e^{2 \pi i n \frac{m}{k}} y^{\frac{n}{k}+1} \bar{y}^{\frac{n}{k}})^2 (1 + \varepsilon e^{2 \pi i n \frac{m}{k}} y^{\frac{n}{k}} \bar{y}^{\frac{n}{k}+1})^2}{(1-e^{2 \pi i n \frac{m}{k}} y^{\frac{n}{k}-1} \bar{y}^{\frac{n}{k}+1})(1-e^{2 \pi i n \frac{m}{k}} y^{\frac{n}{k}+1} \bar{y}^{\frac{n}{k}-1})(1-e^{2 \pi i n \frac{m}{k}} y^{\frac{n}{k}+1} \bar{y}^{\frac{n}{k}+1})(1-e^{2 \pi i n \frac{m}{k}} y^{\frac{n}{k}} \bar{y}^{\frac{n}{k}})^{4}}.
\fe
The first few are given by
\ie \label{eq: degeneracies state counting}
\Zhat_1^{(0,0)} &= 1 + (4 \varepsilon) y + (7 + 6 \varepsilon^2) y^2 + (36 \varepsilon + 4 \varepsilon^3) y^3 + \cdots \\
\Zhat_2^{(0,0)} &= -(1 + 6 \varepsilon^2 + \varepsilon^4) y^2 - (36 \varepsilon + 56 \varepsilon^3 + 4 \varepsilon^5) y^3 - \cdots \\
\Zhat_3^{(0,0)} &= +(\varepsilon^4) y^2 + \left(4 \varepsilon + 56 \varepsilon^3 + 60 \varepsilon^5 + O(\varepsilon^7) \right) y^3 + \cdots \\
\Zhat_4^{(0,0)} &= - \left(4 \varepsilon^3 + 60 \varepsilon^5 + O(\varepsilon^7) \right) y^3 - \cdots \\
\Zhat_5^{(0,0)} &= + \left(4 \varepsilon^5 + O(\varepsilon^7) \right) y^3 + \cdots,
\fe
which, due to negative modes that are present for $|y|<1$, contribute with signs $(-1)^{k-1}$ that alternate with $k$.

The partition function $\Zhat_1^{(0,0)}$ for the large $N$ Kaluza-Klein spectrum on $\adsM$ vastly overcounts the true answer \eqref{eq: n1 result state counting} at finite $N$. To account for its degeneracies in \eqref{eq: degeneracies state counting} from the CFT perspective, define the $S_N$-invariant combinations of chiral primary operators
\ieg
\left| \psi^A \right| = \sum_{i=1}^N \psi_{-\frac{1}{2}}^{(i)+A}, \quad \left| \bar\psi^A \right| = \sum_{i=1}^N \bar\psi_{-\frac{1}{2}}^{(i)+A}, \quad \left| \psi^A \bar\psi^B \right| = \sum_{i=1}^N \psi_{-\frac{1}{2}}^{(i)+A} \bar\psi_{-\frac{1}{2}}^{(i)+B} \\
\left| \psi^- \psi^+ \right| = \sum_{i=1}^N \psi_{-\frac{1}{2}}^{(i)+-} \psi_{-\frac{1}{2}}^{(i)++}, \quad \left| \bar\psi^- \bar\psi^+ \right| = \sum_{i=1}^N \bar\psi_{-\frac{1}{2}}^{(i)+-} \bar\psi_{-\frac{1}{2}}^{(i)++} \\
\left| \psi^- \psi^+ \bar\psi^A \right| = \sum_{i=1}^N \psi_{-\frac{1}{2}}^{(i)+-} \psi_{-\frac{1}{2}}^{(i)++} \bar\psi_{-\frac{1}{2}}^{(i)+A}, \quad \left| \psi^A \bar\psi^{-} \bar\psi^{+} \right| = \sum_{i=1}^N \psi_{-\frac{1}{2}}^{(i)+A} \bar\psi_{-\frac{1}{2}}^{(i)+-} \bar\psi_{-\frac{1}{2}}^{(i)++} \\
\left| \psi^{-} \psi^{+} \bar\psi^{-} \bar\psi^{+} \right| = \sum_{i=1}^N \psi_{-\frac{1}{2}}^{(i)+-} \psi_{-\frac{1}{2}}^{(i)++} \bar\psi_{-\frac{1}{2}}^{(i)+-} \bar\psi_{-\frac{1}{2}}^{(i)++}
\feg
in the $S_N$-untwisted sector. The $S_N$-invariant twist operator for the $2$-cycle twisted sector is
\be
\lab \sigma_2^{--} \rab = \sum_{\substack{i,j=1 \\ i\neq j}}^N (\sigma_2^{--})_{i j},
\ee
where $(\sigma_2^{--})_{i j}$ is a BPS twist operator (defined in Section \ref{sec: review}) that combines the $i$-th and $j$-th $1$-cycles into the $2$-cycle state $| 2_{--} \ra^{\rm BPS}$. The $S_N$-invariant operators in the $2$-cycle twisted sector that we will need are\footnote{Though we will not need them for the charge order considered, it may be useful to note, e.g., that the $2$-cycle BPS twist operators dressed with the chiral primary R-current modes $J_{-1}^+$ or $\Jb_{-1}^+$ are written in this notation as
\ieg
\lab \sigma_2^{+-} \rab = \sum_{i,j: i\neq j} (J_{-1}^{(i)+} + J_{-1}^{(j)+}) (\sigma_2^{--})_{i j} = \sum_{i,j: i\neq j} (\psi_{-\frac{1}{2}}^{(i)+-} \psi_{-\frac{1}{2}}^{(i)++} + \psi_{-\frac{1}{2}}^{(j)+-} \psi_{-\frac{1}{2}}^{(j)++}) (\sigma_2^{--})_{i j} \\
\lab \sigma_2^{-+} \rab = \sum_{i,j: i\neq j} (\Jb_{-1}^{(i)+} + \Jb_{-1}^{(j)+}) (\sigma_2^{--})_{i j} = \sum_{i,j: i\neq j} (\bar\psi_{-\frac{1}{2}}^{(i)+-} \bar\psi_{-\frac{1}{2}}^{(i)++} + \bar\psi_{-\frac{1}{2}}^{(j)+-} \bar\psi_{-\frac{1}{2}}^{(j)++}) (\sigma_2^{--})_{i j}.
\feg
$\lab \sigma_2^{+-} \rab$ has quantum numbers $h = j = 2,\, \hb = \jb = 1$ and $\lab \sigma_2^{-+} \rab$ has $h = j = 1,\, \hb = \jb = 2$.}
\ie
\lab \psi^A \sigma_2^{--} \rab &= \sum_{i,j: i\neq j} (\psi_{-\frac{1}{2}}^{(i)+A} + \psi_{-\frac{1}{2}}^{(j)+A}) (\sigma_2^{--})_{i j} \\
\lab \bar\psi^A \sigma_2^{--} \rab &= \sum_{i,j: i\neq j} (\bar\psi_{-\frac{1}{2}}^{(i)+A} + \bar\psi_{-\frac{1}{2}}^{(j)+A}) (\sigma_2^{--})_{i j}.
\fe
At low charges, the operators counted in $\Zhat_1^{(0,0)}$ are, at $O(y)$,
\be
4 \varepsilon: \quad \lab \psi^A \rab, \, \lab \bar\psi^A \rab
\ee
and, at $O(y^2)$,
\ie
7:& \quad \lab \psi^- \psi^+ \rab, \lab \bar\psi^- \bar\psi^+ \rab,\, \lab \psi^A \bar\psi^B \rab,\, \lab \sigma_2^{--}\rab \\
6 \varepsilon^2:& \quad \lab \psi^- \rab \lab \psi^+ \rab,\, \lab \bar\psi^- \rab \lab \bar\psi^+ \rab,\, \lab \psi^A \rab \lab \bar\psi^B \rab
\fe
and, at $O(y^3)$,
\ie
36 \varepsilon:& \quad \lab \psi^- \psi^+ \rab \lab \psi^A \rab,\, \lab \psi^- \psi^+ \rab \lab \bar\psi^A \rab,\, \lab \bar\psi^- \bar\psi^+ \rab \lab \psi^A \rab,\, \lab \bar\psi^- \bar\psi^+ \rab \lab \bar\psi^A \rab, \\
& \quad \lab \psi^A \bar\psi^B \rab \lab \psi^C \rab,\, \lab \psi^A \bar\psi^B \rab \lab \bar\psi^C \rab,\, \lab \psi^- \psi^+ \bar\psi^A \rab,\, \lab \psi^A \bar\psi^- \bar\psi^+ \rab, \\
& \quad \lab \psi^A \sigma_2^{--}\rab,\, \lab \bar\psi^A \sigma_2^{--} \rab,\, \lab \sigma_2^{--}\rab \lab \psi^A \rab,\, \lab \sigma_2^{--} \rab \lab \bar\psi^A \rab \\
4 \varepsilon^3:& \quad \lab \psi^A \rab \lab \bar\psi^- \rab \lab \bar\psi^+ \rab,\, \lab \psi^- \rab \lab \psi^+ \rab \lab \bar\psi^A \rab \\
\fe
and so on.

In order for the formula \eqref{eq: saddle sum state counting} to hold at $N=1$, the sum
\be \label{eq: null state counting}
\sum_{k=2}^\infty \Zhat_k^{(0,0)} = - (1 + 6 \varepsilon^2) y^2 - (32 \varepsilon + 4 \varepsilon^3) y^3 - \cdots
\ee
must subtract from $\Zhat_1^{(0,0)}$ precisely the counting of finite $N$ null states that exist at $N=1$.\footnote{The partition functions $Z_{k>1}^{(0,0)}$ individually contain many spurious degeneracies that cancel out in the sum. To understand this, recall that the space of null states is not in general an invariant concept; one can always adjoin to the ``extended'' state space of a system (prior to quotienting by the constraints) a state $|\psi\ra$ which has zero overlap with the rest. Nevertheless, the subspace of the large $N$ physical Hilbert space that becomes null when $N$ is taken to be an integer is well-defined. The series coefficients of the sum \eqref{eq: null state counting} counts the charge-graded dimensions of this subspace.} There are two classes of finite $N$ null states in symmetric orbifolds. The first class contains the set of states that involve cycles of length greater than $N$. The second class contains the set of $S_N$-invariant states that become null due to vector-like finite $N$ relations among the (chiral primary) oscillators $\psi_{-\frac{1}{2}}^{(i)+A}, \bar\psi_{-\frac{1}{2}}^{(j)+A}$ with $i,j = 1,2,\cdots, N$. Let us show explicitly that the states counted in \eqref{eq: null state counting} are indeed the finite $N$ null states at $N=1$. The $S_N$-invariant combinations of operators that, by the state-operator correspondence, give rise to finite $N$ null states in $\symT$ at $N=1$ are, at $O(y^2)$,
\ie
-1:& \quad \lab \sigma_2^{--} \rab \\
-6 \varepsilon^2:& \quad \lab \psi^- \psi^+ \rab - \lab \psi^- \rab \lab \psi^+ \rab,\, \lab \bar\psi^- \bar\psi^+ \rab - \lab \bar\psi^- \rab \lab \bar\psi^+ \rab, \\
& \quad \lab \psi^A \bar\psi^B \rab - \lab \psi^A \rab \lab \bar\psi^B \rab
\fe
and, at $O(y^3)$,
\ie
-32 \varepsilon :& \quad \lab \psi^A \sigma_2^{--}\rab, \, \lab \bar\psi^A \sigma_2^{--} \rab, \, \lab \sigma_2^{--}\rab \lab \psi^A \rab, \, \lab \sigma_2^{--} \rab \lab \bar\psi^A \rab, \\
& \quad \lab \psi^- \psi^+ \rab \lab \psi^A \rab, \, \lab \bar\psi^- \bar\psi^+ \rab \lab \bar\psi^A \rab, \lab \psi^A \bar\psi^B \rab \lab \psi^A \rab, \lab \psi^A \bar\psi^B \rab \lab \bar\psi^B \rab, \\
& \quad \lab \psi^- \psi^+ \rab \lab \bar\psi^A \rab - \lab \psi^- \psi^+ \bar\psi^A \rab, \, \lab \bar\psi^- \bar\psi^+ \rab \lab \psi^A \rab - \lab \bar\psi^- \bar\psi^+ \psi^A \rab, \\
& \quad \lab \psi^A \bar\psi^B \rab \lab \psi^{\bar{A}} \rab + \lab \psi^A \psi^{\bar{A}} \bar\psi^B  \rab, \, \lab \psi^A \bar\psi^B \rab \lab \bar\psi^{\bar{B}} \rab - \lab \psi^A \bar\psi^B \bar\psi^{\bar{B}} \rab \\
-4 \varepsilon^3 :& \quad \lab \psi^- \psi^+ \bar\psi^A \rab - \lab \psi^- \rab \lab \psi^+ \rab \lab \bar\psi^A \rab, \, \lab \psi^A \bar\psi^- \bar\psi^+ \rab - \lab \psi^A \rab \lab \bar\psi^- \rab \lab \bar\psi^+ \rab
\fe
where $A,B=\pm$ and $\bar{A},\bar{B}=\mp$, and so on at higher charges. To the order demonstrated, we have shown that the supersymmetric partition functions of $\orbifoldT$ with $k>1$ in \eqref{eq: null state counting} capture precisely the counting of finite $N$ null states that exist in $\symT$ at $N=1$.

It is clear that, if the formula \eqref{eq: saddle sum state counting} is to hold at any integer $N$ as claimed, the sum $\sum_{k=2}^\infty \Zhat_k^{(0,0)}$ over the supersymmetric partition functions $\Zhat_k^{(0,0)}$ of IIB strings on $\orbifoldM$ backgrounds must capture the counting of finite $N$ null states that exist in $\symM$ at any integer $N$. We show in the next section that, depending on the region in the space of fugacities $y,\yb$ under consideration, variants of the formula \eqref{eq: saddle sum state counting} involving the partition functions $\Zhat_k^{(0/2,0/2)}$ of spectrally-flowed $\orbifold$ orbifolds apply. In these cases, one can verify that the sum over the non-trivial orbifold partition functions $\Zhat_k^{(0/2,0/2)}$ captures the counting of the finite $N$ null states of $\symM$.

\section{Saddles contributing to the finite $N$ answer} \label{sec: finite N saddles}

\subsection{Proposal for the sum over geometries} \label{subsec: sum proposal}

Let us finally consider the sum over the grand-canonical residues, interpreted as a sum over one-loop supersymmetric partition functions of IIB strings on $\orbifoldM$ backgrounds and their spectral flows. 

Based on our discussion thus far, one could wonder whether the total contribution from the sum over $\orbifold$ orbifolds to the BPS partition function $Z_N$ is
\be
Z_N(y,\yb) \stackrel{?}{=} \sum_{k=1}^\infty \left( \Zhat_{k}^{(0,0)}(y,\bar{y}) + \Zhat_{k}^{(2,0)}(y,\bar{y}) + \Zhat_{k}^{(0,2)}(y,\bar{y}) + \Zhat_{k}^{(2,2)}(y,\bar{y}) \right).
\ee
However, as explained in Section \ref{sec: grand canonical}, this is not quite true due to the structure of singularities of the grand-canonical partition function $\cZ(p;y,\yb) = \sum_{N=0}^\infty p^N Z_N(y,\yb)$ on the $p$-plane. We cannot apply the residue theorem to $\cZ$ due to fact that its poles form infinite sequences that accumulate on a circle of finite radius on the $p$-plane and that it has a wall of essential singularities at the same radius $|p|=|y^{-1} \yb^{-1}|$. Depending on the (relative) values of fugacities $y,\yb$, one or more of the infinite sequences of grand-canonical poles $(0/2,0/2)$ lie inside the wall.

We find, despite these apparent difficulties, that a version of the residue formula holds for the grand-canonical partition function $\cZ$ of $\symM$:

\paragraph{Proposal} Let
\be
\cS_{y,\yb} \ \subset \ \left\{ (0,0), (2,0), (0,2), (2,2) \right\}
\ee
be a subset of the four infinite sequences of poles $(0/2,0/2)$ that lie inside the wall of essential singularities of $\cZ(p;y,\yb)$ on the $p$-plane for given values of the complex fugacities $y,\yb$. We claim that the formula
\be \label{eq: main proposal}
Z_N(y,\yb) = \sum_{\mu \in \cS_{y,\yb}} \sum_{k=1}^\infty  \Zhat_{k}^{\mu}(y,\bar{y})
\ee
holds. That is, only the residues $\Zhat_{k}^{\mu}(y,\bar{y})$ from poles $\mu \in \cS_{y,\yb}$ located inside the wall of essential singularities of the grand-canonical partition function $\cZ(p;y,\yb)$ contribute to the finite $N$ BPS partition function $Z_N(y,\yb)$.

\textit{The bulk interpretation is that the supersymmetric partition function $Z_N$ of the IIB theory on asymptotically $\adsM$ backgrounds contains different Stokes sectors $\cS_{y,\yb}$ in which different infinite subsets of the spectrally-flowed $\orbifoldM$ saddles contribute to the path integral.}

The grand-canonical BPS partition functions $\cZ_{\Tfour}$ and $\cZ_{\Kthree}$ of $\symT$ and $\symK$ have simple poles \eqref{eq: four towers} at the same locations on the $p$-plane. Also common to both is the wall of essential singularities at radius $|p| = |y^{-1}\yb^{-1}|$ on the $p$-plane. Since, according to our proposal, the set of Stokes sectors $\cS_{y,\yb}$ for the bulk path integral depends only on the structure of singularities of the grand-canonical partition function $\cZ(p;y,\yb)$, we conclude that the BPS partition functions $Z_N$ of the IIB theory on asymptotically $\adsT$ and $\adsK$ backgrounds have the same set of Stokes sectors $\cS_{y,\yb}$ as functions of $y,\yb$.

We propose the following classification for the Stokes sectors $\cS_{y,\yb}$ of the finite $N$ supersymmetric partition function $Z_N(y,\yb)$ of the IIB theory on $\adsM$ backgrounds:

Away from the phase boundaries (i.e. anti-Stokes lines), the dominant saddle in each $\cS_{y,\yb}$ is given by one of the $\adsM$ geometries that are spectrally-flowed by the values $(\eta,\bar\eta) = (0,0), (2,0), (0,2)$, or $(2,2)$ (see Section \ref{sec: large N and spectral flows}). The one-loop supersymmetric partition functions associated to these geometries are $\Zhat_1^{(0,0)}$, $\Zhat_1^{(2,0)}$, $\Zhat_1^{(0,2)}$, and $\Zhat_1^{(2,2)}$, respectively.

The following infinite sets of saddles $\cS_{y,\yb}$ contribute to $Z_N$ in the corresponding region in the space of $y,\yb$-fugacities. In regions dominated by the unflowed $\adsM$ saddle, we have
\begin{itemize}
    \item $|\yb| < |y| \leq 1 : \ \cS_{y,\yb} =  \left\{ (0,0),  (2,0) \right\}$
    \item $|y| < |\yb| \leq 1 : \ \cS_{y,\yb} = \left\{ (0,0), (0,2) \right\}$
    \item $|y| = |\yb| < 1 : \ \cS_{y,\yb} = \left\{ (0,0) \right\}$.
\end{itemize}
In regions dominated by $\adsM$ flowed by $(\eta,\bar\eta) = (2,0)$, we have
\begin{itemize}
    \item $|\yb| < |y^{-1}| \leq 1 : \ \cS_{y,\yb} = \left\{ (2,0), (0,0) \right\}$
    \item $|y^{-1}| < |\yb| \leq 1 : \ \cS_{y,\yb} = \left\{ (2,0), (2,2) \right\}$
    \item $|y^{-1}| = |\yb| < 1 : \  \cS_{y,\yb} =  \left\{ (2,0) \right\}$.
\end{itemize}
In regions dominated by $\adsM$ flowed by $(\eta,\bar\eta) = (0,2)$, we have
\begin{itemize}
    \item $|y| < |\yb^{-1}| \leq 1 : \ \cS_{y,\yb} = \left\{ (0,2), (0,0) \right\}$
    \item $|\yb^{-1}| < |y| \leq 1 : \ \cS_{y,\yb} = \left\{ (0,2), (2,2) \right\}$
    \item $|y| = |\yb^{-1}| < 1 : \ \cS_{y,\yb} = \left\{ (0,2) \right\}$.
\end{itemize}
In regions dominated by $\adsM$ flowed by $(\eta,\bar\eta) = (2,2)$, we have
\begin{itemize}
    \item $|y^{-1}| < |\yb^{-1}| \leq 1 : \ \cS_{y,\yb} = \left\{ (2,2), (2,0) \right\}$
    \item $|\yb^{-1}| < |y^{-1}| \leq 1 : \ \cS_{y,\yb} = \left\{ (2,2), (0,2) \right\}$
    \item $|\yb^{-1}| = |y^{-1}| < 1 : \ \cS_{y,\yb} = \left\{ (2,2) \right\}$.
\end{itemize}
The labels $(0/2,0/2)$ denote the infinite set of correspondingly spectral-flowed $\orbifoldM$ geometries at all $k$ collectively. The fugacities may take values $|y|, |\yb|>1$ because $Z_N$ is a polynomial in $y,\yb$.\footnote{We omit the singular limit $|y|=|\yb|=1$ where all saddles contribute with equal weight. The result for $|y|=|\yb|=1$ can be nonetheless found by working within any of the domains above and then taking $y,\yb \to 1$ in the final answer.}

\begin{figure}[t]
\vspace{-1em}
\centering
\includegraphics[width=9cm]{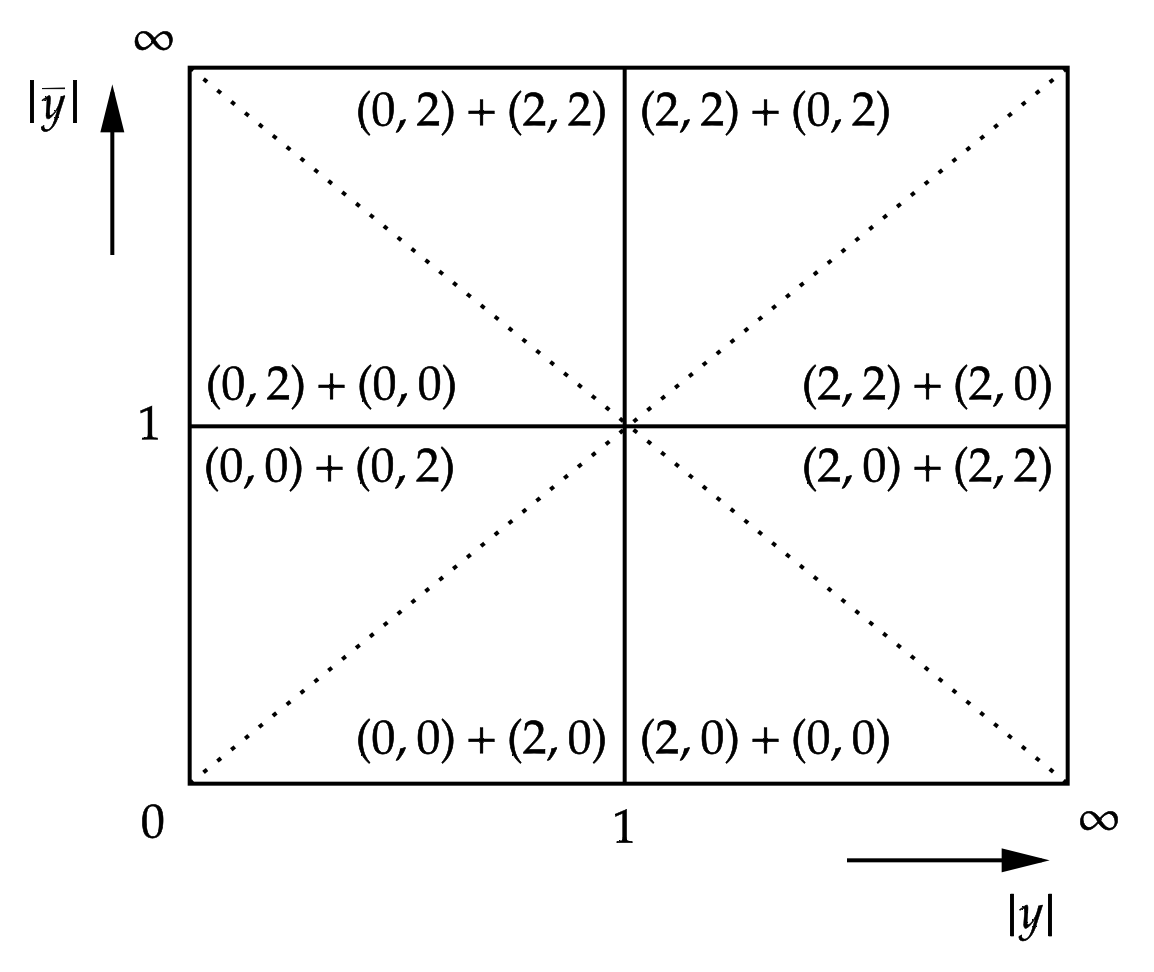}
\vspace{-1em}
\caption{Proposal for the Stokes sectors of the BPS partition function $Z_N(y,\yb)$ of the IIB theory on asymptotically $\adsM$ backgrounds, as functions of the fugacities $y,\yb$. The dotted lines denote Stokes lines and the solid lines denote anti-Stokes lines.}
\label{fig: stokes sectors}
\end{figure}

Figure \ref{fig: stokes sectors} summarizes our classification of the Stokes sectors $\cS_{y,\yb}$. The solid lines are anti-Stokes lines where the leading saddles exchange dominance and a phase transition occurs. The dotted lines are Stokes lines where a discontinuous change occurs in the set of subdominant saddles that contribute to $Z_N$.

\subsection{Deriving the residue formula} \label{subsec: deriving residue formula}

In this subsection, we provide a derivation for the proposed formula
\be \label{eq: proposal repeat}
Z_N(y,\yb) = \sum_{\mu \in \cS_{y,\yb}} \sum_{k=1}^\infty \Zhat_{k}^{\mu}(y,\bar{y})
\ee
for $\Mfour = \Tfour$ and $\Kthree$. Due to terms in the one-loop determinants that we identify as fermionic negative modes, a regularization procedure is required in the $\Tfour$ case. The $\Kthree$ case will be treated first and then describe the treatment of $\Tfour$.

\subsubsection*{Derivation for $\Mfour = \Kthree$}

The grand-canonical partition function
\be
\cZ_{\Mfour}^{\rm R}(p;y,\yb) = \cZ_{\Mfour}(p y^{-1} \yb^{-1}; y, \yb)
\ee
in the Ramond sector has a simpler expression than the NS counterpart $\cZ_{\Mfour}$. The formula \eqref{eq: proposal repeat} will be derived in the Ramond sector for convenience, but the two are related by \eqref{eq: NS R replacement}. Recall that the chiral primary states in the NS sector map under spectral flow to the ground states in the R sector. The grand-canonical partition function $\cZ_{\Kthree}^{\rm R}$ of $\symK$ over the Ramond ground states is
\be
\cZ_{\Kthree}^{\rm R} = \prod_{n=1}^\infty \frac{1}{\left(1 - p^{n} y^{-1} \bar{y}^{-1}\right) \left(1 - p^{n} y \bar{y}^{-1}\right) \left(1 - p^{n} y^{-1} \bar{y}\right) \left(1 - p^{n} y \bar{y}\right) \left(1 - p^{n} \right)^{20}}.
\ee
It has four infinite sequences of simple poles located at
\ie \label{eq: poles R}
(0,0)_k^m:& \quad p \, = \, e^{2 \pi i \frac{m}{k}} y^{\frac{1}{k}} \yb^{\frac{1}{k}} \\
(2,0)_k^m:& \quad p \, = \, e^{2 \pi i \frac{m}{k}} y^{-\frac{1}{k}} \yb^{\frac{1}{k}} \\
(0,2)_k^m:& \quad p \, = \, e^{2 \pi i \frac{m}{k}} y^{\frac{1}{k}} \yb^{-\frac{1}{k}} \\
(2,2)_k^m:& \quad p \, = \, e^{2 \pi i \frac{m}{k}} y^{-\frac{1}{k}} \yb^{-\frac{1}{k}}
\fe
where $k = 1,2,3 \cdots$ and $m =0,1,2 \cdots, k-1$, as well as a wall of essential singularities at radius $|p| = 1$. The residues
\ie \label{eq: Zhat explicit}
&\widehat{Z}_{k}^{(0/2,0/2)} = \sum_{m=0}^{k-1}  \frac{1}{k} e^{-2 \pi i N \frac{m}{k}} y^{\mp \frac{N}{k}} \bar{y}^{\mp \frac{N}{k}} \prod_{\substack{n=1 \\ n \neq k}}^\infty \frac{1}{(1-e^{2 \pi i n \frac{m}{k}} y^{\pm \frac{n}{k} \mp 1} \bar{y}^{\pm \frac{n}{k} \mp 1})} \prod_{n=1}^\infty \bigg[ \frac{1}{(1-e^{2 \pi i n \frac{m}{k}} y^{\pm\frac{n}{k} \pm 1} \bar{y}^{\pm \frac{n}{k} \mp 1})} \\
&\qquad \times \frac{1}{ (1-e^{2 \pi i n \frac{m}{k}} y^{\pm \frac{n}{k} \mp 1} \bar{y}^{\pm \frac{n}{k} \pm 1}) (1-e^{2 \pi i n \frac{m}{k}} y^{\pm \frac{n}{k} \pm 1} \bar{y}^{\pm \frac{n}{k} \pm 1})(1-e^{2 \pi i n \frac{m}{k}} y^{\pm \frac{n}{k}} \bar{y}^{\pm \frac{n}{k}})^{20}} \bigg]
\fe
at simple poles in the R sector are related to their NS counterparts only by the simple replacement
\be \label{eq: NS R replacement}
{\rm NS}: \ y^{N\left(1 \mp \frac{1}{k}\right)} \bar{y}^{N\left(1 \mp \frac{1}{k}\right)} \ \to \ {\rm R}: \ y^{\mp \frac{N}{k}} \bar{y}^{\mp \frac{N}{k}}
\ee
in the classical weights associated to $\orbifold$ and its BPS spectral flows. The non-trivial part of the residue identified as the one-loop determinant around $\orbifold$ remains unchanged as we flow from the NS sector to the R sector.

We are interested in showing that $Z_N(y,\yb)$ receives contributions only from poles $\cS_{y,\yb}$ located inside the wall of essential singularities of $\cZ_{\Kthree}^{\rm R}$. This will be achieved by showing that the following structures yield vanishing contributions to $Z_N$: (1) grand-canonical residues $\Zhat_k^\nu$ (with $\nu \notin \cS_{y,\yb}$) from poles that are not located inside the disk $|p|<1$ and (2) the essential singularities. The grand-canonical partition functions $\cZ_{\Mfour}^{\rm R}(p;y,\yb)$ are invariant under
\be \label{eq: grand canonical symmetry}
y \ \leftrightarrow \ \yb, \qquad y \ \leftrightarrow \ y^{-1}, \qquad \yb \ \leftrightarrow \ \yb^{-1}
\ee
acting individually. It therefore suffices to show that the contributions of (1) and (2) to $Z_N$ vanish in the region $|\yb| < |y| \leq 1$ where $\cS_{y,\yb} =  \left\{ (0,0),  (2,0) \right\}$ as well as in the region $|y| = |\yb| < 1$ where $\cS_{y,\yb} = \left\{ (0,0) \right\}$. The results in other regions are related to these by symmetry.

Our strategy is to apply the usual residue theorem
\be \label{eq: residue theorem}
Z_N[M](y,\yb) = - \sum_{i: \, p_i \neq 0} \,  \Res_{p = p_i} \, p^{-N-1} \cZ_{\Kthree}^{\rm R}[M](p;y,\yb)
\ee
to the grand-canonical partition function with a cutoff $M$
\be
\cZ_{\Kthree}^{\rm R}[M] = \prod_{n=1}^M \frac{1}{\left(1 - p^{n} y^{-1} \bar{y}^{-1}\right) \left(1 - p^{n} y \bar{y}^{-1}\right) \left(1 - p^{n} y^{-1} \bar{y}\right) \left(1 - p^{n} y \bar{y}\right) \left(1 - p^{n} \right)^{20}}
\ee
and then to examine the behavior of the residues in the large $M$ limit. The residues $\Zhat_{k}^{(0/2,0/2)}[M]$ at the simple poles continue to take the form \eqref{eq: Zhat explicit} except instead that the products in their ``one-loop determinants'' are truncated at the cutoff $M$:
\ie \label{eq: Zhat truncated}
&\widehat{Z}_{k}^{(0/2,0/2)}[M] = \sum_{m=0}^{k-1}  \frac{1}{k} e^{-2 \pi i N \frac{m}{k}} y^{\mp \frac{N}{k}} \bar{y}^{\mp \frac{N}{k}} \prod_{\substack{n=1 \\ n \neq k}}^M \frac{1}{(1-e^{2 \pi i n \frac{m}{k}} y^{\pm \frac{n}{k} \mp 1} \bar{y}^{\pm \frac{n}{k} \mp 1})} \prod_{n=1}^M \bigg[ \frac{1}{(1-e^{2 \pi i n \frac{m}{k}} y^{\pm\frac{n}{k} \pm 1} \bar{y}^{\pm \frac{n}{k} \mp 1})} \\
&\qquad \times \frac{1}{ (1-e^{2 \pi i n \frac{m}{k}} y^{\pm \frac{n}{k} \mp 1} \bar{y}^{\pm \frac{n}{k} \pm 1}) (1-e^{2 \pi i n \frac{m}{k}} y^{\pm \frac{n}{k} \pm 1} \bar{y}^{\pm \frac{n}{k} \pm 1})(1-e^{2 \pi i n \frac{m}{k}} y^{\pm \frac{n}{k}} \bar{y}^{\pm \frac{n}{k}})^{20}} \bigg].
\fe
Furthermore, the essential singularities of $\cZ_{\Kthree}^{\rm R}$ at $|p|=1$ are resolved in $\cZ_{\Kthree}^{\rm R}[M]$ into poles of large order at the root-of-unity locations $p = e^{2 \pi i \frac{m}{k}}$ where $m=0,1,\cdots, k-1$ and $k=1,2,\cdots, M$.

Consider first the region $|\yb| < |y| \leq 1$ where $\cS_{y,\yb} = \left\{ (0,0),  (2,0) \right\}$. In this domain, the size of a residue is given by the leading power of $\Zhat_k^{(0/2,0/2)}[M]$ in the fugacity $\yb$ (or better, the ratio $\yb/y$). Besides its classical power $\yb^{\mp \frac{N}{k}}$, a residue \eqref{eq: Zhat truncated} can acquire additional overall powers of $\yb$ if the denominators in its ``one-loop determinant'' involve negative powers of $\yb$, i.e. if the partition function contains modes whose mass-squared becomes negative when $|\yb| < 1$ (see Section \ref{subsec: higher k}).

We find, in particular, that whether a residue \eqref{eq: Zhat truncated} survives depends on whether it has finitely or infinitely many such denominators in the large $M$ limit. For example in
\ie
\Zhat_k^{(0,0)}[M] \ &\sim \ \yb^{-\frac{N}{k}} \prod_{n=1}^{k-1} \frac{1}{(1- y^{\frac{n}{k} - 1} \bar{y}^{\frac{n}{k} - 1})(1- y^{\frac{n}{k} + 1} \bar{y}^{\frac{n}{k} - 1})} \\
&\sim \ O \! \left( \yb^{-\frac{N}{k} + k-1} \right),
\fe
the number of denominators that contain negative powers of $\yb$ stabilizes to a finite value as we take $M \to \infty$ while keeping $k$ fixed. Only terms that are relevant for the overall power in $\yb$ were shown above. Similarly for $\Zhat_k^{(2,0)}[M]$, the number of denominators that contain negative powers of $\yb$ is finite as $M \to \infty$ and we have the behavior
\be
\Zhat_k^{(2,0)}[M] \ \sim \ O \! \left( \yb^{-\frac{N}{k} + k-1} \right).
\ee
Namely, the residues $\Zhat_k^{(0,0)}[M]$ and $\Zhat_k^{(2,0)}[M]$ have magnitudes that are $M$-independent and finite in the $M \to \infty$ limit. On the other hand, in $\Zhat_k^{(0,2)}[M]$ and $\Zhat_k^{(2,2)}[M]$, the number of denominators that contain negative powers of $\yb$ diverges with $M$ in the $M \to \infty$ limit:
\be
\Zhat_k^{(0,2)}[M] \ \sim \ \Zhat_k^{(0,2)}[M] \ \sim \ O \! \left( \yb^{\frac{12}{k} M(M+1) + \frac{N}{k} + k-1} \right).
\ee
It follows that the residues $\Zhat_k^{(0,2)}[M]$ and $\Zhat_k^{(2,2)}[M]$ vanish as we take $M \to \infty$ while keeping $k$ fixed, in the region $|\yb| < |y| \leq 1$. Furthermore, we find that the residues at the root-of-unity poles $p = e^{2\pi i \frac{m}{k}}$ (which become essential singularities when $M=\infty$) scale as
\be
- \sum_{k=1}^M \sum_{\substack{m=0 \\ m,k \, {\rm coprime}}}^{k-1} \, \Res_{p = e^{2\pi i \frac{m}{k}}} \, p^{-N-1} \cZ_{\Kthree}^{\rm R}[M] \ \sim \ O \! \left( \yb^{2M} \right),
\ee
which vanishes for $|\yb|<1$ in the large $M$ limit. It can also be checked that the residue of $\cZ_{\Kthree}^{{\rm R}}[M]$ at $p = \infty$ also vanishes. Therefore, by taking the large cutoff $M$ limit of \eqref{eq: residue theorem}, we establish the result
\be \label{eq: result domain 1}
Z_N(y,\yb) = \sum_{k=1}^\infty \left[ \Zhat_{k}^{(0,0)}(y,\bar{y}) + \Zhat_{k}^{(2,0)}(y,\bar{y}) \right]
\ee
in the region $|\yb|<|y| \leq 1$. 

Now let us turn to the case $|y| = |\yb| < 1$ where $\cS_{y,\yb} = \left\{ (0,0) \right\}$. The magnitude of the residues in this domain can be estimated via the leading power of $\Zhat_k^{(0/2,0/2)}[M]$ in $|y| = |\yb|$. As before, the distinction between the residues that do or do not survive the large $M$ limit comes from the $M$-dependence of the leading power $|y|^{\#}$, which is in turn determined by the growth in the number of ``negative modes'' as a function of the cutoff $M$. Here, we find that the $(0,0)$ residues
\be
\Zhat_k^{(0,0)}[M] \ \sim \ O \! \left( |y|^{-\frac{2N}{k} + k-1} \right)
\ee
are finite as $M \to \infty$, while the $(2,0)$, $(0,2)$, and $(2,2)$ residues
\ieg
\Zhat_k^{(2,0)}[M] \ \sim \ \Zhat_k^{(0,2)}[M] \ \sim \ O \! \left( |y|^{2M} \right) \\
\Zhat_k^{(2,2)}[M] \ \sim \ O \! \left( |y|^{\frac{24}{k} M(M+1) + \frac{2N}{k} + k - 1} \right)
\feg
vanish in the $M \to \infty$ limit. Additionally, we find that the root-of-unity poles $p = e^{2\pi i \frac{m}{k}}$ in the region $|y|=|\yb|<1$ have residues that scale as
\be
- \sum_{k=1}^M \sum_{\substack{m=0 \\ m,k \, {\rm coprime}}}^{k-1} \, \Res_{p = e^{2\pi i \frac{m}{k}}} \, p^{-N-1} \cZ_{\Kthree}^{\rm R}[M] \ \sim \ O \! \left( |y|^{2M} \right),
\ee
which vanishes in the large $M$ limit, and that the residue at $p=\infty$ also vanishes. Taking the large cutoff $M$ limit, we establish the result
\be \label{eq: result domain 2}
Z_N(y,\yb) = \sum_{k=1}^\infty \Zhat_{k}^{(0,0)}(y,\bar{y}).
\ee
in the region $|y|=|\yb| < 1$. 

The symmetries \eqref{eq: grand canonical symmetry} of the grand-canonical BPS partition function $\cZ_{\Mfour}^{\rm R}(p;y,\yb)$ can be used to map the results \eqref{eq: result domain 1} and \eqref{eq: result domain 2} in $|\yb|<|y|\leq 1$ and $|y|=|\yb|<1$ to that in any other region enumerated in Section \ref{subsec: sum proposal}. Doing so, we recover the proposed formula \eqref{eq: proposal repeat} for the case $\Mfour = \Kthree$.

\subsubsection*{Regularization for $\Tfour$}


For $\Mfour = \Tfour$, the pairing between the chiral primary bosonic and fermionic states ensures that the effect of bosonic and fermionic negative modes on the one-loop partition functions $\Zhat_k^{(0/2,0/2)}$ cancel in pairs. As a result, partition functions with infinitely-many bosonic negative modes fail to vanish, and the formula \eqref{eq: proposal repeat} prior to regularization requires one to make sense of an infinite alternating sum of coefficients at each charge. Here we introduce a regularization prescription for $\Zhat_k^{(0/2,0/2)}$ and find that \eqref{eq: proposal repeat} is satisfied upon regularization. We provide further explicit evidence of its validity in the next subsection.

We are interested in the $M \to \infty$ limit of the residue theorem
\be \label{eq: residue theorem T4}
Z_N[M](y,\yb) = - \sum_{i: \, p_i \neq 0} \,  \Res_{p = p_i} \, p^{-N-1} \cZ_{\Tfour}^{\rm R}[M](p;y,\yb)
\ee
applied to the grand-canonical partition function of $\symT$ over the Ramond ground states
\be
\cZ_{\Tfour}^{\rm R}[M] = \prod_{n=1}^M \frac{\left(1 + \varepsilon p^{n} \bar{y}^{-1}\right)^2 \left(1 + \varepsilon p^{n} y^{-1} \right)^2  \left(1 + \varepsilon p^{n} \bar{y}\right)^2 \left(1 + \varepsilon p^{n} y \right)^2}{\left(1 - p^{n} y^{-1} \bar{y}^{-1}\right) \left(1 - p^{n} y \bar{y}^{-1}\right) \left(1 - p^{n} y^{-1} \bar{y}\right) \left(1 - p^{n} y \bar{y}\right) \left(1 - p^{n} \right)^4}
\ee
defined with a cutoff $M$. The residues at the simple poles \eqref{eq: poles R} of $\cZ_{\Tfour}^{\rm R}[M]$ are
\ie \label{eq: Zhat truncated T4}
&\Zhat_{k}^{(0/2,0/2)}[M] = \sum_{m=0}^{k-1}  \frac{1}{k} e^{-2 \pi i N \frac{m}{k}} y^{\mp\frac{N}{k}} \bar{y}^{\mp\frac{N}{k}} \prod_{\substack{n=1 \\ n \neq k}}^M \frac{1}{(1-e^{2 \pi i n \frac{m}{k}} y^{\pm\frac{n}{k}\mp1} \bar{y}^{\pm\frac{n}{k}\mp1})} \prod_{n=1}^M \bigg[ \frac{(1 + \varepsilon e^{2 \pi i n \frac{m}{k}} y^{\pm\frac{n}{k}} \bar{y}^{\pm\frac{n}{k}\mp1})^2}{(1-e^{2 \pi i n \frac{m}{k}} y^{\pm\frac{n}{k}\pm1} \bar{y}^{\pm\frac{n}{k}\mp1})} \\
&\qquad \times \frac{(1 + \varepsilon e^{2 \pi i n \frac{m}{k}} y^{\pm\frac{n}{k}\mp1} \bar{y}^{\pm\frac{n}{k}})^2 (1 + \varepsilon e^{2 \pi i n \frac{m}{k}} y^{\pm\frac{n}{k}} \bar{y}^{\pm\frac{n}{k}\pm1})^2 (1 + \varepsilon e^{2 \pi i n \frac{m}{k}} y^{\pm\frac{n}{k}\pm1} \bar{y}^{\pm\frac{n}{k}})^2}{(1-e^{2 \pi i n \frac{m}{k}} y^{\pm\frac{n}{k}\mp1} \bar{y}^{\pm\frac{n}{k}\pm1})(1-e^{2 \pi i n \frac{m}{k}} y^{\pm\frac{n}{k}\pm1} \bar{y}^{\pm\frac{n}{k}\pm1})(1-e^{2 \pi i n \frac{m}{k}} y^{\pm\frac{n}{k}} \bar{y}^{\pm\frac{n}{k}})^{4}} \bigg].
\fe
While $\varepsilon = \pm 1$ for the partition function and index, respectively, it will be useful to view $\varepsilon$ as a complex variable to regularize the microcanonical contributions of $\orbifold$ orbifolds. Roughly speaking, the power of $\varepsilon$ keeps track of the number of bosonic negative modes that cancel out with the fermionic counterparts. Due to the symmetry \eqref{eq: grand canonical symmetry} of $\cZ_{\Tfour}^{\rm R}$, we may again restrict ourselves to the fugacity regions $|\yb|<|y| \leq 1$ and $|y|=|\yb|<1$.

Our prescription is to initially treat the residues of $\cZ_{\Tfour}^{\rm R}[M]$ as a formal power series in $\varepsilon$. Note that, since $\varepsilon$ did not affect the pole structure of any residue of $\cZ_{\Tfour}^{\rm R}[M]$ prior to the formal power series interpretation, taking the power series of any residue of $\cZ_{\Tfour}^{\rm R}[M]$ in $\varepsilon$ commutes with taking the power series in $y,\yb$. The arguments that led to the formula \eqref{eq: proposal repeat} for $\Mfour = \Kthree$ can then be repeated at each order in the formal $\varepsilon$-series for $\Mfour = \Tfour$.

At each order in $\varepsilon$, one finds that the residues \eqref{eq: Zhat truncated T4} with finitely-many denominators that contain negative powers of the relevant fugacity are finite as $M \to \infty$, while those with infinitely-many such denominators vanish in the $M \to \infty$ limit.\footnote{Alternatively, at each fixed power of the relevant fugacity, the residues with (in)finitely-many such denominators have powers of $\varepsilon$ that stay finite (resp. diverge) as $M \to \infty$.} The residues at the root-of-unity poles and at $p=\infty$ also vanish at each order in $\varepsilon$ in the $M \to \infty$ limit. We thus recover the result \eqref{eq: result domain 1} in $|\yb|<|y| \leq 1$ and \eqref{eq: result domain 2} in $|\yb|<|y|\leq 1$ in the $\Tfour$ case. Using the symmetries \eqref{eq: grand canonical symmetry} to map the results in these regions to that in any other region enumerated in Section \ref{subsec: sum proposal}, we find the proposed formula \eqref{eq: proposal repeat} for $\Mfour = \Tfour$.

Our derivations suggest that, at least in the context under consideration, whether the one-loop partition function $\Zhat_k^\mu$ of a saddle does (not) contribute to $Z_N$ is directly related to whether the number of its one-loop bosonic fluctuations that acquire a negative mass-squared is (in)finite within a given region of the space of chemical potentials. It would be of interest to study how this observation relates to the Kontsevich-Segal-Witten criterion for allowable metrics \cite{Kontsevich:2021dmb,Witten:2021nzp}.

\subsection{Explicit checks}

We conclude this section with a demonstration of some explicit checks of the formula \eqref{eq: main proposal} at low values of $N$, in the NS sector of the AdS$_3$/CFT$_2$ duality.

To make direct contact with the issues discussed in the Introduction, we set $y = \yb$ and $|y| < 1$, for which the relevant Stokes sector is $\cS_{y,\yb} = \{ (0,0) \}$. That is, we verify that
\be \label{eq: check proposal}
Z_N(y,y) = \lim_{K\to \infty} \ \sum_{k=1}^K \Zhat_{k}^{(0,0)}(y,y)
\ee
is satisfied at the level of the microcanonical coefficients as take the cutoff $K$ to infinity. Further checks in different Stokes sectors $\cS_{y,\yb}$ are provided in Appendix \ref{app: further checks}.

The BPS partition function $Z_N$ of the $\symK$ theory at $N=1$ is simply that of the $\Kthree$ sigma model
\be
Z_{N=1} = 1 + 22 y^2 + y^4,
\ee
and contains the Betti numbers of $\Kthree$. This polynomial is reproduced in the bulk, according to our proposal, by summing over one-loop supersymmetric partition functions $\Zhat_{k}^{(0,0)}$ \eqref{eq: Zhat K3 app} of the IIB theory on $\orbifoldK$. At increasing values of the cutoff $K$, we find
\ie
K=1:& \qquad 1+23 y^2+300 y^4+2876 y^6+22450 y^8+150606 y^{10} + \cdots \\
2:& \qquad 1+22 y^2-23 y^4-22450 y^6-1025620 y^8-28954249 y^{10} + \cdots \\
3:& \qquad 1+22 y^2+y^4+3176 y^6+4879226 y^8+604364555 y^{10} + \cdots \\
4:& \qquad 1+22 y^2+y^4-24 y^6-1048070 y^8-2094819829 y^{10} + \cdots \\
5:& \qquad 1+22 y^2+y^4 + 0 y^6 +25626 y^8+609221331 y^{10} + \cdots \\
6:& \qquad 1+22 y^2+y^4+ 0 y^6-24 y^8-30002319 y^{10} + \cdots \\
7:& \qquad 1+22 y^2+y^4+ 0 y^6+ 0 y^8+176232 y^{10} + \cdots \\
8:& \qquad 1+22 y^2+y^4+ 0 y^6+ 0 y^8-24 y^{10} + \cdots \\
9:& \qquad 1+22 y^2+y^4+ 0 y^6+ 0 y^8 + 0 y^{10}+ \cdots
\fe
and so on. The tail of zeros propagates to higher charges as $K \to \infty$.

From the bulk perspective, that the Hilbert space $\cH_{cc}$ of left-right chiral primary states is finite-dimensional at finite $N$ is seen to be the result of large cancellations among the $\orbifold$ partition functions $\Zhat_{k}^{(0,0)}$. The relative signs $(-1)^{k-1}$ in $\Zhat_{k}^{(0,0)}$ that are responsible for the cancellations arise, as we have argued, from the need to Wick rotate the path integral contours for the negative modes that are present in the $\bZ_k$-twisted sector of excitations on $\orbifoldK$ backgrounds. The result is that, non-perturbatively, one finds vastly fewer states than what one would expect from a semiclassical analysis around any given background.

The pattern persists at higher $N$. The BPS partition function $Z_N$ of $\symK$ at $N=2$ is
\be
Z_{N=2} = 1 + 23 y^2 + 276 y^4 + 23 y^6 + y^8.
\ee
Summing again over the supersymmetric partition functions $\Zhat_{k}^{(0,0)}$ \eqref{eq: Zhat K3 app} on $\orbifoldK$ with $N=2$, we find at increasing cutoff $K$
\ie
K=1:& \qquad 1+23 y^2+300 y^4+2876 y^6+22450 y^8+150606 y^{10}+897464 y^{12} + \cdots \\
2:& \qquad 1+23 y^2+276 y^4-300 y^6-150606 y^8-5603634 y^{10}-136356040 y^{12} + \cdots \\
3:& \qquad 1+23 y^2+276 y^4+24 y^6+25326 y^8+24398685 y^{10}+2538580265 y^{12} + \cdots \\
4:& \qquad 1+23 y^2+276 y^4+23 y^6-323 y^8-5754240 y^{10}-8345559094 y^{12} + \cdots \\
5:& \qquad 1+23 y^2+276 y^4+23 y^6+y^8+175932 y^{10}+2562828344 y^{12} + \cdots \\
6:& \qquad 1+23 y^2+276 y^4+23 y^6+y^8-324 y^{10}-142110280 y^{12} + \cdots \\
7:& \qquad 1+23 y^2+276 y^4+23 y^6+y^8+0y^{10}+1073396 y^{12} + \cdots \\
8:& \qquad 1+23 y^2+276 y^4+23 y^6+y^8+0y^{10}-324 y^{12} + \cdots \\
9:& \qquad 1+23 y^2+276 y^4+23 y^6+y^8+0y^{10}+0y^{12} + \cdots
\fe
and so on. For $\symK$ at $N=3$, we have
\be
Z_{N=3} = 1 + 23 y^2 + 299 y^4 + 2554 y^6 + 299 y^8 + 23 y^{10} + y^{12},
\ee
and the bulk sum with $N=3$ gives
\ie
K=1:& \qquad 1+23 y^2+300 y^4+2876 y^6+22450 y^8+150606 y^{10}+897464 \
y^{12}+4856776 y^{14} + \cdots \\
2:& \qquad 1+23 y^2+299 y^4+2553 y^6-2876 y^8-897464 y^{10}-28207391 \
y^{12}-604213949 y^{14} + \cdots \\
3:& \qquad 1+23 y^2+299 y^4+2554 y^6+323 y^8+173056 y^{10}+113902889 \
y^{12}+10166919985 y^{14} + \cdots \\
4:& \qquad 1+23 y^2+299 y^4+2554 y^6+299 y^8-3176 y^{10}-29104855 \
y^{12}-31879707215 y^{14} + \cdots \\
5:& \qquad 1+23 y^2+299 y^4+2554 y^6+299 y^8+24 y^{10}+1070520 \
y^{12}+10279925410 y^{14} + \cdots \\
6:& \qquad 1+23 y^2+299 y^4+2554 y^6+299 y^8+23 y^{10}-3199 y^{12}-633318804 \
y^{14} + \cdots \\
7:& \qquad 1+23 y^2+299 y^4+2554 y^6+299 y^8+23 y^{10}+y^{12}+5927296 y^{14} + \cdots \\
8:& \qquad 1+23 y^2+299 y^4+2554 y^6+299 y^8+23 y^{10}+y^{12}-3200 y^{14} + \cdots \\
9:& \qquad 1+23 y^2+299 y^4+2554 y^6+299 y^8+23 y^{10}+y^{12} + 0 y^{14} + \cdots
\fe
and so on.

Finally, we consider the proposal \eqref{eq: check proposal} for $\Mfour = \Tfour$. The BPS partition function $Z_N$ of $\symT$ at $N=1$
\be
Z_{N=1} = 1 + 4 \varepsilon y + 6 y^2 + 4 \varepsilon y^3 + y^4
\ee
encodes the Betti numbers of $\Tfour$. This polynomial is reproduced in the bulk by summing over the supersymmetric partition functions $\Zhat_{k}^{(0,0)}$ \eqref{eq: Zhat T4 app} on $\orbifoldT$ backgrounds, where $\Zhat_{k}^{(0,0)}$ are treated using the regularization prescription of Section \ref{subsec: deriving residue formula}. We find
\ie
K=1:& \qquad 1+4 \varepsilon y +\left(7+6 \varepsilon ^2\right)y^2 + \left(36 \varepsilon +4 \varepsilon ^3\right)y^3+\left(36+74 \varepsilon ^2+\varepsilon ^4\right)y^4 \\
& \myquad[4] +\left(208 \varepsilon +76 \varepsilon ^3\right)y^5  + \cdots \\
2:& \qquad 1+4 \varepsilon y +\left(6- \varepsilon^4 \right)y^2 -\left(52 \varepsilon^3+ 4 \varepsilon^5\right)y^3 -\left(7+350 \varepsilon^2+ 238 \varepsilon^4 + O(\varepsilon^6)\right)y^4\\
&\myquad[4]  -\left(772 \varepsilon +2448 \varepsilon ^3 + 588 \varepsilon^5 +O(\varepsilon ^7)\right)y^5  + \cdots \\
3:& \qquad 1+4 \varepsilon y +6y^2+ \left(4 \varepsilon + 4 \varepsilon^3+ 56 \varepsilon^5+O(\varepsilon^7)\right)y^3+\left(1+64 \varepsilon ^2+1217 \varepsilon ^4+O(\varepsilon ^6)\right)y^4 \\
&\myquad[4] +\left(208 \varepsilon +8272 \varepsilon^3+15080 \varepsilon^5+O(\varepsilon ^7)\right)y^5 + \cdots \\
4:& \qquad 1+4 \varepsilon y +6y^2+\left(4 \varepsilon - 4 \varepsilon^5 +O(\varepsilon^7)\right)y^3 +\left(1- 223 \varepsilon^4 +O(\varepsilon ^6)\right)y^4 \\
&\myquad[4] -\left(4 \varepsilon +2244 \varepsilon ^3+ 49056 \varepsilon^5+O(\varepsilon^7)\right) y^5 + \cdots \\
5:& \qquad 1+4 \varepsilon y +6y^2+\left(4 \varepsilon +O(\varepsilon ^7)\right)y^3+\left(1+ \varepsilon^4 +O(\varepsilon ^6)\right)y^4 \\
& \myquad[4] +\left(56 \varepsilon^3 + 14240 \varepsilon^5 +O(\varepsilon ^7)\right)y^5 + \cdots \\
6:& \qquad 1+4 \varepsilon y + 6 y^2+\left(4 \varepsilon +O(\varepsilon ^9)\right)y^3 +\left(1+O(\varepsilon^6)\right)y^4 -\left( 492 \varepsilon^5 + O(\varepsilon ^7) \right) y^5 + \cdots \\
7:& \qquad 1+4 \varepsilon y + 6 y^2+\left(4 \varepsilon +O(\varepsilon^{11})\right)y^3 +\left(1+O(\varepsilon^8)\right)y^4 + O(\varepsilon ^7) y^5 + \cdots
\fe
and so on. The powers of $\varepsilon$ in the remaining terms increase linearly with $K$ and we drop these terms in our regularization in the $K \to \infty$ limit.

\section{Discussion}

In this work, we presented a bulk prescription for computing the finite $N$ spectrum $Z_N$ of chiral primary states in $\symM$, in terms of a sum over one-loop supersymmetric partition functions $\Zhat_k^\mu$ of the IIB theory on $\orbifoldM$ orbifolds and their spectral flows, and we proposed a classification for the Stokes sectors of the IIB theory on $\orbifoldM$ backgrounds.

We conclude with a partial list of open questions:
\begin{enumerate}
    \item It would be important to derive the one-loop supersymmetric partition functions of $\orbifoldM$ via the Euclidean gravitational path integral. A path integral treatment would be helpful for elucidating features such as one-loop exactness, for providing a contour treatment of negative modes, and possibly for providing a sum-over-saddles explanation for the Gauss constraint.
    \item It is natural to ask whether our proposal can be generalized to the $\frac{1}{4}$-BPS states below and/or above the black hole threshold captured in the elliptic genus of $\symK$ at finite $N$ \cite{IP}. It would be especially interesting to understand the connection between these ideas and the Farey-tail expansion \cite{Dijkgraaf:2000fq,Manschot:2007ha}.
    \item It would be very interesting to connect  generalizations of our proposal to fortuity in the D1-D5 system \cite{Chang:2024zqi,Chang:2025rqy}.
    \item In higher-dimensional examples of AdS/CFT, the bulk contributions that account for the finite $N$ BPS spectrum are giant graviton branes \cite{Gaiotto:2021xce,Imamura:2021ytr}. It would be nice to know whether the partition functions on $\orbifold$ can also be understood in terms of branes in $\adsS$.
    \item The connection between grand-canonical partition functions and the partition functions of bulk saddles that contribute at finite $N$ appears remarkably rich and general. Can the analytic structure of the grand-canonical partition function be used to provide an organizing principle for the sum over saddles?
    \item Can the considerations here be generalized to systems with large $\cN = 4$ SCA such as strings on AdS$_3$ $\times$ S$^3$ $\times$ S$^3$ $\times$ S$^1$ \cite{Eberhardt:2017fsi,Eberhardt:2017pty,Witten:2024yod,Gaberdiel:2024dva,Heydeman:2025fde,Murthy:2025moj} and does it help eludicate curious features therein?
\end{enumerate}

\section*{Acknowledgements} 

We thank Alexandre Belin, Alejandra Castro, Chi-Ming Chang, Matthias Gaberdiel, Anthony Houppe, Emil Martinec, Kiarash Naderi, Masaki Shigemori, Douglas Stanford, Zhenbin Yang, and Haoyu Zhang for useful discussions. We are especially grateful to Matthias Gaberdiel for comments on the draft and for several helpful conversations during the course of this work.

JHL would like to thank the theory groups at Imperial College London, SISSA, Stanford University, University of Milano-Bicocca, University of Michigan, and University of Turin (Mathematical physics) for the opportunities to present this work. JHL would also like to thank the Workshop on Black Holes, Quantum Chaos and Quantum Information (YITP-I-25-01) at the Yukawa Institute for Theoretical Physics at Kyoto University and the Workshop on Holography, Operators and Finite N at Nordita and Stockholm University for opportunities to present this work. 
WL would like to thank the 2025 Chinese QFT and String meeting and 2025 International Congress of Basic Science (ICBS)  for opportunities to present this work.  

The work of the group at ETH is supported in part by the Simons Foundation grant 994306 (Simons Collaboration on Confinement and QCD Strings) and by the NCCR SwissMAP funded by the Swiss National Science Foundation. 
The work of WL is supported by NSFC  No.\ 12447101, 12247103. We are also grateful for the hospitality of Perimeter Institute, Simons Center for Geometry and Physics (during the 2022 Simons Summer Workshop), and Kavli Institute for Theoretical Physics (during the program ``What is String Theory? Weaving Perspectives Together'' NSF PHY-1748958) when parts of this work were carried out. 
This work was concluded in part at Aspen Center for Physics, which is supported by National Science Foundation grant PHY-2210452.

\appendix

\section{$\mathcal{N}=(4,4)$ SCA and its free field realization}
\label{appssec:N4SCA}

\subsubsection*{$\mathcal{N}=(4,4)$ superconformal algebra}

In this paper, both the dual 2D CFT and the worldsheet CFT have the symmetry given by the $\mathcal{N}=(4,4)$ superconformal algebra (SCA), which we now summarize.

The $\mathcal{N}=4$ SCA (in the left-moving sector) consists of 
\begin{itemize}
\item the stress-energy tensor $T(z)=\sum_{n\in\mathbb{Z}}\frac{L_n}{z^{n+2}}$, with central charge $c$;
\item the four supercharges $G^{\alpha\beta}(z)=\sum_{r\in\mathbb{Z}+1/2}\frac{G^{\alpha\beta}_r}{z^{r+3/2}}$, where the two indices $\alpha,\beta=\pm$ are the spinor indices of the $\mathfrak{su}(2)_{\textrm{R}}$ R-symmetry and the $\mathfrak{su}(2)$ outer-automorphism, respectively, of  the $\mathcal{N}=4$ SCA;\footnote{
We will use $\alpha_i,\beta_i$ to label the  $\mathfrak{su}(2)_{\textrm{R}}$ R-symmetry and the $\mathfrak{su}(2)$ outer-automorphism, respectively. 
}
\item the $\mathfrak{su}(2)_{\textrm{R}}$ current $J^{3,\pm}(z)=\sum_{n\in\mathbb{Z}}\frac{J^{3,\pm}_n}{z^{n+1}}$, with level $k=\frac{c}{6}$.
\end{itemize}

The commutation relations among these fields are\footnote{
We follow the convention of $\mathcal{N}=4$ superconformal algebra in \cite{Gaberdiel:2025eaf}, but with the rescaling of the supercharges $G^{\textrm{here}}=\frac{i}{2}G^{\textrm{there}}$, in order to make the expressions in terms of $\mathbb{T}^4$ fields look nicer; also $J^{\textrm{here}}=K^{\textrm{there}}$.
}
\begin{equation}
\begin{aligned}
[L_m\, , \, L_n]
&=(m-n)L_{m+n}+\tfrac{c}{12}m(m^2-1)\delta_{m+n,0} \,,\\
[L_m\,,\, G^{\alpha\beta}_r]
&=(\tfrac{1}{2}m-r) G^{\alpha\beta}_{m+r}\,, \\ 
[L_m\,,\, J^a_n]
&=-n J^{a}_{m+n} \,, \\
\{G^{\alpha_1\beta_1}_r \,,\, G^{\alpha_2\beta_2}_s  \}
&= \epsilon^{\alpha_1\alpha_2} \epsilon^{\beta_1\beta_2}\left(\tfrac{c}{6}(r^2-\tfrac{1}{4})\delta_{r+s,0}+\tfrac{1}{4}L_{r+s}\right) +\epsilon^{\beta_1\beta_2} D^{\alpha_1\alpha_2}(t_a) (r-s)J^a_{r+s} \,, \\
{}[J^3_m\,,\,J^3_n] 
&= \tfrac{c}{12}m\delta_{m+n,0} \ , 
\\
{} [J^3_m\,,\,J^\pm_n] &= \pm J^\pm_{m+n}\ ,  
\\
{}[J^+_m\,,\,J^-_n] &=  \tfrac{c}{6} m\,\delta_{m+n,0}+2J^3_{m+n}\\
[J^a_m\,,\, G^{\alpha\beta}_r] &= D^{(1/2)}(t^a)^{\alpha}{}_{\alpha'}  \, G^{\alpha'\beta}_{m+r} \ , 
\end{aligned}
\end{equation}
where $\epsilon^{+-} = -\epsilon^{-+} =+1$, the non-zero $D$-matrix elements in the $G$-$G$ and $K$-$G$ commutators are \cite{Gaberdiel:2025eaf}:
\begin{equation}
\begin{aligned}
&D^{+-}(t_3)  =  D^{-+}(t_3)= 1 \,, 
\qquad D^{++}(t_+) = - 1  \,, 
\qquad D^{--}(t_-) =  1 \,, \\
& D^{(1/2)}(t^3)^+{}_{+} = - D^{(1/2)}(t^3)^-{}_{-} 
= \tfrac{1}{2} 
\,, \quad 
D^{(1/2)}(t^+)^{-}{}_{+} = 1\,, \quad
D^{(1/2)}(t^-)^{+}{}_{-}  = 1 \ . 
\end{aligned}
\end{equation}
The $\mathcal{N}=4$ SCA in the right-moving sector will be labeled by the barred fields $(\bar{L},\bar{G}^{\alpha\beta},\bar{J}^{a})$.

The $\mathcal{N}=2$ SCA is a subalgebra of $\mathcal{N}=4$ SCA:
\begin{equation}
\begin{aligned}
[L_m\, , \, L_n]&=(m-n)L_{m+n}+\frac{c}{12}m(m^2-1)\delta_{m+n,0}\\
\{G^{+}_r \,,\, G^{-}_s  \}
&=   \left[\tfrac{c}{6}(r^2-\tfrac{1}{4})\delta_{r+s,0}+L_{r+s}\right] +  \left[(r-s)J_{r+s}\right] \\
{}[J_m\,,\,J_n] &= \tfrac{c}{12}m\delta_{m+n,0}\ , \\
[L_m \,,\, J_n]&=-n J_{m+n}\\
[L_m \,,\, G^{\pm}_r]&=(\tfrac{1}{2}m-r) G^{\pm}_{m+r}\\
[J_m\,,\, G^{\pm}_r] &=\pm \tfrac{1}{2}  \, G^{\pm}_{m+r}
\end{aligned}
\end{equation}
From the generators of the $\mathcal{N}=4$ SCA, we can write two special $\mathcal{N}=2$ SCA:
\be
\{L\,, G^{+}\,,G^{-}\,,J\}_{\mathcal{N}=2}=\{L\,,G^{++}\,,G^{--}\,, J^{3}\}_{\mathcal{N}=4}
\ee
or
\be
\{L\,, G^{+}\,,G^{-}\,,J\}_{\mathcal{N}=2}=\{L\,,G^{+-}\,,-G^{-+}\,, J^{3}\}_{\mathcal{N}=4} \,.
\ee
We can use either of them to define the chiral primary states.

\subsubsection*{$\mathbb{T}^4$ fields}

The seed theory of the symmetric  orbifold of $\mathbb{T}^4$ consists of four scalars and four fermions, and they form an $\mathcal{N}=(4,4)$ superconformal algebra with central charge
\begin{equation}
c_{\textrm{L}}=c_{\textrm{R}}=6\,. \end{equation}
The field content consists of the following.
\begin{itemize}
\item Four real bosons $X^{i=1,\dots,4}$, which generate an $\mathfrak{so}(4)\sim \mathfrak{su}(2)\times \mathfrak{su}(2)$ symmetry.
We will use the spinor indices of the two $\mathfrak{su}(2)$'s to  label the four bosons as $X^{\beta A}$. 
Here $\beta=\pm$ is the spinor index of the first $\mathfrak{su}(2)$, which will be chosen as the $\mathfrak{su}(2)_{\mathrm{o}}$ outer-automorphism of the $\mathcal{N}=4$ SCA; $A=\pm$ is the spinor index of the second $\mathfrak{su}(2)$ and will be a flavor index. 
(In fact, these bosonic $\mathfrak{su}(2)$ symmetries are only present for $\mathbb{R}^4$, rather than $\mathbb{T}^4$, and they are global symmetries that act simultaneously on left- and right-movers.
For the moment it is convenient to introduce independent $\mathfrak{su}(2)$ symmetries for the left- and right-movers; the actual symmetries of the torus theory will then be a certain diagonal discrete subgroup of these different bosonic $\mathfrak{su}(2)$ symmetries.)
These four bosons have the mode expansion 
\begin{equation}
\partial X^{\beta A}(z)=\sum_{n\in\mathbb{Z}}\frac{\mathtt{a}^{\beta A}_n}{z^{n+1}} 
\,, \qquad 
\bar{\partial} X^{\bar{\beta}\bar{ A}}(\bar{z})=\sum_{n\in\mathbb{Z}}\frac{\bar{\mathtt{a}}^{\bar{\beta}\bar{ A}}_n}{\bar{z}^{n+1}} \,.
\end{equation}
The modes $\mathtt{a}^{\beta A}_n$ transform under the left $\mathcal{N}=4$ superconformal algebra as:
\begin{equation}\label{eq:N4a}
\begin{aligned}
[L_m\,,\, \mathtt{a}^{\beta A}_n]&=-n \mathtt{a}^{\beta A}_{m+n} \\
[G^{\alpha\beta_1}_r\,,\, \mathtt{a}^{\beta_2 A}_n]&=\epsilon^{\beta_1\beta_2}n \psi^{\alpha A}_{r+n}\\
[J^{a}_m\,,\, \mathtt{a}^{\beta A}_n]&=0 \,;
\end{aligned}
\end{equation}
and similarly for the barred fields in the right-moving sector.
\item Four complex fermions $(\psi^{i=1,\dots,4},\bar{\psi}^{i=1,\dots,4})$, which forms the $\mathfrak{so}(4)\sim \mathfrak{su}(2)\times\mathfrak{su}(2)$ current algebra at level $1$ in the left- and right-moving sectors, respectively.
Similar to the bosons, we use the spinor indices of the $\mathfrak{su}(2)$ to label the fermions as $(\psi^{\alpha A},\bar{\psi}^{\bar{\alpha}\bar{A}})$, where the first indices $\alpha,\bar{\alpha}=\pm$ are the spinor indices of the $\mathfrak{su}(2)_{\textrm{R}}$ R-symmetry of the left and right $\mathcal{N}=4$ SCA, respectively, and the second indices $A,\bar{A}=\pm$ are the flavor $\mathfrak{(su)}(2)$'s.
These four fermions have the mode expansion in the NS sector
\begin{equation}
\psi^{\alpha A}(z)=\sum_{r\in\mathbb{Z}+1/2}\frac{\psi^{\alpha A}_r}{z^{r+1/2}}
\,, \qquad
\bar{\psi}^{\bar{\alpha}\bar{A}}(\bar{z})=\sum_{r\in\mathbb{Z}+1/2}\frac{\bar{\psi}^{\bar{\alpha}\bar{A}}_r}{\bar{z}^{r+1/2}}\,.
\end{equation} 
The modes
transform under the $\mathcal{N}=4$ SCA as:
\begin{equation}\label{eq:N4psi1}
\begin{aligned}
[L_m\,,\, \psi^{\alpha A}_s]&=(-\tfrac{1}{2}s-m) \psi^{\alpha A}_{m+s} \\
[G^{\alpha_1\beta}_r\,,\, \psi^{\alpha_2 A}_s]&=\epsilon^{\alpha_1\alpha_2} \mathtt{a}^{\beta A}_{r+s} \\
[J^{a}_m\,,\, \psi^{\alpha A}_s]&=
D^{(1/2)}(t^a)^{\alpha}{}_{\alpha'} \psi^{\alpha' A}_{m+s} \,,
\end{aligned}
\end{equation}
and similarly for the barred fields in the right-moving sector. 
\end{itemize}

The commutation relations among these fields are
\begin{equation}
\begin{aligned}
[\mathtt{a}^{\beta_1 A_1}_m\,,\, \mathtt{a}^{\beta_2 A_2}_n ]&=\epsilon^{\beta_1\beta_2}\epsilon^{ A_1 A_2} m \delta_{m+n} 
\,, 
\\
\{\psi^{\alpha_1 A_1}_m\,,\, \psi^{\alpha_2 A_2}_n \}&=\epsilon^{\alpha_1\alpha_2}\epsilon^{A_1 A_2} \delta_{m+n}
\,, 
\\
[\mathtt{a}^{\beta A_1}_m\,,\, \psi^{\alpha A_2}_n ]&=0 \,.   
\end{aligned}
\end{equation}
and similarly for the barred fields.

\subsubsection*{$\mathcal{N}=(4,4)$ superconformal algebra in terms of the $\mathbb{T}^4$ fields}

The $\mathbb{T}^4$ fields generate the left-moving $\mathcal{N}=4$ superconformal algebra via\footnote{Note that there are some freedom in expressing the $\mathcal{N}=4$ SCA fields in terms of the $\mathbb{T}^4$ fields. 
We choose the basis that is kept invariant by the diagonal $\mathfrak{su}(2)$ of the two second (auxilary) $\mathfrak{su}(2)$'s of the bosons and the fermions.}
\begin{equation}\label{eq:N4SCAT4left}
\begin{aligned}
T&=\tfrac{1}{2}\epsilon_{\beta_1\beta_2}\epsilon_{ A_1 A_2}:\partial X^{\beta_1 A_1}\partial X^{\beta_2 A_2}: 
+\tfrac{1}{2}\epsilon_{\alpha_1\alpha_2}\epsilon_{A_1 A_2}:(\partial\psi^{\alpha_1 A_1}) \psi^{\alpha_2 A_2}:\\
G^{\alpha\beta}&= \epsilon_{A_1 A_2}:\psi^{\alpha A_1}\partial X^{\beta A_2}:\\
J^{a}&=\tfrac{1}{2}\epsilon_{\alpha_3\alpha_1}\epsilon_{A_1 A_2}D^{(1/2)}(t^a)^{\alpha_3}{}_{\alpha_2}:\psi^{\alpha_1A_1}\psi^{\alpha_2 A_2}:
\end{aligned}
\end{equation}
where $\epsilon_{+-} = -\epsilon_{-+} =+1$, and the boson and fermions are contracted using the diagnoal $\mathfrak{su}(2)$ of their two second $\mathfrak{su}(2)$'s that commute with the $\mathcal{N}=4$ SCA.
Similarly for the right-moving sector:
\begin{equation}\label{eq:N4SCAT4right}
\begin{aligned}
\bar{T}&=\tfrac{1}{2}\epsilon_{\bar{\beta}_1\bar{\beta}_2}\epsilon_{\bar{A}_1 \bar{A}_2}:\bar{\partial} X^{\bar{\beta}_1 \bar{A}_1}\bar{\partial} X^{\bar{\beta}_2\bar{A}_2}: 
+\tfrac{1}{2}\epsilon_{\bar{\alpha}_1\bar{\alpha}_2}\epsilon_{\bar{A}_1\bar{A}_2}:(\bar{\partial}\bar{\psi}^{\bar{\alpha}_1\bar{A}_1}) \bar{\psi}^{\bar{\alpha}_2\bar{A}_2}:\\
\bar{G}^{\bar{\alpha}\bar{\beta}}&= \epsilon_{\bar{A}_1\bar{A}_2}:\bar{\psi}^{\bar{\alpha}\bar{A}_1}\bar{\partial} X^{\bar{\beta}\bar{A}_2}:\\
\bar{J}^{a}&=\tfrac{1}{2}\epsilon_{\bar{\alpha}_3\bar{\alpha}_1}\epsilon_{\bar{A}_1\bar{A}_2}D^{(1/2)}(t^a)^{\bar{\alpha}_3}{}_{\bar{\alpha}_2}:\bar{\psi}^{\bar{\alpha}_1\bar{A}_1}\bar{\psi}^{\bar{\alpha}_2\bar{A}_2}:
\end{aligned}
\end{equation}

\section{Grand-canonical residues} \label{app: relevant formulas}

Here we provide explicit expressions of the grand-canonical residues. These are identified in the main text with the supersymmetric one-loop partition functions of the IIB theory on $\orbifoldM$ backgrounds with NS-NS boundary conditions around the spatial circle.

There are four classes of contributions $(0,0)$, $(2,0)$, $(0,2)$, $(2,2)$, corresponding to different spectrally-flowed sectors, to the full left-right chiral primary partition function $Z_N$ at finite $N$. For $\Mfour = \Kthree$, we have
\ie \label{eq: Zhat K3 app}
&\widehat{Z}_{k}^{(0/2,0/2)} = \sum_{m=0}^{k-1}  \frac{1}{k} e^{-2 \pi i N \frac{m}{k}} y^{N \left( 1 \mp\frac{1}{k} \right)} \bar{y}^{N \left(1 \mp\frac{1}{k} \right)} \prod_{\substack{n=1 \\ n \neq k}}^\infty \frac{1}{(1-e^{2 \pi i n \frac{m}{k}} y^{\pm \frac{n}{k} \mp 1} \bar{y}^{\pm \frac{n}{k} \mp 1})} \prod_{n=1}^\infty \bigg[ \frac{1}{(1-e^{2 \pi i n \frac{m}{k}} y^{\pm\frac{n}{k} \pm 1} \bar{y}^{\pm \frac{n}{k} \mp 1})} \\
&\qquad \times \frac{1}{ (1-e^{2 \pi i n \frac{m}{k}} y^{\pm \frac{n}{k} \mp 1} \bar{y}^{\pm \frac{n}{k} \pm 1}) (1-e^{2 \pi i n \frac{m}{k}} y^{\pm \frac{n}{k} \pm 1} \bar{y}^{\pm \frac{n}{k} \pm 1})(1-e^{2 \pi i n \frac{m}{k}} y^{\pm \frac{n}{k}} \bar{y}^{\pm \frac{n}{k}})^{20}} \bigg].
\fe For $\Mfour = \Tfour$, we have
\ie \label{eq: Zhat T4 app}
&\Zhat_{k}^{(0/2,0/2)} = \sum_{m=0}^{k-1}  \frac{1}{k} e^{-2 \pi i N \frac{m}{k}} y^{N \left(1 \mp\frac{1}{k} \right)} \bar{y}^{N \left(1 \mp\frac{1}{k} \right)} \prod_{\substack{n=1 \\ n \neq k}}^\infty \frac{1}{(1-e^{2 \pi i n \frac{m}{k}} y^{\pm\frac{n}{k}\mp1} \bar{y}^{\pm\frac{n}{k}\mp1})} \prod_{n=1}^\infty \bigg[ \frac{(1 + \varepsilon e^{2 \pi i n \frac{m}{k}} y^{\pm\frac{n}{k}} \bar{y}^{\pm\frac{n}{k}\mp1})^2}{(1-e^{2 \pi i n \frac{m}{k}} y^{\pm\frac{n}{k}\pm1} \bar{y}^{\pm\frac{n}{k}\mp1})} \\
&\qquad \times \frac{(1 + \varepsilon e^{2 \pi i n \frac{m}{k}} y^{\pm\frac{n}{k}\mp1} \bar{y}^{\pm\frac{n}{k}})^2 (1 + \varepsilon e^{2 \pi i n \frac{m}{k}} y^{\pm\frac{n}{k}} \bar{y}^{\pm\frac{n}{k}\pm1})^2 (1 + \varepsilon e^{2 \pi i n \frac{m}{k}} y^{\pm\frac{n}{k}\pm1} \bar{y}^{\pm\frac{n}{k}})^2}{(1-e^{2 \pi i n \frac{m}{k}} y^{\pm\frac{n}{k}\mp1} \bar{y}^{\pm\frac{n}{k}\pm1})(1-e^{2 \pi i n \frac{m}{k}} y^{\pm\frac{n}{k}\pm1} \bar{y}^{\pm\frac{n}{k}\pm1})(1-e^{2 \pi i n \frac{m}{k}} y^{\pm\frac{n}{k}} \bar{y}^{\pm\frac{n}{k}})^{4}} \bigg],
\fe
where $\varepsilon = \pm 1$ for the partition function and index, respectively. The $0/2$ labels in the left- and right-moving sectors correspond to upper/lower signs in the powers of $y$ and $\yb$, respectively.

The residues in the R sector are related to their counterparts in the NS sector only by the simple replacement
\be
{\rm NS}: \ y^{N\left(1 \mp \frac{1}{k}\right)} \bar{y}^{N\left(1 \mp \frac{1}{k}\right)} \ \to \ {\rm R}: \ y^{\mp \frac{N}{k}} \bar{y}^{\mp \frac{N}{k}}
\ee
in the classical weight associated to $\orbifold$ and its BPS spectral flows. The non-trivial part of the residue that we identified as the one-loop determinant remains unchanged as we flow from the NS sector to the R sector.

\section{Further checks} \label{app: further checks}

Here we provide further checks of our proposal in different Stokes sectors $\cS_{y,\yb}$.

For generic $|y|$ and $|\yb|$, the microcanonical coefficients of $\Zhat_k^\mu$ are defined via the inverse Laplace transform
\be
(d_{n m})_k^\mu = \oint_{|y|}\frac{dy}{2 \pi i y} \oint_{|\yb|} \frac{d\yb}{2 \pi i \yb} \, y^{-n} \, \yb^{-m} \, \Zhat_k^\mu (y,\yb).
\ee
The coefficients $(d_{n m})_k^\mu$ can jump discontinuously as the contour radii $|y|,|\yb|$ cross a pole of $\Zhat_k^\mu$. Our proposal is that the \textit{sum} of the microcanonical contributions $(d_{n m})_k^\mu$ at a given charge $n,m$
\be \label{eq: microcanonical proposal}
D_{n m}^N = \sum_{\mu \in \cS_{y,\yb}} \sum_{k=1}^\infty (d_{n m})_k^\mu,
\ee
remains invariant within any region enumerated in Section \ref{subsec: sum proposal}, where $Z_N(y,\yb) = \sum_{n,m} D_{n m}^N y^n \yb^m$ is a polynomial, i.e. the coefficients $D_{n m}^N$ are unambiguous.

We provide further checks of our proposal in the following regions of the $y,\yb$ fugacity space: (1) the region $|\yb|<|y|\leq 1$ where $\cS_{y,\yb} = \{ (0,0), (2,0) \}$ and we have
\be \label{eq: example 2}
Z_N(y,\yb) = \lim_{K\to \infty} \ \sum_{k=1}^K \left[ \Zhat_{k}^{(0,0)}(y,\bar{y}) + \Zhat_{k-1}^{(2,0)}(y,\bar{y}) \right]
\ee
and (2) the region $|y^{-1}|<|\yb|\leq1$ where $\cS_{y,\yb} = \{ (2,0), (2,2) \}$ and we have
\be \label{eq: example 3}
Z_N(y,\yb) = \lim_{K\to \infty} \ \sum_{k=1}^K \left[ \Zhat_{k}^{(2,0)}(y,\bar{y}) + \Zhat_{k-1}^{(2,2)}(y,\bar{y}) \right]
\ee
where we defined $\Zhat_0^{\mu} \equiv 0$.

Due to the pole structure of $\Zhat_k^\mu$, the microcanonical coefficients $(d_{n m}^{1})_k^\mu$ that one finds by series expanding first in $\yb$ and then in $y$ implicitly amounts to extracting the coefficients in the subregion $|\yb|<|y|=1^-$ of region 1 via
\be
(d_{n m}^{1})_k^\mu = \oint_{|y|=1^-}\frac{dy}{2 \pi i y} \oint_{|\yb|<|y|} \frac{d\yb}{2 \pi i \yb} \, y^{-n} \, \yb^{-m} \, \Zhat_k^\mu (y,\yb).
\ee
Also, the coefficients $(d_{n m}^{2})_k^\mu$ that one finds by series expanding first in $y^{-1}$ and then in $\yb$ implicitly amounts to extracting the coefficients in the subregion $|y^{-1}| < |\yb| = 1^-$ of region 2 via
\be
(d_{n m}^{2})_k^\mu = \oint_{|\yb| = 1^-} \frac{d\yb}{2 \pi i \yb} \oint_{|y^{-1}| < |\yb|} \frac{dy}{2 \pi i y} \, y^{-n} \, \yb^{-m} \, \Zhat_k^\mu (y,\yb).
\ee
The checks below are performed in terms of the series coefficients $(d_{n m}^{1})_k^\mu$ and $(d_{n m}^{2})_k^\mu$ in the respective subregions, but our proposal \eqref{eq: microcanonical proposal} can be checked also with general contours $|y|,|\yb|$.

The BPS partition function of $\symK$ at $N=1$ is
\be
Z_{N=1} = 1 + y^2 + 20 y \yb + \yb^2 + y^2 \yb^2.
\ee
In region 1 for increasing $K$, we find
{\fontsize{8}{6}\selectfont
\ie
K=1:& \quad \left(1+y^2+y^4+\ccdots\right)+ \left(21 y+22y^3+\ccdots \right)\yb + \left(1+254 y^2+276 y^4+\ccdots\right) \yb^2 \\
&\quad + \left(22 y+2278 y^3+ \ccdots \right) \yb^3 + \left(1+276 y^2+16744 y^4+\ccdots\right) \yb^4 +\cdots \\
K=2:& \quad \left(1+y^2+0 y^4+ \ccdots\right)+ \left(20 y+0 y^3+\ccdots\right)\yb + \left(1+2 y^2+23 y^4+\ccdots\right) \yb^2 \\
& \quad + \left(\frac{1}{y}+298 y+2830 y^3+\ccdots\right) \yb^3 + \left(\frac{23}{y^2}+2852+125604 y^2+163876y^4+\ccdots\right) \yb^4 + \cdots \\
K=3:& \quad \left(1+y^2+0 y^4+\ccdots\right)+ \left(20 y+0 y^3+\ccdots\right) \yb + \left(1+y^2+0 y^4+\ccdots\right) \yb^2 \\
& \quad + \left(0 y^1 + y^3+\ccdots\right) \yb^3 + \left(\frac{1}{y^2}+300+3151 y^2+22427 y^4+\ccdots\right) \yb^4 +\cdots \\
K=4:& \quad \left(1+y^2+0 y^4+\ccdots\right)+ \left(20 y+0 y^3+\ccdots\right) \yb + \left(1+y^2+0 y^4+\ccdots\right) \yb^2 \\
&\quad + \left(0 y^1 + 0y^3+\ccdots\right) \yb^3 + \left(0 + 0 y^2 + y^4+\ccdots\right) \yb^4 + \cdots \\
K=5:& \quad \left(1+y^2+0 y^4+\ccdots\right)+ \left(20 y+0 y^3+\ccdots\right) \yb + \left(1+y^2+0 y^4+\ccdots\right) \yb^2 \\
&\quad + \left(0 y^1 + 0y^3+\ccdots\right) \yb^3 + \left(0 + 0 y^2 + 0y^4+\ccdots\right) \yb^4 + \cdots. \\
\fe}
In region 2 for increasing $K$, we find
{\fontsize{8}{6}\selectfont
\ie
K=1:& \quad \left(1+\yb^2+\yb^4+\ccdots\right)y^2+ \left(21 \yb+22 \yb^3+\ccdots\right)y +\left(1+254 \yb^2+276 \yb^4+\ccdots\right) \\
& \quad + \left( 22 \yb+2278 \yb^3+\ccdots \right) \frac{1}{y}+ \left( 1+276 \yb^2+16744 \yb^4+\ccdots \right) \frac{1}{y^2}+\cdots \\
K=2:& \quad \left(1+\yb^2+0 \yb^4 + \ccdots\right) y^2+ \left(20 \yb+0 \yb^3 + \ccdots\right)y +\left(1+2 \yb^2+23\yb^4+\ccdots\right) \\
& \quad +\left( \frac{1}{\yb}+298 \yb+2830\yb^3+\ccdots \right) \frac{1}{y}+\left(\frac{23}{\yb^2}+2852+125604 \yb^2+163876\yb^4+\ccdots \right)\frac{1}{y^2}+ \cdots \\
K=3:& \quad \left(1+\yb^2+ 0 \yb^4 + \ccdots\right)y^2 + \left(20\yb+0 \yb^3 + \ccdots \right)y +\left(1+\yb^2+0 \yb^4 + \ccdots \right) \\
&\quad +\left(0 \yb +  \yb^3+\ccdots \right)\frac{1}{y}+\left( \frac{1}{\yb^2}+300+3151 \yb^2+22427 \yb^4+\ccdots \right)\frac{1}{y^2}+\cdots \\
K=4:& \quad \left(1+\yb^2+0 \yb^4 +\ccdots\right)y^2 + \left(20\yb+0 \yb^3 +\ccdots\right)y +\left(1+\yb^2+0 \yb^4 +\ccdots\right) \\
& \quad +\left(0 \yb +  0 \yb^3+\ccdots \right)\frac{1}{y} +\left(0 + 0 \yb^2 +\yb^4+\ccdots \right)\frac{1}{y^2}+ \cdots \\
K=5:& \quad \left(1+\yb^2+0 \yb^4 +\ccdots\right)y^2 + \left(20\yb+0 \yb^3 +\ccdots\right)y +\left(1+\yb^2+0 \yb^4 +\ccdots\right) \\
& \quad +\left(0 \yb +  0 \yb^3+\ccdots \right)\frac{1}{y} +\left(0 + 0 \yb^2 + 0\yb^4+\ccdots \right)\frac{1}{y^2}+ \cdots.
\fe}

The BPS partition function of $\symK$ at $N=2$ is
\be
Z_{N=2} = 1 + y^2 + 21 y \yb + \yb^2 + y^4 + 21 y^3 \yb + 232 y^2 \yb^2 + 21 y \yb^3 + \yb^4 + y^4 \yb^2 + 21 y^3 \yb^3 + y^2 \yb^4 + y^4 \yb^4.
\ee
In region 1 for increasing $K$, we find
{\fontsize{8}{6}\selectfont
\ie
K=1:& \quad \left(1+y^2+y^4+\ccdots\right)+ \left(21 y+22 y^3+22 y^5+\ccdots\right)\yb+\left(1+254 y^2+276 y^4+\ccdots\right)\yb^2 \\
&\quad + \left(22 y+2278 y^3+2553 y^5+\ccdots\right)\yb^3+ \left(1+276 y^2+16744 y^4+\ccdots\right) \yb^4 \\
&\quad + \left(22 y+2553 y^3+106306 y^5+\ccdots\right)\yb^5+\cdots \\
K=2:& \left(1+y^2+y^4+\ccdots\right)+ \left(21 y+21 y^3+0 y^5 + \ccdots\right)\yb+ \left(1+232y^2+y^4+\ccdots\right)\yb^2 \\
& \quad + \left(22 y+45 y^3+299 y^5+\ccdots\right)\yb^3 + \left(24+2830 y^2+21851 y^4+\ccdots\right) \yb^4 \\
& \quad + \left(\frac{1}{y^3}+\frac{300}{y}+22126 y+727605 y^3+975616 y^5+\ccdots\right)\yb^5 +\cdots \\
K=3:& \quad \left(1+y^2+y^4+\ccdots\right)+ \left(21 y+21 y^3+ 0 y^5 + \ccdots\right)\yb + \left(1+232 y^2+y^4+\ccdots\right)\yb^2 \\
& \quad + \left(21 y+21 y^3+0 y^5+ \ccdots\right)\yb^3 + \left(1+2 y^2+25 y^4+\ccdots\right)\yb^4 \\
& \quad + \left(\frac{24}{y}+2875 y+24978 y^3+150306 y^5+\ccdots\right)\yb^5 + \cdots \\
K=4:& \quad \left(1+y^2+y^4+\ccdots\right)+ \left(21 y+21 y^3+ 0 y^5 + \ccdots\right)\yb+ \left(1+232 y^2+y^4+\ccdots\right)\yb^2 \\
& \quad + \left(21 y+21 y^3+ 0 y^5 +\ccdots\right)\yb^3 + \left(1+y^2+y^4+\ccdots\right)\yb^4 \\
& \quad + \left(0 y + y^3+24 y^5+\ccdots\right)\yb^5 +\cdots \\
K=5:& \quad \left(1+y^2+y^4+\ccdots\right)+ \left(21 y+21 y^3+ 0 y^5 + \ccdots\right)\yb+ \left(1+232 y^2+y^4+\ccdots\right)\yb^2 \\
& \quad + \left(21 y+21 y^3+ 0 y^5 +\ccdots\right)\yb^3 + \left(1+y^2+y^4+\ccdots\right)\yb^4 \\
& \quad + \left(0 y + 0 y^3+0 y^5+\ccdots\right)\yb^5 +\cdots.
\fe}
In region 2 for increasing $K$, we find
{\fontsize{8}{6}\selectfont
\ie
K=1:& \quad \left(1+\yb^2+\yb^4+\ccdots\right)y^4+ \left(21 \yb+22 \yb^3+22 \yb^5+\ccdots\right)y^3+\left(1+254 \yb^2+276 \yb^4+\ccdots\right)y^2 \\
&\quad + \left(22 \yb+2278 \yb^3+2553 \yb^5+\ccdots\right)y+ \left(1+276 \yb^2+16744 \yb^4+\ccdots\right) \\
&\quad + \left(22 \yb+2553 \yb^3+106306 \yb^5+\ccdots\right)\frac{1}{y}+\cdots \\
K=2:& \left(1+\yb^2+\yb^4+\ccdots\right) y^4+ \left(21 \yb+21 \yb^3+0 \yb^5 + \ccdots\right) y^3+ \left(1+232\yb^2+\yb^4+\ccdots\right) y^2 \\
& \quad + \left(22 \yb+45 \yb^3+299 \yb^5+\ccdots\right) y + \left(24+2830 \yb^2+21851 \yb^4+\ccdots\right) \\
& \quad + \left(\frac{1}{\yb^3}+\frac{300}{\yb}+22126 \yb+727605 \yb^3+975616 \yb^5+\ccdots\right) \frac{1}{y} +\cdots \\
K=3:& \quad \left(1+\yb^2+\yb^4+\ccdots\right) y^4+ \left(21 \yb+21 \yb^3+ 0 \yb^5 + \ccdots\right) y^3 + \left(1+232 \yb^2+\yb^4+\ccdots\right) y^2 \\
& \quad + \left(21 \yb+21 \yb^3+0 \yb^5+ \ccdots\right) y + \left(1+2 \yb^2+25 \yb^4+\ccdots\right) \\
& \quad + \left(\frac{24}{\yb}+2875 \yb+24978 \yb^3+150306 \yb^5+\ccdots\right) \frac{1}{y} + \cdots \\
K=4:& \quad \left(1+\yb^2+\yb^4+\ccdots\right)y^4 + \left(21 \yb+21 \yb^3+ 0 \yb^5 + \ccdots\right) y^3+ \left(1+232 \yb^2+\yb^4+\ccdots\right) y^2 \\
& \quad + \left(21 \yb+21 \yb^3+ 0 \yb^5 +\ccdots\right) y + \left(1+\yb^2+\yb^4+\ccdots\right) \\
& \quad + \left(0 \yb + \yb^3+24 \yb^5+\ccdots\right) \frac{1}{y} +\cdots \\
K=5:& \quad \left(1+\yb^2+\yb^4+\ccdots\right) y^4+ \left(21 \yb+21 \yb^3+ 0 \yb^5 + \ccdots\right) y^3+ \left(1+232 \yb^2+\yb^4+\ccdots\right) y^2 \\
& \quad + \left(21 \yb+21 \yb^3+ 0 \yb^5 +\ccdots\right) y + \left(1+\yb^2+\yb^4+\ccdots\right) \\
& \quad + \left(0 \yb + 0 \yb^3+0 \yb^5+\ccdots\right) \frac{1}{y} +\cdots.
\fe}

The BPS partition function of $\symT$ at $N=1$ is
\be
Z_{N=1} = 1 + 2 \varepsilon y + 2 \varepsilon \yb + y^2 + 4 y \yb + \yb^2 + 2 \varepsilon y^2 \yb + 2 \varepsilon y \yb^2 + y^2 \yb^2.
\ee
In region 1 for increasing $K$, we find, up to and including terms of order $\yb^2$, $y^2$, and $\varepsilon^3$,
{\fontsize{8}{6}\selectfont
\ie
K=1:& \quad \left(1+\left(2 \varepsilon \right) y+ \left(1+\varepsilon ^2\right) y^2 +\ccdots\right) +\left(\left(2 \varepsilon \right) + \left(5+4 \varepsilon ^2\right) y + \left(16 \varepsilon +2 \varepsilon^3 \right) y^2 + \ccdots \right) \yb \\
&\quad + \left(\left(1+\varepsilon ^2\right)+\left(16 \varepsilon+2 \varepsilon ^3\right) y+\left(22+38 \varepsilon^2 +\ccdots\right)y^2+\ccdots\right)\yb^2 + \cdots \\
2:& \quad \left(1+\left(2 \varepsilon \right)y + y^2 +\ccdots \right) + \left(\left(2\varepsilon\right) +\left(4+\varepsilon ^2+ \ccdots\right) y+\left(2 \varepsilon +18 \varepsilon^3 + \ccdots\right) y^2+\ccdots\right)\yb \\
& \quad + \left(\left(4 \varepsilon ^3 + \ccdots \right)\frac{1}{y}+\left(1+15 \varepsilon ^2+\ccdots \right)+\left(18 \varepsilon +232 \varepsilon^3+ \ccdots \right) y+ \left(2+296 \varepsilon ^2+ \ccdots \right)y^2 + \ccdots \right)\yb^2 + \cdots \\
3:& \quad \left(1+\left(2 \varepsilon \right) y + y^2 +\ccdots \right)+  \left(\left(2\varepsilon \right)+\left(4+ \ccdots \right) y +\left(2 \varepsilon +\ccdots \right) y^2+\ccdots \right)\yb \\
&\quad + \left(\left( 2 \varepsilon ^3+\ccdots \right)\frac{1}{y} + \left(1+\varepsilon ^2+\ccdots \right)+\left(2 \varepsilon +110 \varepsilon ^3+\ccdots\right) y +\left(1+12 \varepsilon ^2+\ccdots \right)y^2+\ccdots\right)\yb^2 + \cdots \\
4: & \quad \left(1+\left(2 \varepsilon \right) y + y^2 +\ccdots \right)+  \left(\left(2\varepsilon \right)+\left(4+ \ccdots \right) y +\left(2 \varepsilon +\ccdots \right) y^2+\ccdots \right)\yb \\
&\quad + \left( \left(1+\ccdots \right)+\left(2 \varepsilon + \ccdots\right) y +\left(1+\ccdots \right)y^2+\ccdots\right)\yb^2 + \cdots.
\fe}
In region 2 for increasing $K$, we find, up to and including terms of order $(y^{-1})^0$, $\yb^2$, and $\varepsilon^3$,
{\fontsize{8}{6}\selectfont
\ie
K=1:& \quad \left(1+\left(2 \varepsilon \right) \yb+ \left(1+\varepsilon ^2\right) \yb^2 +\ccdots\right) y^2 +\left(\left(2 \varepsilon \right) + \left(5+4 \varepsilon ^2\right) \yb + \left(16 \varepsilon +2 \varepsilon^3 \right) \yb^2 + \ccdots \right) y \\
&\quad + \left(\left(1+\varepsilon ^2\right)+\left(16 \varepsilon+2 \varepsilon ^3\right) \yb+\left(22+38 \varepsilon^2 +\ccdots\right)\yb^2+\ccdots\right) + \cdots \\
2:& \quad \left(1+\left(2 \varepsilon \right)\yb + \yb^2 +\ccdots \right) y^2 + \left(\left(2\varepsilon\right) +\left(4+\varepsilon ^2+ \ccdots\right) \yb+\left(2 \varepsilon +18 \varepsilon^3 + \ccdots\right) \yb^2+\ccdots\right) y \\
& \quad + \left(\left(4 \varepsilon ^3 + \ccdots \right)\frac{1}{\yb}+\left(1+15 \varepsilon ^2+\ccdots \right)+\left(18 \varepsilon +232 \varepsilon^3+ \ccdots \right) \yb+ \left(2+296 \varepsilon ^2+ \ccdots \right)\yb^2 + \ccdots \right) + \cdots \\
3:& \quad \left(1+\left(2 \varepsilon \right) \yb + \yb^2 +\ccdots \right) y^2 +  \left(\left(2\varepsilon \right)+\left(4+ \ccdots \right) \yb +\left(2 \varepsilon +\ccdots \right) \yb^2+\ccdots \right) y \\
&\quad + \left(\left( 2 \varepsilon ^3+\ccdots \right)\frac{1}{\yb} + \left(1+\varepsilon ^2+\ccdots \right)+\left(2 \varepsilon +110 \varepsilon ^3+\ccdots\right) \yb +\left(1+12 \varepsilon ^2+\ccdots \right)\yb^2+\ccdots\right) + \cdots \\
4: & \quad \left(1+\left(2 \varepsilon \right) \yb + \yb^2 +\ccdots \right) y^2 + \left(\left(2\varepsilon \right)+\left(4+ \ccdots \right) \yb +\left(2 \varepsilon +\ccdots \right) \yb^2+\ccdots \right) y \\
&\quad + \left( \left(1+\ccdots \right)+\left(2 \varepsilon + \ccdots\right) \yb +\left(1+\ccdots \right)\yb^2+\ccdots\right) + \cdots.
\fe}

\bibliographystyle{utphys}
\bibliography{references}

\end{document}